\pdfoutput=1
\documentclass[10pt,palatino,code,picins,kaukecopyright,openright,woolshort,dropcaps,verbatim,index,euler]{styles/woosterthesis}
\ifxetex%
	\defaultfontfeatures{Mapping=tex-text}%
		\setromanfont[Swashes=Contextual,Numbers=OldStyle,BoldFont={Adobe Garamond Pro Semibold}]{Adobe Garamond Pro}
\fi
\newcommand{\xt}{\ifthenelse{\boolean{xetex}}{\XeTeX\ }{XeTeX} }

\newcounter{unnumft}
\newsavebox{\fmbox}

 \definecolor{MyBlue}{cmyk}{.13,.0,.0,0}
 \definecolor{MyPurple}{cmyk}{0.07,0.09,0.03,0}
  \definecolor{MyOrange}{cmyk}{0.08,0.09,0.27,0}
    \definecolor{MyGreen}{cmyk}{0.12,0.0,0.16,0}
\newenvironment{boxedaxm}[2]
     {\vspace{0.5 cm} \begin{lrbox}{\fmbox} \begin{lrbox}{\fmbox} \begin{minipage}{0.9 \linewidth} \begin{axm}[#1] \label{#2} \begin{singlespace}}
     {\end{singlespace} \vspace{0.3 cm} \end{axm}\end{minipage} \end{lrbox}%
   \fcolorbox{black}{MyOrange}{\usebox{\fmbox}}\end{lrbox}\fbox{\usebox{\fmbox}} \vspace{0.5 cm}  \linebreak}

\newenvironment{boxeddefn}[2]
     {\vspace{0.5 cm} \begin{lrbox}{\fmbox} \begin{lrbox}{\fmbox}\begin{minipage}{0.9 \linewidth} \begin{defn}[#1] \label{#2} \begin{singlespace}}
     {\end{singlespace} \vspace{0.3 cm} \end{defn} \end{minipage} \end{lrbox}\fcolorbox{black}{MyGreen}{\usebox{\fmbox}}\end{lrbox}\fbox{\usebox{\fmbox}} \vspace{0.5 cm}  \linebreak}
     
     \newenvironment{boxedeqn}[1]
     {\begin{flushleft}\begin{lrbox}{\fmbox}\begin{minipage}{1.0 \linewidth} \begin{equation}\label{#1} }
     {\vspace{0.3 cm} \end{equation} \end{minipage}  \end{lrbox}\colorbox{MyBlue}{\usebox{\fmbox}}\linebreak\end{flushleft}}

     \newenvironment{boxedthm}[2]
     {\vspace{0.5 cm} \begin{lrbox}{\fmbox} \begin{lrbox}{\fmbox}\begin{minipage}{0.9 \linewidth} \begin{thm2}[#1] \label{#2} \begin{singlespace}}
     {\end{singlespace} \vspace{0.3 cm} \end{thm2} \end{minipage}\end{lrbox}\fcolorbox{black}{MyPurple}{\usebox{\fmbox}}\end{lrbox}\fbox{\usebox{\fmbox}} \vspace{0.5 cm}  \linebreak}
     
\graphicspath{{./figures/}}
\newcommand{\mathsym}[1]{{}}

\newtheoremstyle{scthm}{\topsep}{\topsep}{\itshape}{}{\bfseries}{.}{ }{}
\theoremstyle{break}
\newtheorem{thm}{Theorem}[chapter]

\theoremstyle{definition}
\newtheorem{defn}[thm]{Definition}
\newtheorem{thm2}[thm]{Theorem}
\theoremstyle{definition}
\newtheorem{axm}[thm]{Axiom}
\theoremstyle{remark}

\theoremstyle{plain}

\title{Foundations of Quantum Decoherence}
\thesistype{Physics and Mathematics Independent Study Thesis} 
\author{John Gamble}
\degreetoobtain{B.A. in Physics and Mathematics}
\presentschool{The College of Wooster}
\academicprogram{Departments of Mathematics and Physics}
\gradyear{2008}
\advisor{Dr. John Lindner, \\The Moore Professor of Astronomy \text{\&} Professor of Physics}
\secondadvisor{Dr. Derek Newland, \\ Assistant Professor of Mathematics}
\copyrighted   
\makeindex 

\begin{document}


\frontmatter
\maketitle
\ClearShipoutPicture
\disscopyright 

%
%

I gratefully acknowledge the loving help and support of my parents, John and Clare Gamble, and of my fianc\'ee, Katherine Kelley. I extend sincere thanks to my advisors, John Lindner and Derek Newland, for their long hours and dedication to this project. I also thank Jon Breitenbucher for painstakingly assembling and maintaining this \LaTeX \\ template, which made the writing process significantly more enjoyable than it would have been otherwise. Finally, I am grateful to The Gallows program for providing me an environment in which I could grow, learn, and succeed. \pagebreak
%
%

\begin{abstract}
The conventional interpretation of quantum mechanics, though it permits a correspondence to classical physics, leaves the exact mechanism of transition unclear. Though this was only of philosophical importance throughout the twentieth century, over the past decade new technological developments, such as quantum computing, require a more thorough understanding of not just the \textit{result} of quantum emergence, but also its \textit{mechanism}. Quantum decoherence theory is the model that developed out of necessity to deal with the quantum-classical transition explicitly, and without external observers. In this thesis, we present a self-contained and rigorously argued full derivation of the master equation for quantum Brownian motion, one of the key results in quantum decoherence theory. We accomplish this from a foundational perspective, only assuming a few basic axioms of quantum mechanics and deriving their consequences. We then consider a physical example of the master equation and show that quantum decoherence successfully represents the transition from a quantum to classical system.
\end{abstract}

\if@xetex
	\cleardoublepage
	\phantomsection
	\addcontentsline{toc}{chapter}{Contents}
\else
	\ifpdf
		\cleardoublepage
		\phantomsection
		\addcontentsline{toc}{chapter}{Contents}
	\else
		\cleardoublepage
		\addcontentsline{toc}{chapter}{Contents}
	\fi
\fi

\tableofcontents
 \listoffigures 


\chapter{Preface}\label{chap:intro}
 \addcontentsline{toc}{chapter}{Preface}
\lettrine[lines=2, lhang=0.33, loversize=0.1]{T}his thesis is designed to serve a dual purpose. First, it is a stand-alone treatment of contemporary decoherence theory, accomplishing this mostly within a rigorous framework more detailed than is used in typical undergraduate quantum mechanics courses. It assumes no prior knowledge of quantum mechanics, although a basic understanding obtained through a standard introductory quantum mechanics or modern physics course would be helpful for depth of meaning. Although the mathematics used is introduced thoroughly in chapter \ref{chap:math_background}, the linear algebra can get quite complicated. Readers who have not had a formal course in linear algebra would benefit from having ref. \cite{poole} on-hand during some components, especially chapters \ref{chap:quantum_formal} and \ref{chap:dynamics}. The bulk of the work specifically related to decoherence is found in the last three chapters, and readers familiar with quantum mechanics desiring a better grasp of decoherence theory should proceed to the discussion of quantum mechanics in phase-space, found in chapter \ref{chap:wigner}.

Second, this thesis is an introduction to the rigorous study of the foundations of quantum mechanics, and is again stand-alone in this respect. It develops the bulk of quantum mechanics from several standard postulates and the invariance of physics under the Galilei group\index{Group!Galilei} of transformations, outlined in sections \ref{sec:posts} and \ref{sec:galgroup}, respectively. Readers interested in this part of the thesis should study the first three chapters, where many fundamental results of quantum mechanics are developed. We now begin with a motivating discussion of quantum decoherence.

One of the fundamental issues in physics today is the emergence of the familiar macroscopic physics that governs everyday objects from the strange, underlying microscopic laws for the motion of atoms and molecules. This collection of laws governing small bodies is called quantum mechanics, and operates entirely differently than classical Newtonian physics. However, since all macroscopic objects are made from microscopic particles, which obey quantum mechanics, there should be some way to link the two worlds: the macro and the micro. The conventional interpretation of quantum mechanics answers questions about the transition from classical to quantum mechanics, known as quantum emergence\index{Quantum Emergence}, through a special \textit{measurement}\index{Measurement} process, which is distinct from the other rules of quantum mechanics \cite{griffiths}.\footnote{In fact, the motion of a system not being measured is considered \textit{unitary}, and hence reversible, while the measurement process is conventionally considered discontinuous, and hence irreversible. So, not only are they treated separately, but they are considered fundamentally different processes!}

However, when this measurement concept is used, problems arise. The most famous of these problems is known as Schr\"odinger's cat\index{Schr\"odinger's cat}, which asks about the nature of measurement through a paradox \cite{omnes}. The problem creates ambiguity about \begin{enumerate} \item when a measurement occurs, and  \item who (or what) performs it.\end{enumerate} When all is said and done, the conventional interpretation leaves a bitter taste in the mouths of many physicists; what they want is a theory of quantum measurement that does not function due to subjectively defined observation. If no external observers are permitted, how can classical mechanics ever emerge from quantum mechanics? The answer is that complex systems, in essence, measure themselves, which leads us to decoherence.

\section{Decoherence and the Measurement Problem}
\begin{figure}[bt] 
\begin{center} 
\includegraphics[width=0.9 \linewidth]{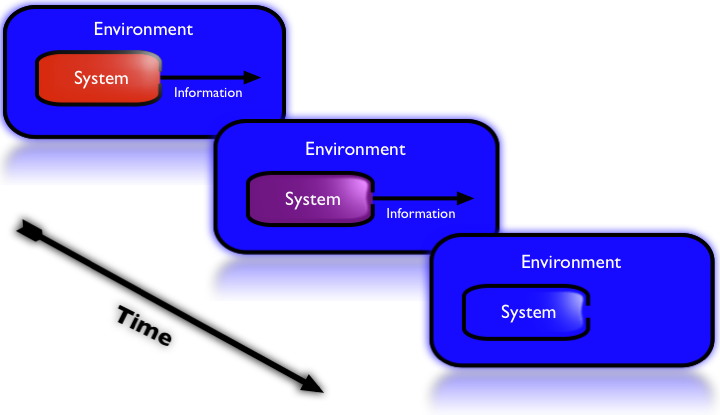}
\end{center} 
\caption[Graphical representation of decoherence]{A graphical representation of decoherence. Here, the environment, which is treated statistically, can be thought of as an information reservoir. It serves to absorb the quantum interference properties of the system, making the system appear as a classical, statistically prepared state.}\label{fig:informationexchange} 
\end{figure}

Quantum decoherence theory is a quantitative model of how this transition from quantum to classical mechanics occurs, which involves systems performing local measurements on themselves. More precisely, we divide our universe into two pieces: a simple system component, which is treated quantum mechanically, and a complex environmental component, which is treated statistically.\footnote{The words statistical and classical are being tossed around here a bit. What we mean is statistical in the thermodynamic sense, for example probability distributions prepared by random coin-tosses. These random, statistical distributions are contrasted against quantum states, which may \textit{appear} to be random when observed, but actually carry quantum interference information.} Since the environment is treated statistically, it obeys the rules of classical (statistical) mechanics, and we call it a \textbf{mixture}\index{Impure State} \cite{ballentine}. When the environment is coupled to the system, any quantum mechanical information that the system transfers to the environment is effectively lost, hence the system becomes a mixture over time, as indicated in figure \ref{fig:informationexchange}.

In the macroscopic world, ordinary forces are huge compared to the subtle effects of quantum mechanics, and thus large systems are very difficult to isolate from their environments. Hence, the time it takes large objects to turn to mixtures, called the \textbf{decoherence time}\index{Decoherence Time}, is very short. It is important to keep in mind that decoherence is inherently local. That is, if we consider our entire universe, the system plus the environment, quantum mechanically, classical effects do not emerge. Rather, we need to ``focus''  on a particular component, and throw away the quantum mechanical information having to do with the environment \cite{omnes}. 

\begin{figure}[t] 
\begin{center} 
\includegraphics[width=0.9 \linewidth]{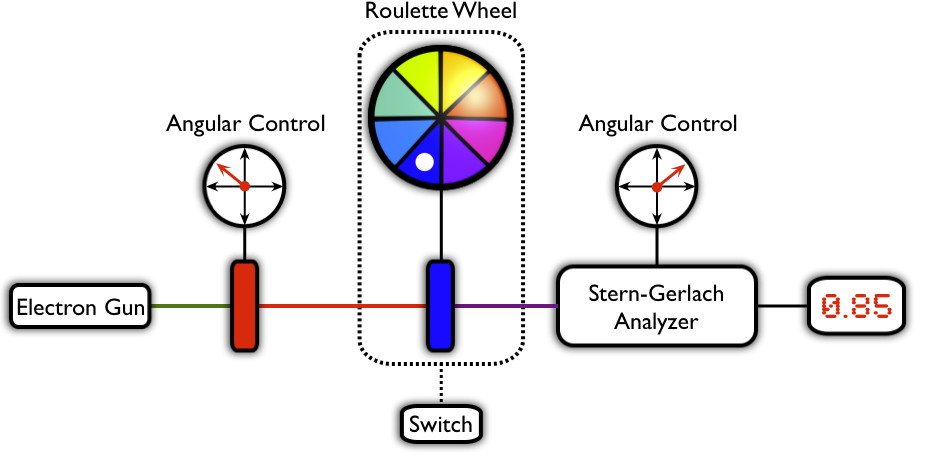}
\end{center} 
\caption[H. J. Bernstein's simple model of decoherence]{A sketch of Bernstein's thought experiment. The electrons with initial random spin are set to a certain angle in the $xy$-plane at the first angular control. A switch determines whether or not an additional phase factor is added using a roulette wheel. Then, a Stern-Gerlach analyzer is used to measure the angle of electron spin.} \label{bernsteindevice}
\end{figure}

In order to clarify this notion of decoherence, we examine the following unpublished example originally devised by Herbert J. Bernstein\index{Bernstein, H. J.} \cite{greenstein}. To start, consider an electron gun, as shown in figure \ref{bernsteindevice}. Electrons are an example of a two-state system, and as such they possess a quantum-mechanical property called spin \cite{nielsenchuang}. As we develop in detail later in section \ref{sec:quantumsup}, the spin of a two-state system\index{Two-State System} can be represented as a vector pointing on the unit two-sphere. Further, any possible spin can be formed as a linear combination of a spin pointing up in the $\hat z$ direction, and a spin pointing down in the $-\hat z$ direction.\footnote{In linear algebra terminology, we call the spin vectors pointing in $+ \hat z$ and $- \hat z$ a \textbf{basis} for the linear vector space of all possible states. We deal with bases precisely in section \ref{sec:linearvecspace}.} 

We suppose that our electron gun fires electrons of random spin, and then we use some angular control device to fix the electron's spin to some angle (that we set) in the $xy$-plane. Then, we use a Stern-Gerlach\index{Stern-Gerlach Analyzer} analyzer adjusted to some angle to measure the resulting electron. The Stern-Gerlach analyzer measures how close its control angle is to the spins of the electrons in the beam passing through it \cite{greenstein}. It reads out a number on a digital display, with $1$ corresponding to perfect alignment and $0$ corresponding to anti-alignment.
So far, we can always use the analyzer to measure the quantum-mechanical spin of each electron in our beam. We simply turn the analyzer's angular control until its digital display reads one, and then read the value of the angular control. Similarly, if we were to turn the analyzer's control to the angle opposite from the beam's angle, the display would read zero. The fact that these two special angles always exist is fundamental to quantum mechanics, resulting from a purely non-classical phenomenon called \textbf{superposition}.\index{Superposition}\footnote{The precise nature of quantum superposition is rather subtle, and we discuss it at length in section \ref{sec:quantumsup}.} We next insert another component into the path of the electron beam. By turning on a switch, we activate a second device that adjusts the angle of our beam in the $xy$-plane by adding $\theta$. The trick is that this device is actually attached to a modified roulette wheel, which we spin every time an electron passes. The roulette wheel is labeled in radians, and determines the value of $\theta$ \cite{greenstein}. 

We now frantically spin the angular control attached to our analyzer, attempting to find the initial angle of our electron beam. However, much to our surprise, the display appears to be stuck on $0.5$ \cite{greenstein}. This reading turns out to be no mistake, since the angles of the electrons that the analyzer is measuring are now randomly distributed (thanks to the randomness of the roulette wheel) throughout the $xy$-plane. No matter how steadfastly we attempt to measure the spin of the electrons in our beam, we cannot while the roulette wheel is active. Essentially, the roulette wheel is absorbing the spin information of the electrons, as we apparently no longer have access to it. 

This absorption of quantum information is the exact process that the environment performs in quantum decoherence theory. In both cases, the information is lost due to statistical randomness, and forces a quantum system to be classically random as well. The roulette wheel in this simplified example, just like the environment in reality, is blocking our access to quantum properties of a system. In chapter \ref{chap:applications}, we return to a more physical example of decoherence using the quantitative tools we develop in this thesis. First, we need to discuss the mathematical underpinnings of quantum mechanics.

\section{Notational Conventions}
Throughout this thesis, we adopt a variety of notational conventions, some more common than others. Here, we list them for clarity.
\begin{itemize}

\item The symbol $(\equiv)$ will always be used in the case of a definition. It indicates that the equality does not follow from previous work. The $(=)$ sign indicates equality that logically follows from previous work.

\item An integral symbol without bounds, \[\left( \int \right), \] is a definite integral from $- \infty$ to $+ \infty$, rather than the antiderivative, unless otherwise noted. 

\item Usually, the differential term in an integrand will be grouped with the integral symbol and separated by $(\cdot)$. This is standard multiplication, and is only included for notational clarity.

\item Vectors are always given in Dirac kets, $\left( \, \left| \cdot \right> \, \right)$, operators on abstract vector or Hilbert spaces are always given with hats, $\left( \, \hat{\cdot}\, \right)$, linear functionals over vector spaces are given in Dirac bras, $\left( \, \left< \cdot \right| \, \right) $, and operators on function spaces are given with checks, $\left( \, \check{\cdot} \, \right)$.
\item Both partial and total derivatives are given using either standard Leibniz or in a contracted form $d_x$, where \[ d_x \equiv \frac{d}{dx}. \]

\item The symbol $(\leftrightarrow )$ is used to denote a special representation of a particular structure. Its precise definition is made clear by context.

\item The symbol $(*)$ is used to denote the complex conjugate of a complex number.

\end{itemize}

%
%
\mainmatter

\chapter{Mathematical background}\label{chap:math_background}
\lettrine[lines=2, lhang=0.33, loversize=0.1]{B}efore we begin our discussion of quantum mechanics, we take this chapter to review the mathematical concepts that might be unfamiliar to the average undergraduate physics major wishing a more detailed understanding quantum mechanics. We begin with a discussion of linear vector spaces and linear operators. We next generalize these basic concepts to product spaces, and finally consider spaces of infinite dimension. Quantum mechanics is much more abstract than other areas of physics, such as classical mechanics, and so the immediate utility of the techniques introduced here is not evident. However, for the treatment in this thesis to be mostly self-contained, we proceed slowly and carefully.

\section{Linear Vector Spaces}\label{sec:linearvecspace}
In this section, we introduce linear vector spaces, which will be the stages for all of our subsequent work.\footnote{Well, actually we will work in a triplet of abstract spaces called a \textbf{rigged Hilbert space}, which is a special type of linear vector space. However, most textbooks on quantum mechanics, and even most physicists, do not bother much with the distinction. We will look at this issue in more detail in section \ref{sec:infdim}.} We begin with the elementary topic of vector spaces \cite{poole}.

\begin{boxeddefn}{Vector space\index{Linear!Vector Space}}{defn:vecspace}
Let $F$ be a field with addition $(+)$ and multiplication $(\cdot)$. A set $V$ is a \textbf{vector space} under the operation $(\oplus)$ over $F$ if for all $\left|  u \right> , \left| v \right>, \left| w \right> \in V$ and $a,b \in F$:
\begin{enumerate}
\item $\left| u \right> \oplus \left|  v \right> = \left| v \right> \oplus \left| u \right> $.
\item $(\left| u \right> \oplus \left| v \right> ) \oplus \left| w \right> = \left| u \right> \oplus (\left| v \right> \oplus \left| w \right> )$.
\item There exists $\left| 0 \right> \in V$ such that $\left| 0 \right> \oplus \left| u \right> = \left| u \right>$.
\item There exists $- \left| u \right> \in V$ such that $ - \left| u \right> \oplus \left| u \right> = \left| 0\right> $.
\item $a \cdot ( b \left| u \right> ) = (a \cdot b ) \left| u\right> $.
\item $(a + b) \left| u \right> = a \left| u \right> + b \left| u\right> $.
\item $a ( \left| u \right> + \left| v \right> ) = a \left| u \right> + a \left| v\right> $.
\item For the unity of $F$, $1$, $1 \left| u \right> = \left| u \right>$.
\end{enumerate}
\end{boxeddefn}

If $V$ satisfies the criteria for a vector space, the members $\left| u \right> \in V$ are called \textbf{vectors}\index{Vector}, and the members $a \in F$ are called \textbf{scalars}\index{Scalar}. For the purposes of quantum mechanics, the field $F$ we are concerned with is almost always $\mathbb C$, the field of complex numbers, and $V$ has the usual (Euclidean) topology.\footnote{The fields\index{Field (algebraic)} we refer to here are those from abstract algebra, and should not be confused with force fields (such as the electric and magnetic fields) used in physics. Loosely speaking, most of the sets of numbers we deal with in physics are algebraic fields, such as the real and complex numbers. For more details, see ref \cite{anderson}.} Since the operation $(\oplus)$ is by definition interchangeable with the field operation $(+)$, it is conventional to use the symbol $(+)$ for both, and we do so henceforth \cite{anderson}.\footnote{In definition \ref{defn:lindep}, we use the notion $\alpha \in \Lambda$, which might be foreign to some readers. $\Lambda$ is considered an index set, or a set of all possible allowed values for $\alpha$. Then, by $\alpha \in \Lambda$, we are letting $\alpha$ run over the entire index set. Using this powerful notation, we can treat almost any type of general sum or integral. For more information, see ref. \cite{gamelin}}

\begin{boxeddefn}{Linear dependence\index{Linear!Dependence}}{defn:lindep}
A collection of vectors $\{ \left| v _{\alpha} \right> \}_{\alpha \in \Lambda}$, where $\Lambda$ is some index set, belonging a vector space $V$ over $F$ is \textbf{linearly dependent} if there exists a set $\{a_{\alpha}\}_{\alpha \in \Lambda}$ such that 
\begin{equation}
\sum_{\alpha \in \Lambda} a_{\alpha} \left| v_{\alpha}\right>  = \left| 0\right> 
\end{equation}
 given that at least one $a_{i} \in \{ a_{\alpha} \} \neq 0$. 
\end{boxeddefn}
This means that, if a set of vectors is linearly dependent, we can express one of the member vectors in terms of the others. If a set of vectors is not linearly dependent, we call it \textbf{linearly independent}\index{Linear!Independence}, in which case we would not be able to express one of the member vectors in terms of the others \cite{poole}.

\begin{boxeddefn}{Dimension\index{Dimension}}{defndimension}
Consider the vector space $V$ and let $\{ \left| v\right>_{\alpha}   \}_{\alpha \in \Lambda} \subseteq V$ be an arbitrary set of linearly independent vectors. Then, if $\Lambda$ is alway finite, the \textbf{dimension} of $V$ is the maximum number of elements in $\Lambda$. If $\Lambda$ is not always finite, then $V$ is said to have \textbf{infinite dimension}.
\end{boxeddefn}

\begin{boxeddefn}{Basis\index{Basis}}{def:basis}
Let $B=\{\left| v_{\alpha}\right> \}_{\alpha \in \Lambda} \subseteq V$, where $V$ is a vector space over the field $F$.  If $\left| v_{\alpha} \right>$ and $\left| v_{\beta} \right>$ when $\alpha \neq \beta$ are linearly independent and an arbitrary vector $\left| u \right> \in V$ can be written as a linear combination of $\left| v _{\alpha}\right>$`s, i.e.
\begin{equation}
\left| u \right>  = \sum_{\alpha \in \Lambda} c_{\alpha} \left| v _{\alpha}\right>,
\end{equation}
with $c_{\alpha} \in F$, we say  $\{\left| v_{\alpha}\right> \}_{\alpha \in \Lambda}$ is a \textbf{basis set} or \textbf{basis} for $V$.
\end{boxeddefn}
It follows directly from this definition that, in any vector space with finite dimension $D$, any basis set will have precisely $D$ members. Because quantum mechanics  deals with a Euclidean vector space over the complex numbers, it is advantageous to precisely define the inner product of two vectors within that special case \cite{ballentine}.

\begin{boxeddefn}{Inner product}{defn:innderproduct}
Let $V$ be a vector space over the field of complex numbers $\mathbb C$. Then, $g:V \times V \rightarrow \mathbb C$ is an \textbf{inner product} if, for all $\left| u \right>, \left| v \right>, \left| w \right> \in V$ and $\alpha, \beta \in \mathbb C$, 
\begin{enumerate}
\item $g \left( \left| u \right> , \left| v \right> \right) = g \left( \left| v \right> , \left| u \right> \right)^ *$,
\item $ g \left( \left| u \right> , a \left| v \right> + b \left| w \right> \right) = a \cdot g \left( \left| u \right> , \left| v \right> \right)+ b \cdot g \left( \left| u \right> , \left| w \right> \right) $,
\item $ g \left( \left| u \right> , \left| u \right> \right) \geq 0$ with $ g \left( \left| u \right> , \left| u \right> \right) = 0 \Leftrightarrow \left| u \right> = \left| 0 \right>$.
\end{enumerate}
\end{boxeddefn}

Although it is not immediately clear, the inner product is closely related to the space of linear functionals on $V$, called the dual space of $V$ and denoted $V^*$. Below, we define these concepts precisely and then show their connection through the Riesz representation theorem \cite{ballentine}.

\begin{boxeddefn}{Linear functional\index{Linear!Functional}}{}
A \textbf{linear functional} on a vector space $V$ over $\mathbb C$ is any function $F: V \rightarrow \mathbb C$ such that for all $\alpha, \beta \in \mathbb C$ and for all $\left| u \right>, \left| v \right> \in V$, 
\begin{equation}
F \left( a \left| u \right>+ b \left| v \right> \right) = a \cdot F \left( \left| u \right> \right) + b \cdot F \left( \left| v \right> \right).
\end{equation}
We say that the space occupied by the linear functionals on $V$ is the \textbf{dual space}\index{Dual Space} of $V$, and we denote it by $V^*$.
\end{boxeddefn}
We connect the inner product with the dual space $V^*$ using the Riesz representation theorem \cite{ballentine}.

\begin{boxedthm}{Riesz representation\index{Riesz Representation Theorem}}{thm:rieszthm}
Let $V$ be a finite-dimensional vector space and $V^*$ be its dual space. Then, there exists a bijection $h: V^* \rightarrow V$ defined by $h(F) = \left| f \right> $ for $F \in V^*$ and $\left| f \right> \in V$ such that $F \left( \left| u \right> \right) = g \left( \left| f \right>, \left| u \right> \right)\, \forall \left| u \right> \in V$, where $g$ is an inner product of $V$ \cite{ballentine}.
\end{boxedthm}
The proof of this theorem is straightforward, but too lengthy for our present discussion, so we will reference a simple proof for the interested reader \cite{ballentine}. The consequences of this theorem are quite drastic. It is obviously true that the inner product of two vectors, which maps them to a scalar, is a linear functional. However, the Riesz theorem asserts that any linear functional can be represented as an inner product. This means that every linear functional has precisely one object in the dual space, corresponding to a vector in the vector space. For this reason, we call the linear functional associated with with $\left| u \right>$ a dual vector and write it as
\begin{equation}
 \left< u \right| \in V^*,
\end{equation}
and we contract our notation for the inner product of two vectors $\left| u \right>$ and $\left| v \right>$ to
\begin{equation}
g\left( \left| u \right>, \left| v \right> \right) \equiv \left< u \big| v \right>,
\end{equation}
a notational convention first established by P. A. M. Dirac.\index{Dirac, P. A. M.} The vectors in $V$ are called \textbf{kets}\index{Ket} and the dual vectors, or linear functionals associated with vectors in $V^*$, are called \textbf{bras}\index{Bra}. Hence, when we adjoin a bra and a ket, we get a bra-ket or bracket, which is an inner product. Note that by the definition of the inner product, we have 
\begin{equation}\label{eqn:diracinner}
\left< u \big | v \right> = \left< v \big | u \right>^*,
\end{equation}
so if we multiply some vector $\left| v \right>$ by a (complex) scalar $\alpha$, the corresponding dual vector is $\alpha^* \left< v \right|$. When we form dual vectors from vectors, we must always remember to conjugate such scalars. As another note, when choosing a basis, we frequently pick it as \textbf{orthonormal}, which we define below \cite{poole}.

\begin{boxeddefn}{Orthonormality of a Basis\index{Basis!Orthonormality of}}{defn:orthobasis}
A basis $B$ for some vector space $V$ is \textbf{orthonormal} if any two vectors $\left| \phi_i \right>$ and $\left| \phi_j \right>$ in $B$ satisfy
\begin{equation}
\left< \phi_i \big | \phi_j \right> = \begin{cases} 1 & \text{if $i=j$} \\ 0&  \text{if $i \neq j$} \end{cases}.
\end{equation}
\end{boxeddefn}
For any vector space, we can always find such a basis, so we do not lose any generality by always choosing to use one.\footnote{The process for finding an orthonormal basis is called the Graham-Schmidt algorithm\index{Graham-Schmidt Algorithm}, and allows us to construct an orthonormal basis from any basis. For details, see ref. \cite{poole}.}

A useful example that illustrates the use of vectors and dual vectors can be found by constraining our vector space to a finite number of dimensions.\index{Matrix Representation!of Vectors}\index{Matrix Representation!of Linear Functionals} Working in such a space, we represent vectors as column matrices and dual vectors as row matrices \cite{nielsenchuang}. For example, in three dimensions we might have
\begin{equation}
\left| e_1 \right> \leftrightarrow \left( \begin{array}{c} 1 \\ 0 \\ 0 \end{array} \right)
\end{equation}
and
\begin{equation}
i \left| e_2 \right> \leftrightarrow \left( \begin{array}{c} 0 \\ i \\0 \end{array} \right),
\end{equation}
where $\left| e_1 \right>$ and $\left| e_2 \right>$ are the unit vectors from basic physics \cite{hrw}. Then, the linear functional corresponding to $\left| e_2 \right>$ is\footnote{Here, notice that to generate the representation for $\left< e_2 \right|$ from $\left| e_2 \right>$, we must take the complex conjugate. This is necessary due to the complex symmetry of the inner product established in eqn. \ref{eqn:diracinner}.}
\begin{equation}
\left< e_2 \right| \leftrightarrow i^* \left( \begin{array}{ccc} 0 & 1 & 0 \end{array} \right) =  \left( \begin{array}{ccc} 0 & -i & 0 \end{array} \right) .
\end{equation}
We represent the inner product as matrix multiplication, so we write
\begin{equation}
- i \left< e_2 \big |e_1 \right> \leftrightarrow \left( \begin{array}{ccc} 0 & -i & 0 \end{array} \right) \left( \begin{array}{c} 1 \\ 0 \\ 0 \end{array} \right) = 0,
\end{equation}
which indicates that $\left| e_1 \right> $ and $\left| e_2\right>$ are orthogonal, as we expect.
\section{Linear Operators}
So far, we have looked at two main types of objects in a vector space: vectors and linear functionals. In this section, we focus on a third: the linear operator. Recall that linear functionals take vectors to numbers. Similarly, linear operators are objects that take vectors to other vectors. Formally, this is the following definition \cite{riley}.

\begin{boxeddefn}{Linear Operator\index{Linear!Operator!Abstract}}{}
Let $\left| u \right>, \left| v \right> \in V$ be vectors and $\alpha, \beta$ be scalars in the field associated with $V$. Then, we say $\hat A$ is a \textbf{linear operator} on $V$ if
\begin{equation}
\hat A \left| v \right> \in V
\end{equation}
and
\begin{equation}
\hat A \left( \alpha \left| u \right> + \beta \left| v \right> \right) = \alpha \hat A \left| u \right> + \beta \hat A \left| v \right>.
\end{equation}
\end{boxeddefn}
Throughout the rest of this thesis, whenever we discuss an operator on a vector space, we will always use a hat to avoid confusion with a scalar. In a finite dimensional vector space, as indicated previously, we often represent vectors by column matrices and dual vectors by row matrices. Similarly, we represent operators by square matrices \cite{nielsenchuang}.\index{Matrix Representation!of Linear Operators} For example, if
\begin{equation}
\hat{A} \leftrightarrow  \left( \begin{array}{ccc} 0 & 0 & 0\\ 1 & 0 & 0\\0 & 0 & 0 \end{array} \right),
\end{equation}
then 
\begin{equation}
\hat{A} \left| e_1 \right> \leftrightarrow  \left( \begin{array}{ccc} 0 & 0 & 0\\ 1 & 0 & 0\\0 & 0 & 0 \end{array} \right) \left( \begin{array}{c} 1 \\ 0 \\ 0 \end{array} \right)  = \left( \begin{array}{c} 0 \\ 1 \\0 \end{array} \right) \leftrightarrow \left| e_2 \right>.
\end{equation}
We can also use our formalism to access individual elements of an operator in its matrix representation. Working in the three-dimensional standard, orthonormal basis from the example above, we specify $\hat B$ as
\begin{equation}
\hat{B} \left| u \right> = \left| v \right>,
\end{equation}
where
\begin{equation}
\left| u \right> = u_1 \left| e_1 \right> + u_2 \left| e_2 \right> + u_3 \left| e_3 \right>
\end{equation}
and
\begin{equation}
\left| v \right> = v_1 \left| e_1 \right> + v_2 \left| e_2 \right> + v_3 \left| e_3 \right>.
\end{equation}
Then, 
\begin{eqnarray}
\left<e_i \right| \hat B\left| u \right>&=&\left< e_i \right| \hat B \left( u_1\left| e_1 \right> +u_2\left| e_2 \right> + u_3\left| e_3 \right> \right)  \nonumber \\
						&=& \left<e_i \right| \hat B \sum_{j=1}^3 u_j \left| e_j \right> \nonumber \\
						&=& \left<e_i \big | v \right> \nonumber \\
						&=&\sum_{j=1}^3 v_j \left<e_i \big | e_j \right> \nonumber \\
						&=& v_i,
\end{eqnarray}
which is just the matrix equation  \cite{ballentine}
\begin{equation}
\sum_{j=1}^3 B(i,j)u_j=v_j,
\end{equation}
where we made the definition
\begin{equation}\label{eqn:matrixelem}
B_{ij}=B(i,j)\equiv \big < e_i \big | \hat B \left| e_j \right>.
\end{equation}
We call $B(i,j)$ the \textbf{matrix element}\index{Matrix Element} corresponding to the the operator $\hat B$. Note that the matrix elements of an operator depend on our choice of basis set. Using this expression for a matrix element, we define the trace of an operator. This definition is very similar to the elementary  notion of the trace of a matrix as the sum of the elements in the main diagonal.\footnote{Since the individual matrix elements of an operator depend on the basis chosen, it might seem as if the trace would vary with basis, as well. However, the trace turns out to be independent of basis choice \cite{ballentine}.}

\begin{boxeddefn}{Trace}{defn:trace}
Let $\hat A$  be an operator on the vector space $V$ and let $B=\{\left| v_{\alpha}\right> \}_{\alpha \in \Lambda} \subseteq V$ be an orthonormal basis for $V$. Then, the \textbf{trace} of $\hat{A}$ is
\begin{equation}
\mathrm{Tr} \left( \hat A \right) \equiv \sum_{\alpha \in \Lambda} \left< v _{\alpha }\right| \hat A \left| v_{\alpha} \right>.
\end{equation}
\end{boxeddefn}
So far, we have defined operators as acting to the right on vectors. However, since the Riesz theorem guarantees a bijection between vectors and dual vectors (linear functionals in the dual space), we expect operators to also act to the left on dual vectors. To make this concept precise, we write a definition.

\begin{boxeddefn}{Adjoint\index{Adjoint}}{def:adjoint}
Suppose $\left| u \right>, \left| v \right> \in V$ such that an operator on $V$, $\hat A$, follows
\begin{equation}
\hat A \left| u \right> = \left| v \right>.
\end{equation}
Then, we define the \textbf{adjoint} of $\hat A$, $\hat A ^{\dagger}$, as 
\begin{equation}
\left< u \right| \hat A^{\dagger} \equiv \left< v \right|.
\end{equation}
\end{boxeddefn}
From this definition, it follows that
\begin{eqnarray} \label{eqn:adjointapp}
\left( \left< u \right| \hat A^{\dagger} \left| w \right> \right)^* &=& \left<v \big | w \right>^* \nonumber \\
&=& \left< w \big| v \right> \nonumber \\
&=& \left< w \right| \hat A \left| u \right>,
\end{eqnarray}
which is an important result involving the adjoint, and is sometimes even used as its definition. This correctly suggests that the adjoint for operators is very similar to the conjugate transpose for square matrices, with the two operations equivalent for the matrix representations of finite vector spaces.\footnote{Many physicists, seeing that linear functionals are represented as row matrices and vectors are represented as column matrices, will write $\left| v \right> = \left< v \right| ^{\dagger}$. This is not \textit{technically} correct, as the formal definition \ref{def:adjoint} only defined the adjoint operation for an operator, not a functional. However, though it is an abuse of notation, it turns out that nothing breaks as a result \cite{ballentine}. For clarity, we will be careful not to use the adjoint in this way.}

Although the matrix representation of an operator is useful, we need to express operators using Dirac's bra-ket notation. To do this, we define the outer product \cite{nielsenchuang}.

\begin{boxeddefn}{Outer Product\index{Outer Product}}{}
Let $ \left| u \right>, \left| v \right> \in V$ be vectors. We define the \textbf{outer product} of $\left| u \right>$ and $\left| v \right>$ as the operator $\hat A$ such that
\begin{equation}
\hat A \equiv \left| u \right> \left< v \right|.
\end{equation}
\end{boxeddefn}
Note that this is clearly linear, and is an operator, as
\begin{equation}
\big( \left| u \right> \left< v \right| \big) \left| w \right>  = \left| u \right> \left< v \big | w \right> = \left< v \big | w \right> \left| u \right> \in V
\end{equation}
for $\left| u \right>, \left| v \right>, \left| w \right> \in V$, a vector space. Further, if an operator is constructed in such a way, eqn. \ref{eqn:adjointapp} tells us that its adjoint is
\begin{equation}
\left( \left| u \right> \left< v \right| \right)^{\dagger} = \left| v \right> \left< u \right|.
\end{equation}

Self-adjoint opeartors\index{Linear!Operator!Self-Adjoint}, i.e. operators such that 
\begin{equation}
\hat A^{\dagger} = \hat A,
\end{equation}
are especially important in quantum mechanics. The main properties that make self-adjoint operators useful concern their eigenvectors and eigenvalues.\footnote{We assume that the reader has seen eigenvalues and eigenvectors. However, if not, see ref. \cite{poole} or any other linear algebra text for a thorough introduction.} We summarize them formally in the following theorem \cite{ballentine}.

\begin{boxedthm}{Eigenvectors and Eigenvalues of Self-adjoint Operators}{}
Let $\hat A$ be a self-adjoint operator. Then, all its eigenvalues are real and any two eigenvectors corresponding to two distinct eigenvalues are orthogonal.
\end{boxedthm}
\begin{proof}
Let $ \hat A \left| u \right> = u \left| u \right>$ and $\hat A \left| v \right> = v \left| v \right>$ so that $\left| u \right>$ and $\left| v \right>$ are arbitrary (nonzero) eigenvectors of $\hat A$ corresponding to the eigenvalues $u$ and $v$.
Then, using eqn. \ref{eqn:adjointapp}, we deduce \cite{ballentine}
\begin{eqnarray}
u \left< u \big | u \right> 	&=& \left< u \right| u \left| u \right> \nonumber \\
					&=&  \left< u \right| \hat A^{\dagger} \left| u \right>^* \nonumber \\
					&=&  \left< u \right| \hat A \left| u \right>^* \nonumber \\
					&=& \left< u \right| u \left| u \right>^* \nonumber \\
					&=& u^* \left< u \big| u \right>^* \nonumber \\
					&=& u^* \left< u \big| u \right>.
\end{eqnarray}
Since $\left| u \right> \neq 0$, we get $u=u^*$, so $u$ is real. Hence, any arbitrary eigenvalue of a self-adjoint operator is real. Next, we consider combinations of two eigenvectors. That is,
\begin{eqnarray}
0 &=& \left< u \right| \hat A \left| v \right> - \left< u \right| \hat A \left| v \right> \nonumber \\
 &=& \left< u \right| \hat A \left| v \right> - \left< v \right| \hat A^{\dagger} \left| u \right>^* \nonumber \\
 &=& \left< u \right| \hat A \left| v \right> - \left< v \right| \hat A \left| u \right>^* \nonumber \\
  &=& \left< u \right| v \left| v \right> - \left< v \right| u \left| u \right>^* \nonumber \\
    &=& \left( v - u \right) \left< u \big | v \right>.
\end{eqnarray}
Thus, if $v \neq u$, $\left< u \big | v \right> = 0$, so $\left| u \right>$ and $\left| v \right>$ are orthogonal as claimed.
\end{proof}
Now that we have shown this orthogonality of distinct eigenvectors or an operator, we would like to claim that these eigenvectors form a basis for the vector space in which the operator works. For finite dimensional spaces, this turns out to be the case, although the proof quite technical, so we omit it with reference \cite{ballentine}. However, infinite dimensional cases produce problems mathematically, hence the eigenvectors of an operator in such a space need not form a basis for that space \cite{ballentine}. For the moment, we will proceed anyway, returning to this issue in section \ref{sec:infdim}.

Suppose that $\{\left| v_{\alpha}\right> \}_{\alpha \in \Lambda}$ is the set of all eigenvectors of the self-adjoint operator $\hat A$. Since eigenvectors are only determinable up to a scaling factor, as long as our vectors are of finite magnitude, we may rescale all of these vectors to be an orthonormal set of basis vectors \cite{poole}. By our assumption, this set forms a basis for our vector space, $V$. Thus, for any $\left| u \right> \in V$, we can write 
\begin{equation}
\left| u \right> = \sum_{\alpha \in \Lambda} u_{\alpha} \left| v_{\alpha} \right> =  \sum_{\alpha \in \Lambda} \left| v_{\alpha} \right>u_{\alpha} .
\end{equation}
Noting that, since the basis vectors are orthonormal,
\begin{equation}
\left<v_i \big | u \right> = \sum_{\alpha \in \Lambda} u_{\alpha} \left< v_i \big| v_\alpha \right> = u_i,
\end{equation}
we get
\begin{equation}
\left| u \right> = \sum_{\alpha \in \Lambda} \left| v_{\alpha} \right> \left< v_{\alpha} \big | u \right> =  \left(\sum_{\alpha \in \Lambda}  \left| v_{\alpha} \right> \left< v_{\alpha}\right| \right) \left| u \right> .
\end{equation}
It follows immediately that 
\begin{boxedeqn}{eqn:projector}
\sum_{\alpha \in \Lambda}  \left| v_{\alpha} \right> \left< v_{\alpha}\right|  = \hat 1,
\end{boxedeqn}
which is called the \textbf{resolution of the identity}\index{Resolution of the Identity}. This leads us to a result that allows us to represent self-adjoint operators in terms of their eigenvector bases, the spectral theorem \cite{ballentine}.

\begin{boxedthm}{Spectral Theorem\index{Spectral Theorem}}{thm:spectral}
Let $\hat A$ be an operator on the vector space $V$. Assuming that the spectrum of eigenvectors of $\hat A$, $\{\left| v_{\alpha}\right> \}_{\alpha \in \Lambda}$, forms a basis for $V$, $\hat A$ can be expressed as
\begin{equation}
\hat A = \sum_{\alpha \in \Lambda} a_{\alpha} \left| v_{\alpha} \right> \left< v_{\alpha} \right|,
\end{equation}
where $\{ a_{\alpha} \}_{\alpha \in \Lambda}$ are the eigenvalues of $\hat A$.
\end{boxedthm}
\begin{proof}
Let $\left| u \right> \in V$ be an arbitrary vector. Then, since $\{\left| v_{\alpha}\right> \}_{\alpha \in \Lambda}$ is a basis for $V$, we can write
\begin{equation}
\left| u \right> = \sum_{\alpha \in \Lambda} u_{\alpha} \left| v_{\alpha} \right>.
\end{equation}
Hence, 
\begin{equation}
\hat A \left| u \right> = \sum_{\alpha \in \Lambda} u_{\alpha} \hat A \left| v_{\alpha} \right> = \sum_{\alpha \in \Lambda} u_{\alpha} a_{\alpha} \left| v_{\alpha} \right>.
\end{equation}
Now, we consider the other side of the equation. We get \cite{ballentine}
\begin{eqnarray}
\left( \sum_{\alpha \in \Lambda} a_{\alpha} \left| v_{\alpha} \right> \left< v_{\alpha} \right| \right) \left| u \right> 
&=& \left( \sum_{\alpha \in \Lambda} a_{\alpha} \left| v_{\alpha} \right> \left< v_{\alpha} \right| \right) \sum_{\beta \in \Lambda} u_{\beta} \left| v_{\beta} \right> \nonumber \\
&=&  \sum_{\alpha \in \Lambda} \sum_{\beta \in \Lambda} a_{\alpha}u_{\beta}  \left| v_{\alpha} \right> \left< v_{\alpha} \big| v_{\beta} \right> \nonumber \\
&=& \sum_{\alpha \in \Lambda} a_{\alpha}u_{\alpha}  \left| v_{\alpha} \right> \nonumber \\
&=& \hat A \left| u \right>,
\end{eqnarray}
where we used the orthonormality of our basis vectors. This holds for arbitrary $\left| u \right> \in V$, so \cite{ballentine}
\begin{equation}
\hat A = \sum_{\alpha \in \Lambda} a_{\alpha} \left| v_{\alpha} \right> \left< v_{\alpha} \right| ,
\end{equation}
as desired.
\end{proof}
Since we assumed that the eigenvectors for any self-adjoint operator formed a basis for the operator's space, we may use the spectral theorem to decompose self-adjoint operators into basis elements, which we make use of later.

\section{The Tensor Product}
So far, we have discussed two types of products in vector spaces: inner and outer. The tensor product falls into the same category as the outer product in that it involves arraying all possible combinations of two sets, and is sometimes referred to as the cartesian or direct product \cite{anderson}. We formally define the tensor product operation $( \otimes )$ below \cite{nielsenchuang}.

\begin{boxeddefn}{Tensor Product\index{Tensor Product}}{defn:tensor}
Suppose $V$ and $W$ are two vector spaces spanned by the orthonormal bases $\{\left| v_{\alpha}\right> \}_{\alpha \in \Lambda}$ and $\big\{\left| w_{\beta}\right> \big \}_{\beta \in \Gamma}$, respectively. Then, we define the \textbf{tensor product space}, or product space, as the space spanned by the basis set 
\begin{equation}
\big \{ \left( \left| x \right>, \left| y \right> \right) \, : \, \left| x \right> \in \{\left| v_{\alpha}\right> \}_{\alpha \in \Lambda}, \left| y \right> \in \big\{\left| w_{\beta}\right> \big \}_{\beta \in \Gamma} \big \}
\end{equation}
and denote the space as $V \otimes W$. We call each ordered pair of vectors a \textbf{tensor product} of the two vectors and denote it as $\left| x \right> \otimes \left| y \right>$. We require
\begin{equation}
\left< \left( \left| x_1 \right> \otimes \left| y_1 \right> \right) \big| \left( \left| x_2 \right> \otimes \left| y_2 \right>  \right) \right> \equiv \left< x_1 \big | x_2 \right> \otimes \left< y_1 \big | y_2 \right>.
\end{equation}
\end{boxeddefn}
The tensor product is linear in the normal sense, in that it is distributive and can absorb scalar constants \cite{nielsenchuang}. Further, we define linear operators on a product space by
\begin{equation} \label{eqn:tensorop}
\left( \hat A \otimes \hat B \right) \left| v \right> \otimes  \left| w \right> \equiv \hat A \left| v \right> \otimes \hat B \left| w \right>.
\end{equation}
The definition for the tensor product is quite abstract, so we now consider a special case in a matrix representation for clarity. Consider a a two-dimensional vector space, $V$, and a three-dimensional vector space $W$. We let the operator
\begin{equation}
\hat A \leftrightarrow \left( \begin{array}{cc} 1 & -i \\ 0 & 2 \end{array} \right)
\end{equation}
act over $V$, and the operator
\begin{equation}
\hat B \leftrightarrow \left( \begin{array}{ccc} i & 2 & -1 \\ 0 & 1 & -2 \\ 2i & -1 & 0 \end{array} \right)
\end{equation}
act over $W$. Then, operating on arbitrary vectors, we find
\begin{equation}
\hat A \left| v \right> \leftrightarrow \left( \begin{array}{cc} 1 & -i \\ 0 & 2 \end{array} \right) \left( \begin{array}{c} v_1 \\ v_2 \end{array} \right) = \left( \begin{array}{c} v_1-i v_2 \\ 2 v_2 \end{array} \right) 
\end{equation}
and
\begin{equation}
\hat B \left| w \right>  \leftrightarrow  \left( \begin{array}{ccc} i & 2 & -1 \\ 0 & 1 & -2 \\ 2i & -1 & 0 \end{array} \right) \left( \begin{array}{c} w_1 \\ w_2 \\ w_3 \end{array} \right) = \left( \begin{array}{c} iw_1+2w_2-w_3 \\ w_2 -2 w_3\\ 2iw_1 - w_2 \end{array} \right). 
\end{equation}
The representation of the tensor product as a matrix operation is called the \textbf{Kronecker product}, and is formed by nesting matrices from right to left and distributing via standard multiplication \cite{nielsenchuang}. We now illustrate it by working our example.
\begin{eqnarray}
\hat A \left| v \right> \otimes \hat B \left| w \right>
&\leftrightarrow& \left( \begin{array}{c} v_1-i v_2 \\ 2 v_2 \end{array} \right)  \otimes \left( \begin{array}{c} iw_1+2w_2-w_3 \\ w_2 -2 w_3\\ 2iw_1 - w_2 \end{array} \right) \nonumber \\
&=& \left( \begin{array}{c} \left( v_1-i v_2 \right)\left( \begin{array}{c} iw_1+2w_2-w_3 \\ w_2 -2 w_3\\ 2iw_1 - w_2 \end{array} \right) \\ 2 v_2 \left( \begin{array}{c} iw_1+2w_2-w_3 \\ w_2 -2 w_3\\ 2iw_1 - w_2 \end{array} \right)\end{array} \right) \nonumber \\
&=& \left( \begin{array}{c} \left( v_1-i v_2 \right) \left(iw_1+2w_2-w_3 \right) \\ \left( v_1-i v_2 \right)\left(w_2 -2 w_3 \right) \\ \left( v_1-i v_2 \right) \left( 2iw_1 - w_2 \right) \\ 2 v_2  \left( iw_1+2w_2-w_3 \right) \\ 2 v_2 \left( w_2 -2 w_3 \right) \\ 2 v_2 \left( 2iw_1 - w_2 \right) \end{array} \right) .
\end{eqnarray}
But by eqn. \ref{eqn:tensorop}, we should be able to first construct the tensor product of the of the operators $\hat A$ and $\hat B$ and apply the resulting operator to the tensor product of $\left| v \right>$ and $\left| w \right>$. Working this out using the Kronecker product\index{Kronecker Product}, we have
\begin{eqnarray}
\hat A \otimes \hat B 
&\leftrightarrow&\left( \begin{array}{cc} 1 & -i \\ 0 & 2 \end{array} \right) \otimes  \left( \begin{array}{ccc} i & 2 & -1 \\ 0 & 1 & -2 \\ 2i & -1 & 0 \end{array} \right) \nonumber \\
&=& \left( \begin{array}{cc} 1 \left( \begin{array}{ccc} i & 2 & -1 \\ 0 & 1 & -2 \\ 2i & -1 & 0 \end{array} \right) & -i  \left( \begin{array}{ccc} i & 2 & -1 \\ 0 & 1 & -2 \\ 2i & -1 & 0 \end{array} \right) \\ 0 \left( \begin{array}{ccc} i & 2 & -1 \\ 0 & 1 & -2 \\ 2i & -1 & 0 \end{array} \right) & 2  \left( \begin{array}{ccc} i & 2 & -1 \\ 0 & 1 & -2 \\ 2i & -1 & 0 \end{array} \right)\end{array} \right) \nonumber \\
&=& \left( \begin{array}{cccccc}  i & 2 & -1 & 1 & -2i & i \\ 0 & 1 & -2 & 0 & -i & 2i\\ 2i & -1 & 0 & 2 & i & 0 \\ 0 & 0 & 0 & 2i & 4 & -2 \\ 0 & 0 & 0 & 0 & 2 & -4 \\ 0 & 0 & 0 & 4i & -2 & 0 \end{array} \right) \nonumber \\
\end{eqnarray}
and
\begin{equation}
\left| v \right> \otimes \left| w \right> \leftrightarrow  \left( \begin{array}{c} v_1 \\ v_2 \end{array} \right)  \otimes  \left( \begin{array}{c} w_1 \\ w_2 \\ w_3 \end{array} \right) = \left( \begin{array}{c} v_1 w_1 \\ v_2 w_1 \\ v_1 w_2 \\v_2 w_2\\ v_1w_3 \\v_2 w_3 \end{array} \right),
\end{equation}
so
\begin{eqnarray}
\hat A \otimes \hat B\left( \left| v \right> \otimes \left| w \right> \right) 
&\leftrightarrow & \left( \begin{array}{cccccc}  i & 2 & -1 & 1 & -2i & i \\ 0 & 1 & -2 & 0 & -i & 2i\\ 2i & -1 & 0 & 2 & i & 0 \\ 0 & 0 & 0 & 2i & 4 & -2 \\ 0 & 0 & 0 & 0 & 2 & -4 \\ 0 & 0 & 0 & 4i & -2 & 0 \end{array} \right)  \left( \begin{array}{c} v_1 w_1 \\ v_2 w_1 \\ v_1 w_2 \\v_2 w_2\\ v_1w_3 \\v_2 w_3 \end{array} \right) \nonumber \\
&=& \left( \begin{array}{c} i v_1 w_1+ 2 v_2 w_1 -v_1w_2+v_2w_2-2iv_1w_3+iv_2w_3 \\ v_1w_2-iv_2w_2-2v_1w_3+2iv_2w_3 \\ 2iv_1w_1+2v_2w_1-v_1w_2+iv_2w_2 \\ 2iv_2w_1+4v_2w_2-2v_2w_3\\ 2v_2w_2-4v_2w_3\\4iv_2w_1-2v_2w_2 \end{array} \right) \nonumber \\
&=& \left( \begin{array}{c} \left( v_1-i v_2 \right) \left(iw_1+2w_2-w_3 \right) \\ \left( v_1-i v_2 \right)\left(w_2 -2 w_3 \right) \\ \left( v_1-i v_2 \right) \left( 2iw_1 - w_2 \right) \\ 2 v_2  \left( iw_1+2w_2-w_3 \right) \\ 2 v_2 \left( w_2 -2 w_3 \right) \\ 2 v_2 \left( 2iw_1 - w_2 \right) \end{array} \right) \nonumber  \\
&\leftrightarrow& \hat A \left| v \right> \otimes \hat B \left| w \right>,
\end{eqnarray}
and we confirm that this example follows
\begin{equation}
\left( \hat A \otimes \hat B\right) \left| v \right> \otimes \left| w \right>  = \hat A \left| v \right> \otimes \hat B \left| w \right>
\end{equation}
when we use the Kronecker product representation for the tensor product. Since the matrix representation is very convenient for finite dimensional vector spaces, we frequently use the Kronecker product to calculate the tensor product and then shift back to the abstract Dirac notation.

\section{Infinite Dimensional Spaces}\label{sec:infdim}
So far, we have largely ignored the main complication that arises when we move from a finite dimensional space to an infinite one: the spectrum of eigenvectors for a self-adjoint operator is no longer guaranteed to form a basis for the space. To deal with this problem, we will have to work in a slightly more specific kind of vector space, called a Hilbert space, denoted $\mathcal H$. A Hilbert space is defined below \cite{ballentine}.

\begin{boxeddefn}{Hilbert Space\index{Hilbert Space}}{}
Let $W$ be a general linear vector space and suppose that $V\subseteq W$ is a vector space formed by any finite linear combinations of the basis set $\{\left| v_{\alpha}\right> \}_{\alpha \in \Lambda}$. That is, if
\begin{equation}
\left| u \right> = \sum_{i=1}^n u_{\alpha_i} \left| v \right>_{\alpha_i},
\end{equation}
for some finite $n$, then  $\left| u \right> \in V$. We say the \textbf{Hilbert space} $\mathcal H$ formed by completing $V$ contains any vector that can be written as
\begin{equation}
\left| u \right> = \lim_{n \rightarrow \infty} \sum_{i=1}^n u_{\alpha_i} \left| v \right>_{\alpha_i},
\end{equation}
provided
\begin{equation}
\sum_{i=1}^{\infty}  \left | u_{\alpha_i} \right|^2
\end{equation}
exists and is finite.
\end{boxeddefn}
Note that for the vector spaces described in the above definition, the Hilbert space associated with them always follows $ V \subseteq \mathcal H \subseteq W$, and that $W=\mathcal H = V$ holds if (but \textit{not} only if) $W$ has finite dimension. Without spending too much time on the technicalities, there is a generalized spectral theorem that applies to spaces very closely related to, but larger than, Hilbert spaces \cite{ballentine}. To determine precisely what this space should be, we must first develop a certain subspace of a Hilbert space, which we define by including all vectors $\left| u \right>$ subject to
\begin{equation}
\left< u \big | u\right> = \sum_{n=1}^{\infty} \left| u_{\alpha_n} \right|^2 n^{m}
\end{equation}
converging for all $m \in \mathbb N$. For a Hilbert space, we require a much weaker condition, as we do not have the rapidly increasing $n^{m}$ in each term of the summand. We define this space as $\Omega$, and note that always $\Omega \subseteq \mathcal H$ \cite{ballentine}. The ramifications of the extra normalization requirement for a vector to be in $\Omega$ can be thought of as a requirement for an extremely fast decay as $n \rightarrow \infty$. We now define the space of interest, called the conjugate space\index{Conjugate Space} of $\Omega$, and written as $\Omega^{\times}$ in terms of its member vectors \cite{gamelin}. Any vector $\left| w \right>$ belongs to $\Omega^{\times}$ if
\begin{figure}[tb] 
\begin{center} 
\includegraphics[width=0.9 \linewidth]{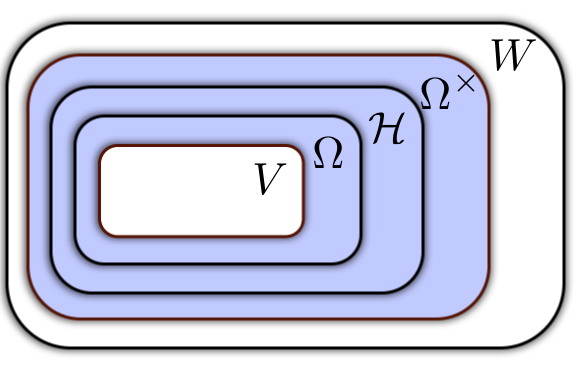}
\end{center} 
\caption[Venn diagram of a rigged Hilbert space triplet]{The spaces $V \subseteq \Omega \subseteq \mathcal H \subseteq \Omega^{\times} \subseteq W$. The area shaded blue is the rigged Hilbert space triplet.}\label{fig:riggedhilbertspace} 
\end{figure}
\begin{equation}
\left< w \big | u \right>= \sum_{n=1}^{\infty} w_n^* u_n 
\end{equation}
converges for all $\left| u \right> \in \Omega$ and $\left< w \right|$ is continuous on $\Omega$. Since we noted that for a vector $\left| u \right>$ to be in $\Omega$, it must vanish very quickly at infinity, $\left| w \right>$ is not nearly as restricted as a vector in $\mathcal H$. Thus, we have the triplet
\begin{equation}
\Omega \subseteq \mathcal H \subseteq \Omega^{\times},
\end{equation}
which is called a \textbf{rigged Hilbert Space triplet}\index{Rigged Hilbert Space Triplet}, and is shown in figure \ref{fig:riggedhilbertspace} \cite{ballentine}.\footnote{The argument used here is rather subtle. If the reader is not clear on the details, it will not impair the comprehension of later sections. To thoroughly understand this material, we recommend first reading the treatment of normed linear spaces in ref. \cite{gamelin}, and then the discussion of rigged Hilbert spaces in refs. \cite{ballentine} and \cite{sudbery}.} We noted earlier that the set of eigenvectors of a self-adjoint operator need not form a basis for that operator's space if the space has infinite dimension. This means that the spectral theorem would break down, which is what we wish to avoid. Fortunately, a generalized spectral theorem\index{Spectral Theorem!Generalized} has been proven for rigged Hilbert space triplets, which states that any self adjoint operator in $\mathcal H$ has eigenvectors in $\Omega^{\times}$ that form a basis for $\mathcal H$ \cite{ballentine}. Due to this, we will work in a rigged Hilbert space triplet, which we will normally denote by the corresponding Hilbert space, $\mathcal H$. We do this with the understanding that to be completely rigorous, it might be necessary to switch between the component sets of the triplet on a case-by-case basis.

Now that we have outlined the space in which we will be working, there is an important special case of an infinite dimensional basis that we need to examine. If our basis is \textbf{continuous}\index{Basis!Continuous}, then we can convert all of our abstract summation formulas into integral forms, which are used very frequently in quantum mechanics, since the two most popular bases (position and momentum) are usually continuous.\footnote{A common form of confusion when first studying quantum mechanics is the abstract notion of vectors. In classical mechanics, a vector might point to a particular spot in a physical space. However, in quantum mechanics, a vector can have infinite dimensionality, and so can effectively point to every point in a configuration space simultaneously, with varying magnitude. For this reason, a very clear distinction must be drawn between the vectors used in the formalism of quantum mechanics and the everyday vectors used in classical mechanics. } Specifically, suppose we have a continuous, orthonormal basis for a rigged Hilbert space $\mathcal H$ given by $\big\{ \left| \phi \right> \big \}_{\phi \in \Phi}$, where $\Phi$ is a real interval. Then, if we have \cite{ballentine} 
\begin{equation}
\left| u \right> = \sum_{\phi \in \Phi} u_{\phi} \left| \phi \right>, \,\, \left| v \right> = \sum_{\phi \in \Phi} v_{\phi} \left| \phi \right>,
\end{equation}
we find a special case of eqn. \ref{eqn:diracinner}. This is\index{Integral Form!Inner Product}\index{Integral Form!Trace}\index{Integral Form!Spectral Theorem}
\begin{equation}
\left< u \big| v \right> = \int_{\Phi}d \phi \cdot  u^*_{\phi} v_{\phi},
\end{equation} 
where the integral is taken over the real interval $\Phi$. Similarly, for an operator $\hat A$, definition \ref{defn:trace} becomes \cite{ballentine}
\begin{equation}
\mathrm{Tr} \left( \hat A \right) = \int_{\Phi} d \phi \cdot \left< \phi \right| \hat A \left| \phi \right>,
\end{equation}
and for self-adjoint $\hat A$, theorem \ref{thm:spectral} is
\begin{equation}
\hat A = \int_{\Phi}d \phi \cdot  a_{\phi} \left| \phi \right> \left< \phi \right|.
\end{equation}
When working in a continuous basis, these integral forms of the inner product, trace, and spectral theorem will often be more useful in calculations than their abstract sum counterparts, and we make extensive use of them in chapter 
\ref{chap:dynamics}.

\chapter{Formal Structure of Quantum Mechanics}\label{chap:quantum_formal}
\lettrine[lines=2, lhang=0.33, loversize=0.1]{W}e now use the mathematical tools developed last chapter to set the stage for quantum mechanics. We begin by listing the correspondence rules that tell us how to represent physical objects mathematically. Then, we develop the fundamental quantum mechanical concept of the state and its associated operator. Next, we investigate the treatment of composite quantum mechanical systems. Throughout this chapter, we work in discrete bases to simplify our calculations and improve clarity. However, following the rigged Hibert space formalism developed in section \ref{sec:infdim}, translating the definitions in this section to an infinite-dimensional space is straightforward both mathematically and physically. 

\section{Fundamental Correspondence Rules of Quantum Mechanics} \label{sec:posts}
At the core of the foundation of quantum mechanics are three rules. The first two tell us how to represent a physical object and describe its physical properties mathematically, and the  third tells us how the the object and properties are connected. These three rules permit us to state a physical problem mathematically, work the problem mathematically, and then interpret the mathematical result physically \cite{ballentine}.

The first physical object of concern is the \textbf{state}, which completely describes the physical aspects of some system \cite{ballentine}. For instance, we might speak of the state of a hydrogen atom, the state of a photon, or a state of thermal equilibrium between two thermal baths. 

\begin{boxedaxm}{State Operator\index{State Operator}}{axm:state}
We represent each physical state as a unique linear operator that is self-adjoint, nonnegative, and of unit trace, which acts on a Rigged Hilbert Space $\mathcal H$. We write this operator $\hat{\rho}$ and call it the \textbf{state operator}.
\end{boxedaxm}
Now that we have introduced the state, we can discuss the physical concepts used to describe states. These concepts include momentum, energy, and position, and are collectively known as dynamical variables \cite{ballentine}.

\begin{boxedaxm}{Observable\index{Observable}}{}
We represent each dynamical variable as a Hermitian linear operator acting on a rigged Hilbert space $\mathcal H$ whose eigenvalues represent all possible values of the dynamical variable. We write this operator using our hat $\left(\, \hat{ }\, \right)$ notation, and call it an \textbf{observable}.
\end{boxedaxm}
We now link the first two axioms with the third \cite{ballentine}.

\begin{boxedaxm}{Expectation Value\index{Expectation Value}}{axm:expectation}
The \textbf{expectation value}, or average measurement\index{Measurement} of the value of an observable $\hat{\mathcal O}$ over infinitely many identically prepared states (called a virtual ensemble of states) is written as $\left< \hat{\mathcal O} \right>$ and given by
\begin{equation}
\left< \hat{\mathcal O} \right> \equiv \mathrm{Tr} \left( \hat{\rho} \hat{\mathcal O } \right).
\end{equation}
\end{boxedaxm}

Though we claimed that these three axioms form the fundamental framework of modern quantum mechanics, they most likely seem foreign to the reader who has seen undergraduate material. In the next section, we work with the state operator and show that, in a special case, the formalism following from the correspondence rules outlined above is identical to that used in introductory quantum mechanics courses.

\section{The State Operator}
In axiom \ref{axm:state}, we defined $\hat{\rho}$, the state operator. However, the formal definition is very abstract, so in this section we investigate some of the properties of the state operator in an attempt to solidify its meaning. Physicists divide quantum mechanical states, and thus state operators, into two broad categories. Any given state is either called \textbf{pure} or \textbf{impure}. Sometimes, impure states are also referred to as mixtures or mixed states. We now precisely define a pure state \cite{ballentine}.

\begin{boxeddefn}{Pure State\index{State!Pure}\index{State!Impure}\index{State!Vector}}{defn:pure}
A given state is called \textbf{pure} if its corresponding unique state operator, $\hat{\rho}$, can be written as
\begin{equation}
\hat{\rho} \equiv \left| \psi \right> \left< \psi \right|,
\end{equation}
where $\left| \psi \right> \in \mathcal H$ is called the state vector in a rigged Hilbert space $\mathcal H$, $\left< \psi \right| \in \mathcal H^*$ is the linear functional corresponding to $\left| \psi \right>$, and $\left< \psi \big| \psi \right>=1$. If a state cannot be so represented, it is called \textbf{impure}.
\end{boxeddefn}

Although the importance of pure and impure states is not yet evident, we will eventually need an efficient method of distinguishing between them. The definition, which is phrased as an existence argument, is not well-suited to this purpose. To generate a more useful relationship, consider a pure state. We have
\begin{equation}
\hat{\rho}^2=\hat{\rho} \hat{\rho} = \left( \left| \psi \right> \left< \psi \right| \right) \left( \left| \psi \right> \left< \psi \right| \right)  = \left| \psi \right> \left( \left< \psi \big| \psi \right> \right) \left< \psi \right| = \left| \psi \right>( 1 )\left< \psi \right| = \left| \psi \right> \left< \psi \right| = \hat{ \rho}.
\end{equation}
Thus, if a state is pure, it necessarily follows \cite{ballentine}
\begin{equation}
\hat{\rho}^2 = \hat{\rho}.
\end{equation}
Although seemingly a weaker condition, this result turns out to also be sufficient to describe a pure state. To show this, we suppose that our state space is discrete and has dimension $D$.\footnote{This is mainly for our convenience. The argument for an infinite-dimensional space is similar, but involves the generalized spectral theorem on our rigged Hilbert space.} Invoking the spectral theorem, theorem \ref{thm:spectral}, we write
\begin{equation}
\hat{ \rho} =  \sum_{n=1}^{D} \rho_n \left| \phi_n \right> \left< \phi_n \right|, \label{eqn:specrho1}
\end{equation}
where $\{\rho_n\}_{n=1}^D$ is the spectrum of eigenvalues for $\hat{\rho}$, corresponding to the unit-normed eigenvectors of $\hat{\rho}$, $\big\{\left| \phi_n \right> \big \}_{n=1}^D$. If we consider some $1 \leq j \leq D$ with $ j,D\in \mathbb Z$ and let $\hat{\rho} = \hat{\rho}^2$, we have
\begin{equation}
\hat{\rho} \left| \phi_{j} \right> = \hat{\rho}^2 \left| \phi_{j} \right>, 
\end{equation}
which is
\begin{equation}
\rho_{j} \left| \phi_{j} \right> = \rho_{j}^2 \left| \phi_{j} \right>,
\end{equation}
so 
\begin{equation}
\rho_{j}=\rho_{j}^2
\end{equation}
or
\begin{equation}
\rho_j \left(1-  \rho_j  \right) = 0.
\end{equation}
Since all of the eigenvalues of $\hat{\rho}$ must also follow this relationship, they must all either be one or zero. But by axiom \ref{axm:state}, $\mathrm{Tr}\left( \hat{\rho} \right) = 1$, so exactly one of the eigenvalues must be one, while all the others are zero. Thus, eqn. \ref{eqn:specrho1} becomes
\begin{equation}
\hat{ \rho} =   \left| \phi_{q_1} \right> \left< \phi_{q_1} \right|,
\end{equation}
where we have taken $q_1=1$. Evidently, $\hat{ \rho}$ is a pure state, and we have shown sufficiency \cite{ballentine}. 

At this point, it is logical to inquire about the necessity of the state operator, as opposed to a state vector alone. After all, most states treated in introductory quantum mechanics are readily represented as state vectors. However, there are many states that are prepared statistically, and so cannot be represented as a state vector. An example of one of these cases is found in section \ref{sec:bellstate}. These impure states or mixtures\index{State!Impure} turn out to be of the utmost importance when we begin to discuss quantum decoherence, the main focus of this thesis \cite{zurek}.

We now turn our attention to the properties of pure states, and illustrate that the state vectors defining pure state operators behave as expected under our correspondence rules. By axiom \ref{axm:expectation}, we know that the expectation value of the dynamical variable (observable) $\hat{A}$ of a state $\hat{\rho}$ is
\begin{equation}
\left< \hat  A \right> = \mathrm{Tr}\left(  \hat{\rho} \hat{A} \right).
\end{equation}
If $\hat{\rho}$ is a pure state, then we can write
\begin{equation}
\hat{\rho} = \left| \psi \right> \left< \psi \right|.
\end{equation}
Hence, $\left<\hat A \right>$ becomes
\begin{equation}
\left< \hat A \right> =  \mathrm{Tr}\left(   \left| \psi \right> \left< \psi \right| \hat{A} \right),
\end{equation}
which, by definition \ref{defn:trace}, is \cite{ballentine}
\begin{eqnarray} \label{eqn:recoverexp}
\left< \hat  A \right> &=& \sum_{n=1}^{D} \left< \phi_n \right| \left( \left| \psi \right> \left< \psi \right| \hat{A} \right) \left| \phi_n \right> \nonumber\\ 
&=& \sum_{n=1}^{D}\left(  \left< \phi_n \big | \psi \right>\right) \left( \left< \psi \right| \hat{A}  \left| \phi_n \right> \right) \nonumber\\ 
&=& \left< \psi \right| \hat{A} \left| \psi \right>,
\end{eqnarray}
where we have used definition \ref{defn:orthobasis} to pick the basis $\big \{ \left| \phi_{\alpha}\right> \big \}_{\alpha \in \mathbb R}$ to be orthonormal and contain the vector $\left| \psi \right>$.\footnote{This works since $\left| \psi \right>$ is guaranteed to have unit magnitude by definition \ref{defn:pure}.} This is the standard definition for an expectation value in introductory quantum mechanics, which we recover by letting $\hat{\rho}$ be pure \cite{griffiths, cohtan}.

\section{Composite Systems}\label{sec:composite}
In order to model complex physical situations, we will often have to consider multiple, non-isolated states. To facilitate this, we need to develop a method for calculating the state operator of a composite, or combined, quantum system \cite{ballentine}. 

\begin{boxedaxm}{Composite State\index{Composite!State Operator}}{}
Suppose we had a pure composite system composed of $n$ substates, $\big \{ \hat{\rho}_i \big\}_{i=1}^n$. Then, the \textbf{composite state operator} $\hat{\rho}$ of this combined system is given by\index{Composite!State Vector}
\begin{equation}
\hat{\rho} \equiv \hat{\rho}_1 \otimes \hat{\rho}_2 \otimes \cdot \cdot \cdot \otimes \hat{\rho}_n,
\end{equation}
where $(\otimes)$ is the tensor product, given in definition \ref{defn:tensor}.
\end{boxedaxm}
Note that if $\hat{\rho}$ is pure, there exists some characteristic state vector $\left| \psi \right>$ of $\hat{ \rho}$ where
\begin{equation}
\left| \psi \right> = \left| \psi_1 \right> \otimes \left| \psi_2 \right> \otimes \cdot \cdot \cdot \otimes \left| \psi_n \right> \label{eqn:stateveccomp}
\end{equation}
and each $\left| \psi_i \right>$ corresponds to $\hat{ \rho}_i$. As an important notational aside, eqn \ref{eqn:stateveccomp} is frequently shortened to \cite{nielsenchuang}
\begin{equation}
\left| \psi \right> = \left| \psi_1\psi_2 ... \psi_n \right> ,
\end{equation}
where the tensor products are taken as implicit in the notation. Just as we discussed dynamical variables associated with certain states, so can we associate dynamical variables with composite systems. In general, an observable of a composite system with $n$ substates is formed by \cite{nielsenchuang}\index{Composite!Observable}
\begin{equation}
\hat{\mathcal O} \equiv \hat{\mathcal O}_1 \otimes \hat{\mathcal O}_2 \otimes \cdot \cdot \cdot \otimes \hat{\mathcal O}_n,
\end{equation}
where each $\hat{\mathcal O}_i$ is an observable of the $i$th substate. We have now extended the concepts of state and dynamical variable to composite systems, so it is logical to treat an expectation value of a composite system. Of course, since a composite system is a state, axiom \ref{axm:expectation} applies, so we have
\begin{equation}
\left< \hat{ \mathcal O } \right> = \mathrm{Tr} \left(\hat{\rho} \hat{ \mathcal O } \right).
\end{equation}
However, composite systems afford us opportunities that single systems do not. Namely, just as we trace over the degrees of freedom of a system to calculate expectation values on that system, we can trace over some of the degrees of freedom of a composite state to focus on a specific subsystem.\footnote{Here, a degree of freedom of a state can be thought of as its dimensionality. It is used analogously with the notion in a general system in classical mechanics, where the dimensionality of a system's configuration space corresponds to the number of degrees of freedom it possesses. For more on this, see ref. \cite{thornton}.} We call this operation the partial trace over a composite system, and we define it precisely below \cite{nielsenchuang}.

\begin{boxeddefn}{Partial Trace\index{Partial Trace}}{def:partialtrace}
Suppose we have an operator
\begin{equation}
\hat{ \mathcal Q} = \hat{\mathcal Q}_1 \otimes \hat{\mathcal Q}_2 \otimes \cdot \cdot \cdot \otimes \hat{\mathcal Q}_n.
\end{equation}
The \textbf{partial trace} of $\hat{\mathcal Q}$ over $\hat{\mathcal Q}_i$ is defined by
\begin{equation}
\mathrm{Tr}_i \left( \hat{ \mathcal Q} \right) \equiv \hat{\mathcal Q}_1 \otimes \hat{\mathcal Q}_2 \otimes \cdot \cdot \cdot \otimes \hat{\mathcal Q}_{i-1}  \cdot \mathrm{Tr} \left( \hat{\mathcal Q}_i \right) \cdot \hat{\mathcal Q}_{i+1} \otimes \cdot \cdot \cdot \otimes \hat{\mathcal Q}_n.
\end{equation}
\end{boxeddefn}
If the partial trace is applied to a composite system repeatedly such that all but one of the subsystem state operators are traced out, the remaining operator is called a reduced state operator \cite{nielsenchuang}. 

\begin{boxeddefn}{Reduced State Operator\index{Reduced State Operator}}{def:redstate}
Suppose we have a composite system $\hat{\rho}$ with $n$ subsystems. The \textbf{reduced state operator for subsystem i} is defined by
\begin{equation}
\hat{\rho}^{(i)} = \mathrm{Tr}_1 \circ \mathrm{Tr}_2 \circ \cdot \cdot \cdot  \circ \mathrm{Tr}_{i-1} \circ \mathrm{Tr}_{i+1} \circ \cdot \cdot \cdot  \circ \mathrm{Tr}_n \left( \hat{\rho} \right).
\end{equation}
\end{boxeddefn}
The partial trace and reduced state operator turn out to be essential in the analysis of composite systems, although that fact is not immediately obvious. To illustrate this, we consider some observable $\hat{\mathcal O}_m$ that acts only on the $k_m$th subsystem of a composite system. We choose a basis $\big\{\left| \Phi_k \right> \big \}_{k=1}^n$, where each element is formed by the Kronecker product of the basis elements of the corresponding subsystems. That is, each basis vector has the form $\left| \Phi_k \right> = \left| \phi_1 \phi_2 ... \phi_n \right>$, where each $\phi_l$ is one of the orthonormal basis vectors of the $l$th substate space. Then, from axiom \ref{axm:expectation}, we have
\begin{eqnarray}
\left< \hat{\mathcal O}_m  \right> &=& \mathrm{Tr} \left( \hat{ \rho} \hat{ \mathcal O}_m \right) \nonumber \\
&=& \sum_{k=1}^n \left< \Phi_k \right|\hat{ \rho} \hat{ \mathcal O}_m \left| \Phi_k \right> \nonumber \\
&=& \sum_{k_1,k_2,...,k_n} \left< \phi_{k_1} \phi_{k_2} ... \phi_{k_n} \right| \hat{ \rho}\hat{ \mathcal O}_m \left| \phi_{k_1} \phi_{k_2} ... \phi_{k_n}\right>.
\end{eqnarray}
We use the resolution of the identity, eqn. \ref{eqn:projector}, to write our expectation value as
\begin{equation}
 \sum_{k_1,k_2,...,k_n} \left< \phi_{k_1} \phi_{k_2} ... \phi_{k_n} \right| \hat{ \rho} \left( \sum_{j_1,j_2,...,j_n} \left| \phi_{j_1} \phi_{j_2} ... \phi_{j_n} \right> \left< \phi_{j_1} \phi_{j_2} ... \phi_{j_n}\right| \right)\hat{ \mathcal O}_m \left| \phi_{k_1} \phi_{k_2} ... \phi_{k_n}\right>,
\end{equation}
where $\left| \phi_{j_1} \phi_{j_2} ... \phi_{j_n} \right>$ corresponds to a basis vector. This becomes
\begin{equation}
 \sum_{k,j}  \left< \phi_{k_1} \phi_{k_2} ... \phi_{k_n} \right| \hat{ \rho} \left| \phi_{j_1} \phi_{j_2} ... \phi_{j_n} \right>  \left< \phi_{j_1} \phi_{j_2} ... \phi_{j_n}\right| \hat{ \mathcal O}_m \left| \phi_{k_1} \phi_{k_2} ... \phi_{k_n}\right>.
\end{equation}
If the observable $\hat{\mathcal O}$ acts as identity on all but the $m$th subsystem, by eqn. \ref{eqn:tensorop}, we have
\begin{equation}
 \sum_{k,j}  \left< \phi_{k_1} \phi_{k_2} ... \phi_{k_n} \right| \hat{ \rho} \left| \phi_{j_1} \phi_{j_2} ... \phi_{j_n} \right> \left< \phi_{j_m} \right| \hat{ \mathcal O}_m \left| \phi_{k_m} \right>\left< \phi_{j_1} ... \phi_{j_{m-1}} \phi_{j_{m+1}}...\phi_{j_n} \big | \phi_{k_1} ... \phi_{k_{m-1}} \phi_{k_{m+1}}...\phi_{k_n} \right>.
\end{equation}
Since our chosen basis is orthonormal, for any non-zero term in the sum, we must have $j=k$ (except for $j_m$ and $k_m$), in which case the final inner produce is unity. Hence, we get
\begin{equation}
\sum_{k_1,k_2,...,k_n,j_m}  \left< \phi_{k_1} \phi_{k_2} ... \phi_{k_n} \right| \hat{ \rho} \left| \phi_{k_1}  ...\phi_{k_m-1} \phi_{j_m} \phi_{k_m+1}... \phi_{k_n} \right> \left< \phi_{j_m} \right| \hat{ \mathcal O}_m \left| \phi_{k_m} \right>.
\end{equation}
If we apply eqn. \ref{eqn:tensorop}, letting $\hat{\rho}=\hat{\rho}_1 \otimes \hat{\rho}_2 \otimes \cdot \cdot \cdot  \otimes \hat{\rho}_n$, we have
\begin{equation}
\sum_{k_1,k_2,...,k_n,j_m} \left< \phi_{k_1} \right| \hat{\rho}_1 \left| \phi_{k_1} \right> \left< \phi_{k_2} \right| \hat{\rho}_2 \left| \phi_{k_2}\right> \cdot \cdot \cdot  \left< \phi_{k_m} \right| \hat{\rho}_m \left| \phi_{j_m} \right> \cdot \cdot \cdot  \left< \phi_{k_n} \right| \hat{\rho}_n \left| \phi_{k_n} \right>\left< \phi_{j_m} \right| \hat{ \mathcal O}_m \left| \phi_{k_m} \right>,
\end{equation}
or
\begin{equation}
\sum_{k_m,j_m} \mathrm{Tr}\left( \hat{\rho}_1 \right) \mathrm{Tr}\left( \hat{\rho}_2 \right) \cdot \cdot \cdot  \left< \phi_{k_m} \right| \hat{\rho}_m \left| \phi_{j_m} \right> \cdot \cdot \cdot  \mathrm{Tr}\left( \hat{\rho}_n \right) \left< \phi_{j_m} \right| \hat{ \mathcal O}_m \left| \phi_{k_m} \right>.
\end{equation}
Since each trace is just a scalar, we can write
\begin{equation}
\sum_{k_m}\left< \phi_{k_m}\right|  \mathrm{Tr}\left( \hat{\rho}_1 \right) \mathrm{Tr}\left( \hat{\rho}_2 \right) \cdot \cdot \cdot  \hat{\rho}_m \cdot \cdot \cdot  \mathrm{Tr}\left( \hat{\rho}_n \right) \left( \sum_{j_m} \left| \phi_{j_m} \right>  \left< \phi_{j_m} \right| \right) \hat{ \mathcal O}_m \left| \phi_{k_m} \right>.
\end{equation}
Recognizing the definition \ref{def:redstate} for the reduced state operator and the resolution of the identity from eqn. \ref{eqn:projector}, we find \cite{ballentine}
\begin{boxedeqn}{}
\left< \hat{\mathcal O}_m  \right> = \sum_{k_m} \left< \phi_{k_m}\right|  \hat{\rho}^{(m)} \left( \, \hat 1 \, \right) \hat{\mathcal O}_m \left| \phi_{k_m} \right> = \mathrm{Tr} \left( \hat{\rho}^{(m)} \hat{\mathcal O}_m \right).
\end{boxedeqn}
Due to this remarkable result, we know that the reduced state operator for a particular subsystem is enough to tell us about any observable that only depends on the subsystem. Further, we end up with a formula for the expectation value of a component observable very similar to axiom \ref{axm:expectation} for observables of the full system.

\section{Quantum Superposition}\label{sec:quantumsup}
Though we have introduced some of the basic formalism of the state, we are still missing one of the key facets of quantum mechanics. This piece is the superposition principle, which, at the time of this writing, is one of the core aspects of quantum mechanics that no one fully understands. However, due to repeated experimental evidence, we take it as an axiom. 

\begin{figure}[t] 
\begin{center} 
\includegraphics[width=0.7 \linewidth]{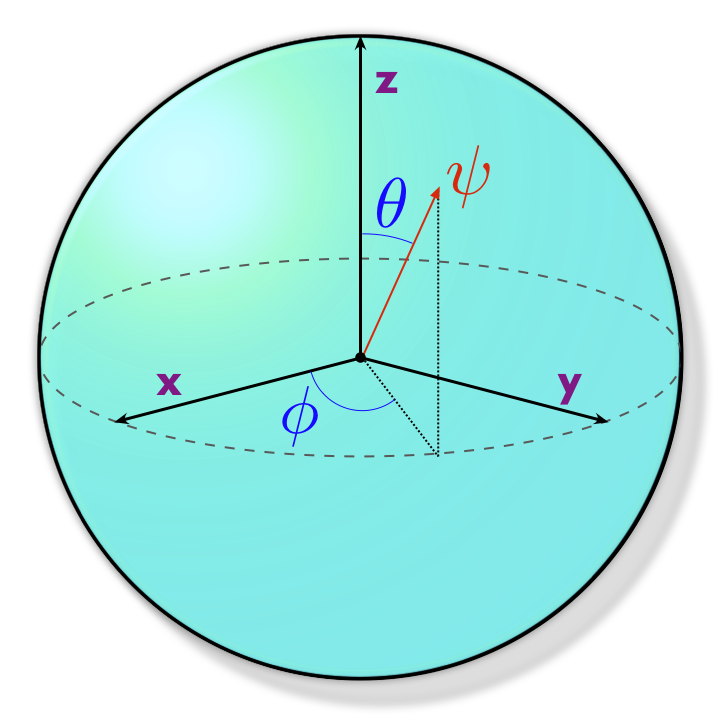}
\end{center} 
\caption[The Bloch sphere representation]{Two-state systems can be visualized as being vectors on a two-sphere, known in quantum physics as the Bloch sphere. The angles $\phi$ and $\theta$ are defined in eqn. \ref{eqn:bloch_def} for pure states, and the axes x, y, and z are defined in eqn. \ref{eqn:paulivec} for all states. \label{fig:bloch_sphere}}
\end{figure}

\begin{boxedaxm}{Superposition Principle\index{Superposition Principle}}{axm:sup}
Suppose that a system can be in two possible states, represented by the state vectors $\left| 0 \right>$ and $\left| 1 \right>$. Then, 
\begin{equation}
\left| \psi \right> = \alpha \left| 0 \right> + \beta \left| 1 \right>,
\end{equation}
where $\alpha, \beta \in \mathbb C$, is also a valid state of the system, provided that $\left| \alpha \right|^2 + \left| \beta \right|^2 = 1$.
\end{boxedaxm}
The superposition principle allows us to create new and intriguing states that we would not have access to otherwise. In fact, if we have $n$ linearly independent states of a system, any point on the unit n-sphere corresponds to a valid state of the system.\footnote{The reader might wonder why the superposition principle is necessary, after all, we know that state vectors exist in a Hilbert space, and Hilbert spaces act linearly. However, we were not guaranteed until now that any vector of unit norm in Hilbert space represents a valid physical situation. The superposition principle gives us this, which allows us great freedom in constructing states.} If we consider a two-state system with an orthonormal basis $\big \{ \left| 0 \right> , \left| 1 \right> \big \}$, the 2-sphere of possible states guaranteed by the superposition principle is conveniently visualized imbedded in 3-space. This visualization of a two-state system\index{Two-State System} is called the \textbf{Bloch sphere representation}\index{Bloch!Sphere}, and is pictured in figure \ref{fig:bloch_sphere} \cite{nielsenchuang}. To calculate the position of a system in Bloch space, we use the formula
\begin{equation} \label{eqn:blochvec}
\hat{\rho} \leftrightarrow r_0 1 + \left< r \big|  \sigma \right>, 
\end{equation}
where $\left| r \right>$ is the 3-vector,
\begin{equation}
\left| r \right> \equiv r_1 \left| e_1 \right> + r_2 \left| e_2 \right> + r_3 \left| e_3 \right>,
\end{equation}
 and $\vec{\sigma}$ is the vector of Pauli spin matrices,
\begin{equation} 
\left| \sigma \right> \equiv  \sigma_x \left| e_1 \right> + \sigma_y \left| e_2 \right> + \sigma_z \left| e_3 \right>. \label{eqn:paulivec}
\end{equation}
The Pauli matrices are

\begin{equation}
\hat{\sigma}_x \leftrightarrow \left(\begin{array}{cc}0 & 1 \\1 & 0\end{array}\right),
\end{equation}
\begin{equation}
\hat{\sigma}_y \leftrightarrow \left(\begin{array}{cc}0 & -i \\i & 0\end{array}\right),
\end{equation}
and
\begin{equation}
\hat{\sigma}_z \leftrightarrow \left(\begin{array}{cc}1 & 0 \\0 & -1\end{array}\right).
\end{equation}
Writing eqn. \ref{eqn:blochvec} explicitly, we find
\begin{equation}
\hat \rho \leftrightarrow \left( \begin{array}{cc} r_0 + r_3 & r_1 -i r_2 \\ r_1+i r_2 & r_0 - r_3 \end{array} \right).
\end{equation}
This is trivially a basis for all two by two matrices, so we can indeed represent any $\hat \rho$ by eqn. \ref{eqn:blochvec}. Further, if we use the fact that $\mathrm{Tr}\left(\hat \rho \right) = 1$, we know
\begin{equation}
\mathrm{Tr}\left(\hat \rho \right)  \leftrightarrow ( r_0 + r_3) + (r_0 - r_3) = 2r_0 = 1,
\end{equation}
so $r_0=1/2$. With this constraint in mind, it is conventional to write eqn. \ref{eqn:blochvec} as  \cite{nielsenchuang}
\begin{boxedeqn}{eqn:bloch2}
\hat{\rho} \leftrightarrow \frac{1 + \left< r \big|  \sigma \right>}{2}.
\end{boxedeqn}
Also, since $\hat \rho$ is self-adjoint, the diagonal entries must all be real, so $r_3 \in \mathbb R$. By the same reasoning,
\begin{equation}
r_1+ir_2 = (r_1 - ir_2)^*.
\end{equation}
Since $r_1$ and $r_2$ are arbitrary, we can choose either of them to be zero, and the resulting equation must hold for all values of the other. Hence, $r_1 = r_1^*$ and $r_2 = r_2^*$, so both $r_1$ and $r_2$ are real, and $\left| r \right>$ is a real-valued vector. Since $\left| r \right>$ is real, we use it as a position vector that tells us the location of the system in Bloch space and call it the Bloch vector.\index{Bloch!Vector} If we have a pure state
\begin{equation}
\left| \psi \right> = \alpha \left| 0 \right> + \beta \left| 1 \right>,
\end{equation}
we can express the location of the state in terms of the familiar polar and azimuthal angles of polar-spherical coordinates. Taking into account our redefined, conventional $\left| r \right>$, eqn. \ref{eqn:bloch2} is 
\begin{equation}
 \frac{1 + \left< r \big| \sigma \right>}{2} \leftrightarrow
 \frac{1}{2} \left(\begin{array}{cc}1+r_z & r_x-i r_y \\r_x+ir_y & 1-r_z\end{array}\right).
\end{equation}
We use the polar-spherical coordinate identities for unit vectors
\begin{eqnarray}
r_x &=& \sin \theta \cos \phi, \nonumber \\
r_y  &=& \sin \theta \sin \phi, \nonumber \\
r_z &=& \cos \theta,
\end{eqnarray}
to determine
\begin{eqnarray}
 \frac{1}{2} \left(\begin{array}{cc}1+r_z & r_x-i r_y \\r_x+ir_y & 1-r_z\end{array}\right) &=&  \frac{1}{2} \left(\begin{array}{cc}1+\cos \theta & \sin \theta \cos \phi-i \sin \theta \sin \phi  \\ \sin \theta \cos \phi+i \sin \theta \sin \phi  & 1- \cos \theta \end{array}\right) \nonumber \\
 &=& \frac{1}{2} \left(\begin{array}{cc}1+\cos \theta & \sin \theta e^{-i\phi} \\ \sin \theta e^{i \phi}  & 1- \cos \theta \end{array}\right) \nonumber \\
  &=& \frac{1}{2} \left(\begin{array}{cc} 2 \cos^2 \left( \frac{\theta}{2} \right) & 2 \sin\left( \frac{\theta}{2} \right) \cos \left( \frac{\theta}{2} \right) e^{-i\phi} \\ 2 \sin\left( \frac{\theta}{2} \right) \cos \left( \frac{\theta}{2} \right) e^{i \phi}  & 2 \sin^2 \left( \frac{\theta}{2} \right) \end{array}\right) \nonumber \\
    &=& \left(\begin{array}{cc}  \cos^2 \left( \frac{\theta}{2} \right) & \sin\left( \frac{\theta}{2} \right) \cos \left( \frac{\theta}{2} \right) e^{-i\phi} \\ \sin\left( \frac{\theta}{2} \right) \cos \left( \frac{\theta}{2} \right) e^{i \phi}  &  \sin^2 \left( \frac{\theta}{2} \right) \end{array}\right).
\end{eqnarray}
If we let $\alpha \equiv \cos \left( \theta / 2 \right)$ and $\beta \equiv e^{i \phi} \sin \left( \theta / 2 \right)$, the right side of eqn. \ref{eqn:blochvec} becomes
\begin{equation}
\left(\begin{array}{cc}  \left| \alpha \right| ^2 & \alpha \beta^* \\ \beta \alpha^* & \left| \beta \right|^2 \end{array}\right) \leftrightarrow \left| \psi \right> \left< \psi \right| = \hat{\rho}.
\end{equation}
Hence, the state vector of the pure state is \cite{nielsenchuang}
\begin{boxedeqn}{}
\left| \psi \right> =\cos \left( \frac{\theta}{2} \right) \left| 0 \right> + e^{i \phi} \sin \left( \frac{\theta}{2} \right) \left| 1 \right>. \label{eqn:bloch_def}
\end{boxedeqn}
We note that the coefficient on $\left| 0 \right>$ is apparently restricted to be real. However, unlike state operators, state vectors are not unique; physically identical state vectors may differ by a phase factor $e^{i \gamma}$ \cite{ballentine}.

The notion of superposition\index{Superposition} also enables us to refine our classification of composite systems. Besides distinguishing between pure and impure states, physicists subdivide composite pure states into two categories: entangled states and product states. 

\begin{boxeddefn}{Product State\index{Product State}\index{Entangled State}}{}
Suppose $\hat{\rho}$ is a pure composite quantum system with associated state vector $\left| \psi \right>$. If there exist state vectors $\left| \phi_1 \right>$ and $\left| \phi_2 \right> $ such that
\begin{equation}
\left| \psi \right> = \left| \phi_1 \right> \otimes \left| \phi_2 \right>,
\end{equation}
then we call $\left| \psi \right>$ a \textbf{product state}. If no such vectors exist, then we say $\left| \psi \right>$ is \textbf{entangled}.
\end{boxeddefn}

To construct entangled states, we take product states and put them into superposition. In illustration of this concept, we consider the following example.

\section{Example: The Bell State}\label{sec:bellstate}
An important example of an entangled state of two two-state systems is called the \textbf{Bell State}\index{Bell State}. Before we define this system, we need to develop some machinery to work with two-state\index{Two-State System} systems. We use the orthonormal basis set introduced previously for a single, pure, two-state system, $\big \{ \left| 0 \right> , \left| 1 \right> \big \}$, which we represent as column matrices by
\begin{eqnarray}
\left| 0 \right>&\leftrightarrow& \left( \begin{array}{c} 1 \\ 0 \end{array} \right), \nonumber \\
\left| 1 \right> &\leftrightarrow& \left( \begin{array}{c} 0 \\ 1 \end{array} \right). 
\end{eqnarray}
In this representation, we define an orthonormal basis for two of these two-state systems as \cite{nielsenchuang}
\begin{equation}
\big \{ \left| 0 \right>\otimes\left| 0 \right> , \left| 0 \right>\otimes \left| 1 \right>,\left| 1 \right> \otimes \left| 0 \right> ,\left| 1 \right>\otimes\left| 1 \right>\big \} = \big \{ \left| 00 \right>,\left| 01 \right>,\left| 10 \right>,\left| 11 \right> \big \},
\end{equation}
which have matrix representations
\begin{eqnarray}
&\left| 00 \right>& \leftrightarrow \left( \begin{array}{c} 1 \\ 0 \\0 \\0 \end{array} \right), \, \, \,
\left| 01 \right> \leftrightarrow \left( \begin{array}{c} 0 \\ 1\\0\\0 \end{array} \right),  \nonumber \\
&\left| 10 \right>& \leftrightarrow \left( \begin{array}{c} 0 \\ 0 \\1\\0 \end{array} \right), \, \, \,
\left| 11 \right> \leftrightarrow \left( \begin{array}{c} 0 \\ 0\\0\\1 \end{array} \right). 
\end{eqnarray}
By the superposition principle, we define the state 
\begin{equation}
\left| \psi_B \right> \equiv \frac{\left| 00 \right> + \left| 11 \right> }{\sqrt{2}} \leftrightarrow \left(\begin{array}{c} \frac{1}{\sqrt 2} \\ 0 \\ 0\\ \frac{1}{\sqrt 2} \end{array}\right),
\end{equation}
which is the Bell state. To check if this state is entangled, we see if we can write $\left| \psi_B \right> = \left| \phi_A \right> \otimes \left| \phi_B \right>$ for some vectors $\left| \phi_A \right>$ and $\left| \phi_B \right>$. As matrices, this equation is
\begin{equation}
\left(\begin{array}{c} \frac{1}{\sqrt 2} \\ 0 \\ 0\\ \frac{1}{\sqrt 2} \end{array}\right) = \left(\begin{array}{c} a_1\\ a_2  \end{array}\right) \otimes \left(\begin{array}{c} b_1 \\ b_2 \end{array}\right) = \left(\begin{array}{c}a_1 b_1 \\ a_1 b_2 \\ a_2 b_1\\ a_2 b_2 \end{array}\right).
\end{equation}
This is a system of four simultaneous equations, $\frac{1}{ \sqrt 2} =  a_1 b_1$, $0 = a_1 b_2$, $0 = a_2 b_1$, and $\frac{1}{ \sqrt 2} = a_2 b_2$. Since $\frac{1}{ \sqrt 2} =  a_1 b_1$, $a_1\neq 0$ and $b_1 \neq 0$. Then, since $a_1 b_2 = 0$, $b_2=0$. But $\frac{1}{ \sqrt 2} = a_2 b_2$, so $b_2 \neq 0 $, which is a contradiction. Hence, $\left| \phi_A \right> $ and $\left| \phi_B \right>$ do not exist, so $\left| \psi_B \right>$ is entangled. \cite{nielsenchuang}

Next, we compute the state operator corresponding to $\left| \psi_B \right>$. By definition \ref{defn:pure}, since the Bell state is pure by construction, its state operator is
\begin{eqnarray}
\hat{ \rho} &=& \left| \psi_B \right> \left< \psi_B \right| \nonumber \\
		&=& \left( \frac{\left| 00 \right> + \left| 11 \right> }{\sqrt{2}} \right) \left( \frac{\left< 00 \right| + \left< 11 \right| }{\sqrt{2}} \right) \nonumber \\
		&=& \frac{\left| 00 \right> \left< 00 \right| +\left| 00 \right> \left< 11\right| +\left| 11 \right> \left< 00 \right| +\left| 11 \right> \left< 11 \right|}{2} \nonumber \\
		&\leftrightarrow& \left( 
		\begin{array}{cccc}
		\frac{1}{2} & 0 & 0 & \frac{1}{2} \\
		0 & 0 & 0 & 0 \\
		0 & 0 & 0 & 0 \\
		\frac{1}{2} & 0 & 0 & \frac{1}{2} \\
		\end{array}
		\right).
\end{eqnarray}
Even though we constructed the Bell state from a state vector, we will explicitly verify its purity as an example. We find
\begin{eqnarray}
\left( \hat{ \rho} \right)^2 &=& \left(\frac{\left| 00 \right> \left< 00 \right| +\left| 00 \right> \left< 11\right| +\left| 11 \right> \left< 00 \right| +\left| 11 \right> \left< 11 \right|}{2} \right) \left(\frac{\left| 00 \right> \left< 00 \right| +\left| 00 \right> \left< 11\right| +\left| 11 \right> \left< 00 \right| +\left| 11 \right> \left< 11 \right|}{2} \right) \nonumber \\
&=& \frac{2 \left(\left| 00 \right> \left< 00 \right| +\left| 00 \right> \left< 11\right| +\left| 11 \right> \left< 00 \right| +\left| 11 \right> \left< 11 \right| \right)}{4} \nonumber \\
&=& \frac{ \left(\left| 00 \right> \left< 00 \right| +\left| 00 \right> \left< 11\right| +\left| 11 \right> \left< 00 \right| +\left| 11 \right> \left< 11 \right| \right)}{2} \nonumber \\
&=& \rho,
\end{eqnarray}
which confirms that the Bell state is pure. 

Next, suppose we want to measure some particular facet of the first subsystem. Since the Bell state is entangled, we cannot ``eyeball" the result, but rather we need to use the reduced state machinery we developed in definition \ref{def:redstate}. The reduced state operator for the first subsystem is

\begin{eqnarray}\label{eqn:thisisamixture}
\hat{\rho}^{(1)} &=& \mathrm{Tr}_2 \left( \hat{\rho} \right) \nonumber \\
			&=&  \mathrm{Tr}_2 \left( \frac{\left| 00 \right> \left< 00 \right| +\left| 00 \right> \left< 11\right| +\left| 11 \right> \left< 00 \right| +\left| 11 \right> \left< 11 \right|}{2} \right) \nonumber \\
			&=& \frac{1}{2}  \mathrm{Tr}_2 \left( \left| 0 \right> \left< 0 \right|  \otimes \left| 0 \right> \left< 0 \right|  + \left| 1 \right> \left< 0 \right|  \otimes \left| 1 \right> \left< 0 \right| + \left| 0 \right> \left< 1 \right|  \otimes \left| 0 \right> \left< 1 \right| + \left| 1 \right> \left< 1 \right|  \otimes \left| 1 \right> \left< 1 \right| \right) \nonumber \\
			&=& \frac{1}{2}  \big ( \left| 0 \right> \left< 0 \right| \cdot \mathrm{Tr}\left(  \left| 0 \right> \left< 0 \right| \right) + \left| 1 \right> \left< 0 \right| \cdot \mathrm{Tr}\left( \left| 1 \right> \left< 0 \right|\right) + \left| 0 \right> \left< 1 \right|  \cdot \mathrm{Tr}\left( \left| 0 \right> \left< 1 \right| \right)+ \left| 1 \right> \left< 1 \right| \cdot \mathrm{Tr}\left( \left| 1 \right> \left< 1 \right| \right) \big ) \nonumber \\
			&=& \frac{1}{2}  \big ( \left| 0 \right> \left< 0 \right| \cdot 1 + \left| 1 \right> \left< 0 \right| \cdot 0+ \left| 0 \right> \left< 1 \right| \cdot 0+ \left| 1 \right> \left< 1 \right| \cdot 1 \big ) \nonumber \\			&=& \frac{ \left| 0 \right> \left< 0 \right| + \left| 1 \right> \left< 1 \right| }{2} \nonumber \\
			&\leftrightarrow& \left(\begin{array}{cc}\frac{1}{2} & 0 \\0 & \frac{1}{2}\end{array}\right).
\end{eqnarray}
Oddly enough, 
\begin{equation}
\left( \hat{\rho}^{(1)} \right) ^2 = \frac{ \left| 0 \right> \left< 0 \right| + \left| 1 \right> \left< 1 \right| }{4} \neq \hat{\rho}^{(1)}, \nonumber \\
\end{equation}
so $\hat{\rho}^{(1)}$ is impure \cite{nielsenchuang}. Surprisingly, a pure composite system does not necessarily contain pure subsystems. If we express $\hat{\rho}^{(1)}$ in terms of the Pauli matrices and the identity as in eqn. \ref{eqn:blochvec}, we find that the Bloch vector corresponding to $\hat{\rho}^{(1)}$ is $\left| r\right> = 0$. We already noted that in Bloch space, the unit two-sphere represents all the possible pure state configurations for a two-state system. However, the unit ball represents all state configurations; the impure states have $\left< r \big| r \right> < 1$ \cite{nielsenchuang}. The Bell state, with $\left<  r\big| r \right>=0$, is a special case of a totally mixed or impure state, meaning that the subsystem is entirely statistical (classical). By symmetry, if we had traced out the first subsystem rather than the second, we find $\hat{\rho}^{(1)}=\hat{\rho}^{(2)}$, so we actually have an entangled state composed of totally classical subsystems. 
\section{Projection Onto a Basis}\label{sec:projonbasis}
So far, we have worked mostly in an abstract Hilbert space, although we have taken brief forays into matrix representations of states and observables. In this section, we formalize the notion of a representation of an operator in a basis. We are mainly interested in infinite and continuous bases\index{Basis!Continuous}, which we use to define a very useful structure \cite{cohtan}.

\begin{boxeddefn}{Wavefuntion\index{Wavefunction}}{defn:wavefunction}
Suppose that we have an infinite and continuous basis for $\mathcal H$, $\{ x \}_{x \in \mathbb R}$. Then, for some pure state vector $\left| \psi \right> \in \mathcal H$, we form the \textbf{wavefunction}
\begin{equation}
\psi : \mathbb R \rightarrow \mathbb C
\end{equation}
defined by
\begin{equation}
\psi(x) \equiv \left< x \big | \psi \right>.
\end{equation}
\end{boxeddefn}
We note that if 
\begin{equation}
\left| \psi \right> = \sum_{x \in \mathbb R} a_x \left| x \right>,
\end{equation}
\begin{equation}
\left< \psi \right| = \sum_{x \in \mathbb R} a_x^* \left< x \right|,
\end{equation}
so
\begin{equation}
\mathrm{Tr}\left( \hat{\rho} \right) = \sum_{x \in \mathbb R} \left< x \big | \psi \right> \left< \psi \big| x \right> = \sum_{x \in \mathbb R } \psi^*(x) \psi(x),
\end{equation}
where we have used the complex symmetry of the inner product given by eqn. \ref{eqn:diracinner}. But since this is a sum over a continuous interval, it can be written as an integral. We obtain
\begin{equation}
\mathrm{Tr}\left( \hat{\rho} \right) = \int dx \cdot  \psi^*(x) \psi(x) = \int dx \cdot  \left| \psi(x)\right|^2 = 1,
\end{equation}
as the state operator has unit trace. Since our sum is infinite, it must be that $\psi(x)$ decays at infinity sufficiently fast such that the integral converges. This special class of functions is known as the set of square-normalizable functions, and is often denoted as $L^2$. Physically, this means that the wavefunction must be \textit{localized} in some sense, so that at extreme distances it is effectively zero. 

Just as we projected a vector into a basis and obtained a function, we can project a linear operator acting in Hilbert space onto a basis to obtain a linear operator in function space. We denote such operator with a check $\left( \, \check{ } \, \right)$, and define it by \cite{cohtan}\index{Linear!Operator!on a Function Space}
\begin{equation}\label{eqn:checkop}
\check{\mathcal O} \psi(x) \equiv \left< x \right| \hat{\mathcal O} \left| \psi \right>,
\end{equation}
where $\hat{\mathcal O}$ is an operator on a Hilbert space. An interesting application of this considers the matrix elements, given by eqn. \ref{eqn:matrixelem}, of a state operator $\hat{\rho}$ in the position basis. If $\hat \rho$ is pure, then
\begin{boxedeqn}{}
\rho(x,y) = \left< x \right| \hat \rho \left| y \right> = \left< x \big| \psi \right> \left< \psi \big| y \right> = \psi(x) \psi^*(y).
\end{boxedeqn}
Since we previously established that every valid wavefunction must vanish quickly at infinity, it follows that sufficiently off-diagonal elements of the state operator must vanish quickly, as well as distant points along the diagonal.

\chapter{Dynamics}\label{chap:dynamics}
\lettrine[lines=2, lhang=0.33, loversize=0.1]{Q}uantum dynamics is the framework that evolves a quantum state forward in time. We begin by considering the Galilei group\index{Group!Galilei} of transformations, under which all non-relativistic physics is believed to be invariant. We show that this group leads to the fundamental commutator relations that govern quantum dynamics, and then use them do derive the famous Schr\"odinger equation. Finally, we consider the free particle in the position basis.

\section{The Galilei Group}\label{sec:galgroup}
Fundamental to the notion of dynamics is the physical assumption that certain transformations will not change the physics of a situation \cite{ballentine}. All known experimental evidence supports this assumption, and it seems reasonable mathematically. This set of transformations forms a \textit{group}, called the Poincar\'e group\index{Group!Poincar\'e} of space translations, time translations, and Lorentz transformations.\footnote{The term group\index{Group} here is used in the formal, mathematical sense. We will not dwell on many of the subtleties that arise due to this here, and the interested reader is directed to ref. \cite{jones}.} However, for our purposes we take $v \ll c$, so the Poincar\'e group becomes the classical Galilei group, which we take as an axiom. For clarity, we assume a pure state in one temporal and one spacial dimension, but this treatment can be readily extended to impure states in three-dimensional space \cite{lindner}.

\begin{boxedaxm}{Invariance Under the Galilei Group\index{Group!Galilei}}{}
Let $G$ be the Galilei group, which contains elements generated by the composition of the operators
\begin{eqnarray}
\check S_{\epsilon} \psi(x,t) &=& \psi(x +\epsilon,t)  \nonumber \\
\check{T}_{\epsilon}  \psi(x,t)  &=&  \psi(x,t+\epsilon)  \nonumber \\
\check{L}_{\epsilon}  \psi(x,t)  &=&  \psi(x+\epsilon t, t) ,
\end{eqnarray}
where $\psi (x,t)$, given by definition \ref{defn:wavefunction}, is a function of position and time, and $(\check{\,})$ represents an operator on the space of such functions, as defined in eqn. \ref{eqn:checkop}. Let $\check g \in G$ and let $\hat A$ be an observable of the state $\left| \psi \right>$ with eigenvectors $\big\{ \left| \phi_n \right> \big\}_{n \in \mathbb R}$ and eigenvalues $\{ a_n \}_{n \in \mathbb R}$. Then,  if $\hat A \left| \phi_n \right> = a_n \left| \phi_n \right>$ and for all wavefunctions $v(x,t)$, 
$\check g v = v'$, we assert
\begin{equation}\label{eqn:galobservables}
\hat A ' \left| \phi_n' \right> \equiv a_n \left| \phi_n' \right>
\end{equation}
and
\begin{equation}\label{eqn:galstates}
\left| \left< \phi_n \big | \psi \right> \right|^2 \equiv \left| \left< \phi_n' \big| \psi' \right> \right|^2.
\end{equation}
\end{boxedaxm}
In essence, eqns. \ref{eqn:galobservables} and \ref{eqn:galstates} refer to the invariance of possible measurement\index{Measurement} and invariance of probable outcome, and thus the invariance of all physics, under the Galilei group. We now write a motivating identity using the Galilei group. Considering the state wavefunction $\psi(x,t)$, we find \cite{lindner}
\begin{eqnarray} \label{eqn:comm1}
 \check L_{\epsilon}^{-1}  \check T_{\epsilon}^{-1}  \check L_{\epsilon}  \check T_{\epsilon}   \psi(x,t)   &=&  \check L_{-\epsilon} \check T_{-\epsilon}  \check L_{\epsilon}  \check T_{\epsilon}   \psi(x,t)    \nonumber \\
&=&  \check L_{-\epsilon} \check T_{-\epsilon}  \check L_{\epsilon}   \psi(x,t+\epsilon)    \nonumber \\
&=&  \check L_{-\epsilon} \check T_{-\epsilon}    \psi(x+\epsilon(t+\epsilon),t+\epsilon)    \nonumber \\
&=&  \check L_{-\epsilon}   \psi(x+\epsilon(t+\epsilon),t)    \nonumber \\
&=&    \psi(x+\epsilon(t+\epsilon)-\epsilon t,t)    \nonumber \\
&=&    \psi(x+ \epsilon^2,t)    \nonumber \\
&=&  \check S_{\epsilon^2}   \psi (x,t)  .
\end{eqnarray}
We conclude that these transformations do not commute, which will play a major role in the dynamics of quantum mechanics. Before we move to a Hilbert space, we need to convert our Galilei group into a more useful form. Due to eqn. \ref{eqn:galstates}, we can make use of Wigner's theorem, which guarantees that any Galilei transformation corresponds to a \textbf{unitary} operator $\hat U$ on a Hilbert space that obeys\footnote{Wigner's theorem is complicated to prove. See ref. \cite{bargmann} for a thorough treatment.}
\begin{equation}
\hat U \hat U^{\dagger} = \hat U^{\dagger} \hat U= \hat 1.
\end{equation}
Thus, if $\hat U$ is a unitary representative of a Galilei group member and $\hat A$ is an observable. If we take that
\begin{equation}
\left| u' \right> \equiv \hat U \left| u \right>
\end{equation}
for all $\left| u \right> \in \mathcal H$, we have
\begin{equation}
\hat A ' \left| \phi_n '\right> = a_n \left| \phi_n ' \right> \Rightarrow \hat A ' \hat U \left| \phi_n \right> = a_n \hat U \left| \phi_n \right>,
\end{equation}
so
\begin{equation}
\hat U^{\dagger} \hat A' \hat U \left| \phi_n \right> = \left( \hat U^{\dagger} \hat U \right) a_n \left| \phi_n \right> = a_n \left| \phi_n \right>. 
\end{equation}
Hence, we get
\begin{equation}
\hat A \left| \phi_n \right> - \hat U^{\dagger} \hat A' \hat U \left| \phi_n \right> = a_n \left| \phi_n \right> - a_n \left| \phi_n \right> = \hat 0.
\end{equation}
Since this equation holds for all eigenvectors of $\hat A$, we have \cite{ballentine}
\begin{equation} \label{eqn:operatortrans}
\hat A - \hat U^{\dagger} \hat A' \hat U = 0 \Rightarrow \hat A = \hat U^{\dagger} \hat A' \hat U \Rightarrow \hat A' = \hat U \hat A \hat U^{\dagger}.
\end{equation}
We now take our unitary transformation to be a function of a single parameter, $t$, subject to $\hat U(t_1+t_2)  = \hat U(t_1) \hat U(t_2)$ and $\hat U(0)=\hat 1$. Then, for small $t$, we take the Taylor expansion of $\hat U$ about $t=0$ to get\footnote{We will be making frequent use of the Taylor expansion. Readers unfamiliar with it are advised to see ref. \cite{riley}.}
\begin{equation}
\hat U(t) = \hat 1 + t \frac{d \hat U}{dt}\Big | _{t=0}+... \, .
\end{equation}
Similarly, we know that \cite{lindner}
\begin{eqnarray}
\hat 1 &=&\hat  U \hat U^{\dagger} \nonumber \\
&=& \hat 1  +  t \frac{d\hat U \hat U^{\dagger}}{dt}\Big|_{t=0} + ... \nonumber \\
&=& \hat 1  +  t \left( \frac{d\hat U }{dt}\hat U^{\dagger} +\hat U\frac{d\hat U^{\dagger}  }{dt} \right) _{t=0}+ ... \nonumber \\	
&\sim& \hat 1  +  t \left( \frac{d\hat U }{dt} +\frac{d\hat U^{\dagger}  }{dt} \right) _{t=0}+ ... \, , \nonumber \\
\end{eqnarray}
as $t \sim 0$ and $\hat U(0)=\hat U^{\dagger}(0)=\hat 1$. Since $\hat 1  = \hat U \hat U^{\dagger}$ for all $t$, it must be that
\begin{equation}
 \left( \frac{d\hat U }{dt}+\frac{d\hat U^{\dagger}  }{dt} \right) _{t=0} = \hat 0.
\end{equation}
We now let 
\begin{equation}
\frac{d\hat U }{dt}\Big |_{t=0} \equiv i \hat K,
\end{equation}
which is well-defined so long as $\hat K$ is self-adjoint. We impose the boundary condition $\hat U (0) = \hat 1$ to find the solution to this first order differential equation,
\begin{equation}
\hat U(s) = e^{i \hat K t}.
\end{equation}
Since any unitary operator can be represented in this form, we now define the three generating operators of the Galilei group\index{Group!Galilei!Unitary Representatives}. They are \cite{lindner}
\begin{eqnarray}\label{eqn:galdefexp}
\check S_{x} \psi(x) &=& \left< x \right| \hat S_x \left| \psi \right> \equiv \left< x \right| e^{-i x \hat p} \left| \psi \right> \nonumber \\
\check T_t \psi(x)&=& \left< x \right|  \hat T_t \left| \psi \right>\equiv \left< x \right| e^{-i t \hat h} \left| \psi \right>\nonumber \\
\check L_v \psi (x)& = & \left< x \right| \hat L_v \left| \psi \right>\equiv \left< x \right| e^{i v \hat f  }\left| \psi \right>,
\end{eqnarray}
where $\hat f$, $\hat h$, and $\hat p$ are self-adjoint, and the particular signs and parameters associated with the transformations are matters of convention. \section{Commutator Relationships}
We next introduce three particular observables. First, the position operator, $\hat Q$\index{Position Operator}, obeys the eigenvalue equation
\begin{equation}
\hat Q \left| x \right> = x \left| x \right>,
\end{equation}
where $\left| x \right>$ is an eigenvector of the position, i.e. a state of definite position. Second, the momentum operator, $\hat P$\index{Momentum Operator}, follows
\begin{equation}
\hat P \left| p \right> = p \left| p \right>.
\end{equation}
We require that the expectation values of these operators follow the classical relationship \cite{griffiths}
\begin{equation}\label{eqn:classcorsp}
\left< \hat P \right> \equiv  \frac{d \left<\hat Q\right> }{dt}.
\end{equation}
Further, we define the energy operator $\hat H$, also known as the \textbf{Hamiltonian}\index{Hamiltonian}, in analogy to the classical total energy of a system, which is the kinetic energy $P^2/(2m)$ plus some potential energy $V$. It is
\begin{equation}\label{eqn:eop}
\hat H \equiv \frac{1}{2m} \hat P^2 + V.
\end{equation}
First, note that
\begin{equation}
\hat H \hat P = \left( \frac{1}{2m} \hat P ^2 + V\right)  \hat P = \hat P \left(\frac{1}{2m} \hat P ^2 + V \right) = \hat P \hat H,
\end{equation}
so $\left[ \hat H , \hat P \right] = 0$. Next, recall that
\begin{equation}\label{eqn:schropic}
\left| \psi(t + \epsilon) \right> =  \hat T_{\epsilon} \left| \psi(t) \right>.
\end{equation}
By the definition of the derivative, we have \cite{lindner}
\begin{eqnarray}
\frac{d}{dt} \left| \psi(t) \right> 
&=& \lim_{\epsilon \rightarrow 0} \frac{\left| \psi(t+\epsilon) \right> - \left| \psi(t) \right>}{\epsilon} \nonumber \\
&=& \lim_{\epsilon \rightarrow 0}  \frac{e^{-i \epsilon \hat h} \left| \psi(t) \right> - \left| \psi(t) \right>}{\epsilon} \nonumber \\
&=& \lim_{\epsilon \rightarrow 0}  \frac{\left(1-i \epsilon \hat h+\left(-i\epsilon \hat h \right)^2/2 +... \right) \left| \psi(t) \right> - \left| \psi(t) \right>}{\epsilon} \nonumber \\
&=& \lim_{\epsilon \rightarrow 0} \left( -i  \hat h \left| \psi(t)\right> - \epsilon \hat h^2 \left| \psi(t) \right> + ... \right) \nonumber \\
&=&  -i  \hat h \left| \psi(t)\right> .
\end{eqnarray}
Following identical logic, we find \cite{lindner}
\begin{equation}
\frac{d}{dt} \left< \psi(t) \right|  =  +i   \left< \psi(t)\right| \hat h.
\end{equation}
Since $\left| \psi(t) \right>$ is pure, we use eqn. \ref{eqn:recoverexp} to write
\begin{eqnarray}
\frac{d}{dt} \left< \hat Q \right>(t) 
&=& \frac{d}{dt} \left< \psi(t) \right| \hat Q \left| \psi(t) \right> \nonumber \\
&=& \left( \frac{d}{dt} \left< \psi(t) \right|\right)  \hat Q \left| \psi(t) \right> +\left< \psi(t) \right|  \hat Q \left( \frac{d}{dt} \left| \psi(t) \right> \right) \nonumber \\
&=& i   \left< \psi(t)\right| \hat h  \hat Q \left| \psi(t) \right> -\left< \psi(t) \right|  \hat Q  i  \hat h \left| \psi(t)\right> \nonumber \\
&=&    \left< \psi(t)\right| i \left( \hat h  \hat Q -  \hat Q    \hat h \right) \left| \psi(t) \right> \nonumber \\
&=&    \left< \psi(t)\right| i \left[ \hat h,  \hat Q \right] \left| \psi(t) \right>,
\end{eqnarray}
so
\begin{equation}
\frac{d}{dt} \left< \hat Q \right>(t) = \left<  i \left[ \hat h,  \hat Q \right]  \right>.
\end{equation}
Then, by eqn. \ref{eqn:classcorsp}, we have
\begin{equation}
\frac{1}{m} \left< \hat P \right> =  \left<  i \left[ \hat h,  \hat Q \right]  \right> \Leftrightarrow \left< \psi(t) \right|  \frac{1}{m} \hat P \left| \psi(t) \right> = \left< \psi(t) \right| i \left[ \hat h, \hat Q \right] \left| \psi(t) \right>.
\end{equation}
Since this result holds for arbitrary $\left| \psi(t) \right>$, we get \begin{equation}
\frac{1}{m}\hat P  =   i \left[ \hat h,  \hat Q \right] ,
\end{equation}
or \cite{lindner}
\begin{equation}
\left[ \hat Q,  \hat h \right] = i \frac{1}{m} \hat P.
\end{equation}

We next continue working with the position operator to derive a second relation. Recall that from eqn. \ref{eqn:operatortrans}, a unitary transformation  defined by 
\begin{equation}
\left| \psi ' \right> = \hat U \left| \psi \right>
\end{equation}
transforms an operator as
\begin{equation}
\hat A ' = \hat U \hat A \hat U^{\dagger}.
\end{equation}
So, if our unitary operator is $\hat S_{x_0} = e^{-i x_0 \hat p}$, we can transform the position operator $\hat Q$ to $\hat Q '$ according to
\begin{equation}\label{eqn:posexps}
\hat Q ' = \hat S_{x_0} \hat Q \hat S_{x_0}^{\dagger} =e^{-i x_0 \hat p} \hat Q e^{+i x_0 \hat p}.
\end{equation}
By our definition of $\hat Q$, we know\footnote{This is because $\left| x \right>$ and $\left| x' \right>$ are valid eigenvectors of $\hat Q$, as the spectrum of allowed positions (the eigenvalues for $\hat Q$) is the entire real line.}
\begin{equation}
\hat Q \left| x \right> = x \left| x \right> \Rightarrow \hat Q \left| x' \right> = x' \left| x' \right>.
\end{equation}
Further, eqn. \ref{eqn:galobservables} tells us
\begin{equation}
\hat Q ' \left| x ' \right> = x \left| x ' \right>.
\end{equation}
Thus,
\begin{equation}
\left( \hat Q ' - \hat Q \right) \left| x' \right> = (x-x') \left| x' \right> = \left( x - (x+x_0) \right) \left| x ' \right> = - x_0 \left| x' \right>.\footnote{Note that $x'=x+x_0$, since $\check S_{x_0} \psi(x) = \psi(x_0+x) = \psi(x')$.}
\end{equation}
Note that this relationship holds for arbitrary $x_0$, and hence for all $\left| x' \right>$. This implies \cite{lindner}
\begin{equation}
\hat Q ' = \hat Q - x_0 .
\end{equation}
Recalling our definition for $\hat Q '$, we have
\begin{equation} \label{eqn:qpcomm3}
 e^{-i x_0 \hat p} \hat Q e^{+i x_0 \hat p} = \hat Q - x_0 .
\end{equation}
As before, we expand the exponential terms in a Taylor series to obtain
\begin{eqnarray}
\left( \sum_{n=1}^{\infty}\frac{ \left( -i x_0 \hat p \right)^n}{n!} \right) \hat Q \left( \sum_{n=1}^{\infty}\frac{ \left( i x_0 \hat p \right)^n}{n!} \right)
 &=& \left( 1 - i x_0 \hat p + ... \right) \hat Q  \left( 1 + i x_0 \hat p + ... \right) \nonumber \\
&=& \hat Q - i x_0 \hat p \hat Q + i x_0 \hat Q \hat p +... \nonumber \\
&=& \hat Q + i x_0 \left( \hat Q \hat p - \hat p \hat Q \right) + ... \nonumber \\
&=& \hat Q + i x_0 \left[ \hat Q , \hat p \right] +... \nonumber \\
&=& \hat Q - x_0 .
\end{eqnarray}
Hence, in the limit as $x_0 \rightarrow 0$, eqn. \ref{eqn:qpcomm3} is \cite{lindner}
\begin{equation}
i \left[ \hat Q, \hat p \right] = -1 \Rightarrow \left[ \hat Q, \hat p \right] = i
\end{equation}
Next, we examine the momentum operator. Taking our unitary operator to be $\hat L_{v_0}=e^{+i v \hat f}$, we get
\begin{equation}
\hat P ' = e^{+i v_0 \hat f} \hat  P e^{-iv_0  \hat f}.
\end{equation}
If we operate on states of definite momentum, we know
\begin{equation}
\hat P \left| p \right> = p \left| p \right> = m v \left| p \right>.
\end{equation}
By direct analogy with the states of definite position above, we find \cite{lindner}
\begin{equation}
\hat P ' = e^{+i v_0 \hat f} \hat  P e^{-iv_0  \hat f} = \hat P - m v_0 .
\end{equation}
As above, we find the Taylor expansion of the exponentials to obtain
\begin{eqnarray}
\left( \sum_{n=1}^{\infty}\frac{ \left( +i v_0 \hat f \right)^n}{n!} \right) \hat P \left( \sum_{n=1}^{\infty}\frac{ \left( i v_0 \hat f \right)^n}{n!} \right)
 &=& \left( 1 + i v_0 \hat f + ... \right) \hat P \left( 1 - i v_0 \hat f + ... \right) \nonumber \\
&=& \hat P + i v_0 \hat f \hat P - i v_0 \hat P \hat f +... \nonumber \\
&=& \hat P + i v_0 \left( \hat f \hat P - \hat P \hat f \right) + ... \nonumber \\
&=& \hat P + i v_0 \left[ \hat f , \hat P \right] +... \nonumber \\
&=& \hat P - mv_0 .
\end{eqnarray}
In the limit as $v_0 \rightarrow 0$, we have
\begin{equation}
i \left[ \hat f , \hat P \right] = - m \Rightarrow \left[ \hat f , \hat P \right] = i m .
\end{equation}
It is a convention to define $\hat f \equiv m \hat q$, in which case we have \cite{lindner}
\begin{equation}
\left[ \hat q , \hat P \right] = i .
\end{equation}
We now have
\begin{eqnarray} \label{eqn:commsecondset}
\left[ \hat H, \hat P \right] &=& 0, \nonumber \\
\left[ \hat Q,  \hat h \right] &=& i \frac{1}{m} \hat P, \nonumber \\
\left[ \hat Q, \hat p \right] &=& i, \nonumber \\
\left[ \hat q , \hat P \right] &=& i.
\end{eqnarray}
We make the standard definition for the position, momentum, and energy operators in terms of the Galilei\index{Group!Galilei!Generators} group generators. It is \cite{lindner}
\begin{equation}
\hat Q \equiv \hbar \hat q, \, \,\, \hat P \equiv \hbar \hat p , \, \, \, \hat H \equiv \hbar \hat h,
\end{equation}
where $\hbar$ is a proportionality constant known as Planck's reduced constant, and is experimentally determined to be
\begin{equation}
\hbar \approx 10^{-34} \, \mathrm{joule-seconds}
\end{equation}
in SI units. Then, eqn. \ref{eqn:commsecondset} reads \cite{lindner}
\begin{eqnarray}\label{eqn:goodcomm}
\left[ \hat P , \hat H \right] &=& 0, \nonumber \\
\left[ \hat Q, \hat H \right] &=& i \hbar \frac{1}{m} \hat P, \nonumber \\
\left[ \hat Q, \hat P \right] &=& i \hbar,
\end{eqnarray} 
where
\begin{boxedeqn}{}
\left[ \hat Q, \hat P \right] = i \hbar
\end{boxedeqn}is especially important, and is called the \textbf{canonical commutator}\index{Canonical Commutator}. 

As a  consequence of our work so far this chapter, we now are in the position to evolve a state operator $\hat \rho$ in time. From eqn. \ref{eqn:operatortrans}, we have
\begin{equation}
\hat A' = \hat U \hat A \hat U^{\dagger}
\end{equation}
for an arbitrary observable $\hat A$. Letting $\hat A = \hat{\rho}$ , the state operator, and $\hat U = \hat T_t= e^{-i t \hat H/\hbar}$, we have
\begin{equation}\label{eqn:freestateop}
\hat{\rho}' =  e^{-i t \hat H/\hbar} \hat{\rho} e^{+i t \hat H/\hbar}.
\end{equation}
Thus, by the definition of the derivative,
\begin{equation}
\hat{\partial}_t \hat{\rho} = \lim_{t \rightarrow 0} \frac{\hat{\rho}' - \hat{\rho}}{t} = \lim_{t \rightarrow 0} \frac{e^{-i t \hat H/\hbar} \hat{\rho} e^{i t \hat H/\hbar}- \hat{\rho}}{t}.
\end{equation}
Expanding the exponential terms in a Taylor series, we get\index{Equation of Motion!of the State Operator}
\begin{eqnarray}
\hat{\partial}_t \hat{\rho} &=& \lim_{t \rightarrow 0} \frac{\left(1 - \frac{it \hat H}{\hbar}+... \right) \hat{\rho} \left(1 + \frac{it \hat H}{\hbar}+... \right) - \hat{\rho}}{t} \nonumber \\
&=& \lim_{t \rightarrow 0}\left(  -\frac{i \hat H}{\hbar} \hat {\rho} + \hat{\rho} \frac{i \hat H}{\hbar} + ... \right) \nonumber \\
&=&-  \frac{i \hat H}{\hbar} \hat {\rho} + \hat{\rho} \frac{i \hat H}{\hbar} \nonumber \\
&=&  \frac{i}{\hbar} \left[ \hat{\rho} , \hat H \right] ,
\end{eqnarray}  
so the equation of motion for the state operator is
\begin{boxedeqn}{eqn:heispic}
\hat{\partial}_t \hat{\rho}= \frac{i}{\hbar} \left[ \hat{\rho} , \hat H \right] .
\end{boxedeqn}

\section{The Schr\" odinger wave equation}
Now that we have the commutator relations in eqn. \ref{eqn:goodcomm}, we can touch base with elementary quantum mechanics by deriving the Schr\"odinger wave equation. We work in the position basis, where our basis vectors follow
\begin{equation}
\hat Q \left| x \right> = x \left| x \right>.
\end{equation}
Considering some state vector 
\begin{equation}
 \left| \psi \right> = \sum_{x \in \mathbb R} a_x \left| x \right>,
 \end{equation}
its wavefunction, given by definition \ref{defn:wavefunction}, is
 \begin{equation}\label{eqn:wavefunction}
 \psi (x) = \left< x \big | \psi \right> = \left< x \right|  \left( \sum_{x \in \mathbb R} a_x  \left| x \right> \right)= a_x.
 \end{equation}
Considering $\hat Q$, we find by eqn. \ref{eqn:checkop} that
\begin{equation}
\check Q \psi (x) = \left< x \right| \hat Q \left| \psi \right>= \left< x \right| x \left| \psi \right> = x \left< x \big| \psi \right> = x \psi(x).
\end{equation}
So, in the position basis, $\check Q$ turns out to be multiplication by $x$. Using this result with eqn. \ref{eqn:recoverexp}, we find \cite{griffiths}
\begin{eqnarray}
\left< \hat Q \right> &=& \left< \psi \right| \hat Q \left| \psi \right> \nonumber \\
&=& \left( \sum_{x \in \mathbb R} a_x^*  \left< x \right| \right)\hat Q  \left( \sum_{y \in \mathbb R} a_y  \left| y \right> \right) \nonumber \\
&=& \left( \sum_{x \in \mathbb R} a_x^*  \left< x \right| \right) \left( \sum_{y \in \mathbb R} a_y \hat Q  \left| y \right> \right) \nonumber \\
&=& \left( \sum_{x \in \mathbb R} a_x^*  \left< x \right| \right) \left( \sum_{y \in \mathbb R} a_y y  \left| y \right> \right) \nonumber \\
&=&  \sum_{x \in \mathbb R}  a_x^* x a_x   \nonumber \\
&=&  \int dx  \cdot  \psi(x)^* x \psi(x)   \nonumber \\
&=&  \int dx  \cdot  \psi(x)^* \check Q \psi(x).  
\end{eqnarray}
We would like to find a similar expression for momentum within the position basis. To do this, we consider the canonical commutator from eqn. \ref{eqn:commsecondset}, $\left[ \hat Q, \hat P \right] = i \hbar$, which corresponds to
\begin{equation}
\left< x \right| \left( \left[ \hat Q, \hat P \right] = i \hbar \right) \Rightarrow \left[ \check Q, \check P \right] \psi(x) = i \hbar \psi(x) 
\end{equation}
 Considering some dummy function $f(x)$, we have \cite{griffiths}
\begin{eqnarray}
i \hbar f(x) &=& x \frac{\hbar}{i} \frac{d f}{dx} - x \frac{\hbar}{i} \frac{df}{dx} + i \hbar f \nonumber \\
&=& x \frac{\hbar}{i} \frac{d f}{dx} - x \frac{\hbar}{i} \frac{df}{dx} - \frac{ \hbar}{i} f \nonumber \\
&=& x \frac{\hbar}{i} \frac{d f}{dx} -  \frac{\hbar}{i} \frac{d }{dx}\left( x \cdot f \right) \nonumber \\
&=&   \left( x \frac{\hbar}{i} \frac{d}{dx} -  \frac{\hbar}{i} \frac{d  }{dx} x \right)f \nonumber \\
&=& \left( \check Q \check P - \check P \check Q \right) f.
\end{eqnarray}
Since we know $\check Qf =xf$ in the position basis, 
\begin{equation} \label{eqn:mominpos}
\check P f= \frac{\hbar}{i} \frac{d}{dx}f.
\end{equation}
We now drop our test function to obtain the famous operator relationship\index{Momentum Operator!in Position Basis}
\begin{boxedeqn}{}
\check P =\frac{\hbar}{i} \frac{d}{dx}.
\end{boxedeqn}
Now, recall from eqn. \ref{eqn:schropic},
\begin{equation}
\left| \psi_{t=\epsilon} \right> = \hat T_{\epsilon} \left| \psi_{t=0} \right> = e^{-i \epsilon \hat H/{\hbar}} \left| \psi_0 \right>.
\end{equation}
It follows that
\begin{equation}
\hat{ \partial_t} \left| \psi_t \right> = \hat{\partial_t} \left( e^{-i \epsilon \hat H/{\hbar}} \left| \psi_0 \right> \right) = -\frac{i \hat H}{\hbar} e^{-i \epsilon \hat H/{\hbar}} \left| \psi_0 \right> = -\frac{i \hat H}{\hbar}\left| \psi_{t} \right>,
\end{equation}
so
\begin{equation}
\check{\partial_t} \psi_t(x) = \left< x \right| \hat{\partial_t} \left| \psi_t \right> =  \left< x \right| -\frac{i \hat H}{\hbar} \left| \psi_t \right> = -\frac{i}{\hbar} \left< x \right| \hat H \left| \psi_t \right> = -\frac{i}{\hbar} \check H \psi_t(x).
\end{equation}
But by eqn. \ref{eqn:eop},
\begin{equation} \label{eqn:hcheck}
\check H \psi_t(x) = \left< x \right| \hat H \left| \psi_t \right> =\left< x \right| \left( \frac{1}{2m} \hat P^2 + V \right) \left| \psi_t \right> = \frac{1}{2m} \check p ^2 \psi_t(x) + V \psi_t(x).
\end{equation}
Hence, we have
\begin{equation}
\check{\partial}_t \psi_t(x) = -\frac{i}{\hbar}  \frac{1}{2m} \check P ^2 \psi_t(x) - V\psi_t(x) = -\frac{i}{\hbar}  \frac{1}{2m} \left( \frac{\hbar}{i} \check{\partial}_x \right)^2 \psi_t(x)- V\psi_t(x).
\end{equation}
This is rewritten as
\begin{boxedeqn}{}
i \hbar \frac{\partial \psi(x,t)}{\partial t} = - \frac{\hbar^2}{2m} \frac{\partial^2 \psi(x,t)}{\partial x ^2} + V \psi (x,t),
\end{boxedeqn}
and is the \textbf{time-dependent Schr\"odinger equation}\index{Schr\"odinger Equation!Time Dependent} \cite{griffiths}.

Remarkably, so long as $V$ is time-independent, this equation turns out to be separable, so we can effectively pull off the time-dependence. To do this, we suppose \cite{griffiths}
\begin{equation}
\psi(x,t) \equiv \psi(x) \varphi(t),
\end{equation}
and substitute into the Schr\"odinger equation. We have
\begin{equation}
i \hbar \frac{\partial \psi(x) \varphi(t)}{\partial t} = - \frac{\hbar^2}{2m} \frac{\partial^2 \psi(x) \varphi(t)}{\partial x ^2} + V \psi(x) \varphi(t),
\end{equation}
which is
\begin{equation}
i \hbar \psi(x) \frac{\partial \varphi(t)}{\partial t} + \frac{\hbar^2}{2m}\varphi(t) \frac{\partial^2 \psi(x)}{\partial x ^2} = V \psi(x) \varphi(t),
\end{equation}
or
\begin{equation}
 i \hbar \frac{1}{\varphi(t)} \frac{\partial \varphi(t)}{\partial t}  = - \frac{\hbar^2}{2m}\frac{1}{\psi(x)} \frac{\partial^2 \psi(x)}{\partial x ^2} + V,
\end{equation}
provided $\varphi(t), \psi(x) \neq 0$. We now have two independent, single-variable functions set equal , so we know each of the functions must be equal to some constant, which we name $E$. That is, we have \cite{griffiths}
\begin{eqnarray}
E &=& i \hbar \frac{1}{\varphi(t)} \check d_t \varphi(t), \nonumber \\
E &=& - \frac{\hbar^2}{2m}\frac{1}{\psi(x)} \check {d}_x^2 \psi(x)+ V ,
\end{eqnarray}
where we have let the partial derivatives go to normal derivatives, since we now have single-variable functions. The time-dependent piece has the solution
\begin{equation}
\varphi(t) = e^{-i E t / \hbar},
\end{equation}
and the time-independent piece is usually written as \cite{griffiths}
\begin{boxedeqn}{eqn:schrotimeind}
 - \frac{\hbar^2}{2m} \check d _x^2 \psi (x) + V \psi(x) = E \psi (x),
\end{boxedeqn}
which is the\textbf{ time-independent Schr\"odinger equation}\index{Schr\"odinger Equation!Time Independent}. Although this result cannot be reduced further without specifying $V$, we can use eqn. \ref{eqn:hcheck} to find
\begin{equation}
\check H \psi(x) = E \psi(x).
\end{equation}
This means that the values for the separation constant $E$ are actually the possible eigenvalues for $\check H$, the position representation of the Hamiltonian (energy operator). Further, if we find $\psi(x)$, we can construct $\psi(x,t)$ by 
\begin{equation}
\psi(x,t) = \varphi(t) \psi(x) = e^{-i E t / \hbar}\psi(x) .
\end{equation}
If we compare this to eqn. \ref{eqn:schropic}, 
\begin{equation}
 \left| \psi_t \right>=\hat T_{t} \left| \psi_{t=0} \right> = e^{-i t \hat H/{\hbar}} \left| \psi_0 \right>,
\end{equation}
we find a distinct similarity between the form of time evolution in Hilbert space  using the unitary $\hat T_t$ operator and time evolution in position space using the complex exponential of the eigenvalues of the associated $\check H$ operator on function space.
\section{The Free Particle}\label{sec:freeparticle}\index{Free Particle!in Position Basis}
Now that we have derived the Schr\"odinger equation, we will put it to use by treating the case of a free particle, when the potential $V=0$. In this case, the time-independent Schr\"odinger equation (eqn. \ref{eqn:schrotimeind}) reads
\begin{equation}
 - \frac{\hbar^2}{2m} \check d _x^2 \psi (x)  = E \psi (x),
\end{equation}
which we write as
\begin{equation}
 \check d _x^2 \psi (x)  = -k^2 \psi (x),
\end{equation}
where 
\begin{equation}
k \equiv \frac{\sqrt{2E}}{\hbar}.
\end{equation}
This equation has a solution \cite{cohtan}
\begin{equation}\label{eqn:planewave1}
\psi(x) = Ae^{  ik x},
\end{equation}
which is sinusoidal with amplitude $A$. Note that we identified the constants in our equation as $k$ with some foresight, as it turns out to be the wave number, $k=2\pi/\lambda$, of the solution.  However, this solution does not decay at infinity, so the condition imposed by definition \ref{defn:pure} is violated. That is \cite{griffiths},
\begin{eqnarray}
\left< \psi \big| \psi \right> 
&=& \left( \sum_{x \in \mathbb R }a^*_x \left<x \right|  \right) \left( \sum_{y \in \mathbb R} a_y \left| y \right> \right) \nonumber \\
&=& \sum_{x \in \mathbb R } \sum_{y \in \mathbb R} a^*_x a_y\left<x \big| y \right> \nonumber \\
&=& \sum_{x \in \mathbb R } a^*_x a_x  \nonumber \\
&=& \int dx \cdot \psi^*(x) \psi(x)  \nonumber \\
&=& \int dx \cdot A^* e^{-ikx} A e^{ikx} \nonumber \\
&=& \left| A \right| ^2 \int dx  \nonumber \\
&=& \infty,
\end{eqnarray}
so we cannot pick appropriate $A$ such that $\left< \psi \big| \psi \right> =1$. Hence, $\left| \psi \right> $ must not be a physically realizable state. The resolution to this problem is to use a linear combination of states with different values for $A$. The general formula for this linear combination is \cite{griffiths}
\begin{boxedeqn}{}
\psi(x) = \int dk \cdot \phi(k) e^{ikx},
\end{boxedeqn}
where $\phi(k)$ is the coefficient that replaces $A$ in our linear combination. Each of the component states of this integral are called \textbf{plane waves}\index{Plane Wave}, while the linear combination is called a \textbf{wave packet}\index{Wave Packet}. We will make use of the plane wave components for free particles later, so we need to investigate their form further. Consider the eigenvalue problem 
\begin{equation}
\check P f_p(x) = p f_p(x),
\end{equation}
where $f_p$ is an eigenfunction and $p$ is an eigenvalue of the momentum operator in the position basis. Using eqn. \ref{eqn:mominpos}, we write
\begin{equation}
\frac{\hbar}{i} \check d_x f_p(x) = p f_p(x).
\end{equation}
This has a solution \cite{griffiths}
\begin{equation}\label{eqn:planewave}
f_p(x) = Ae^{ipx/\hbar},
\end{equation}
which is of identical form to eqn. \ref{eqn:planewave1}. If we identify the eigenfunctions of the position operator with the plane wave states, we get the famous de Broglie relation\index{De Broglie Relation} \cite{griffiths, ballentine, sudbery, cohtan},
\begin{equation}
p = \hbar k.
\end{equation}
Recall that plane wave states are not normalizable, and thus cannot be physically realizable states. This means that in the position basis, states of definite momentum are not permissible, which is a famous consequence of the Heisenberg uncertainty principle.\footnote{The uncertainty principle reads $\Delta x \Delta p \geq \hbar /2$ \cite{griffiths}. If we have a state of definite position, $\Delta p = 0$, so, roughly, $\Delta x = \infty$. This is the result that we have already seen; states of definite momentum are not square-normalizable in the position basis.} The wave packet, then, can be thought of as a superposition\index{Superposition} of states of definite momentum, giving rise to a state of definite position. That is \cite{griffiths},
\begin{equation}
\psi(x) = \int dp \cdot \phi(p) e^{ipx},
\end{equation}
where we have switched to units in which $\hbar \equiv 1$, as we will do for the remainder of this thesis.

\chapter{The Wigner Distribution}\label{chap:wigner}
\lettrine[lines=2, lhang=0.33, loversize=0.1]{T}he Wigner distribution was the first quasi-probability distribution used in physics. Invented in 1932 by E.P. Wigner\index{Wigner, E.P.}, it remains in wide use today in many areas, especially quantum mechanics and signal analysis \cite{wigner}. The Wigner distribution has been used to develop an entirely new formalism of quantum mechanics in phase-space, a space of position vs. momentum, which we touch on briefly in section \ref{sec:wig_harmonic} \cite{zachos2}. 

In the following chapter, we first define the Wigner distribution and derive some of its fundamental properties. Next, we discuss the Wigner distribution of a combined system and treat a free particle. Following that, we extend the distribution to its associated transform. We then create a table of useful inverse relationships between the state operator and Wigner distribution required in subsequent sections. Finally, we construct the Wigner distribution for a simple harmonic oscillator as an example and observe its correspondence to a classical phase-space probability distribution.


\section{Definition and Fundamental Properitees}
In this section, we explore the basic properties of the Wigner distribution, starting with its definition, which is stated below \cite{zachos}

\begin{boxeddefn}{The Wigner distribution\index{Wigner Distribution!Defintion}}{def:WignerDist}
Consider the matrix elements of some state operator, given by $\rho(x,y)=\left< x \right| \hat \rho \left| y \right>$. Then, the Wigner distribution $W$ associated with $\rho$ is given by
\begin{equation}\label{eqn:wigdef1}
W(\bar x,p,t ) \equiv \frac{1}{2 \pi} \int d \delta \cdot e^{-i p \delta } \rho(x,y,t), 
\end{equation}
where $\bar x = (x+y)/2$ and $\delta = x-y$. This is also usefully written
\begin{equation}\label{eqn:wigalt}
W(\bar x,p) = \frac{1}{2 \pi} \int d \delta \cdot e^{-ip \delta}\rho \left( \bar x +\frac{1}{2} \delta, \bar x - \frac{1}{2} \delta \right),
\end{equation}
where time dependence is understood and not written explicitly.
\end{boxeddefn}
Note that the Wigner distribution is given by a special case of the \textbf{Fourier transform} of the state operator with respect to the mean ($\bar x$) and difference ($\delta$) coordinates.\footnote{We assume some basic familiarity with the Fourier transform. If this topic is unfamiliar, the reader is advised to see ref. \cite{riley}.} Using this definition, we now list and verify some of the well known properties of the Wigner distribution. 
\subsection{Inverse Distribution}
As one might guess, just as the Wigner distribution is defined in terms of the state operator, it is possible to define the state operator in terms of the Wigner distribution. This distinction is arbitrary: valid formulations of quantum mechanics have been made with the Wigner distribution as the primary object, while the state operator takes a secondary seat. However, historically the state operator and its associated vector have been the objects of primary importance in the development of quantum mechanics \cite{styer}. If we wish to express a state operator in terms of an associated Wigner distribution, we can make use of the relation \cite{halliwell}\index{Wigner Distribution!Inverse}
\begin{equation}
\rho(x,y) = \int d p \cdot e^{i p \delta} W \left( p, \bar x \right).
\end{equation}
In order to show that this is well-defined, we note that  the  Plancherel theorem states \cite{griffiths}
\begin{equation}
f( p ) = \frac{1}{2 \pi} \int d \delta \cdot e^{- i p \delta} \mathcal{F}\left( f(\delta) \right)  \Leftrightarrow  \mathcal{F}\left( f(\delta) \right)  = \int d p \cdot e^{i p \delta} f( p),
\end{equation}
for some function $f$ and its Fourier transform, $\mathcal F (f) $, so long as the functions decay sufficiently fast at infinity. From this, is evident that the state operator is a kind of Fourier transform of the Wigner distribution, as we claimed in the previous section, so our inverse relationship is indeed appropriate.

\subsection{Reality of the Wigner Distribution}\index{Wigner Distribution!Reality}
One of the most important of the basic properties we will cover is that the Wigner distribution is always real-valued.\footnote{Although it is real-valued, the Wigner distribution is \textit{not} always positive. It is called a quasi-probability distribution since it is analogous to a true probability distribution, but has negative regions. We will deal these apparent negative probabilities more in section \ref{sec:wig_harmonic}.} That is,
\begin{boxedeqn}{}
W(\bar x, p, t) \in \mathbb R. \label{eqn:reality}
\end{boxedeqn}
In order to show this, we will take the complex conjugate of $W(\bar x, p)$. This gives us \cite{cohen}
\begin{eqnarray}
W^*(\bar x, p)	&=&  \frac{1}{2 \pi} \int_{\delta=- \infty}^{\infty} d \delta \cdot e^{ip \delta}\rho^* \left( \bar x +\frac{1}{2} \delta, \bar x - \frac{1}{2} \delta \right) \nonumber \\
			&=& \frac{1}{2 \pi} \int_{\delta= \infty}^{-\infty} \left( - d \delta \right) \cdot e^{-ip \delta}\rho^* \left( \bar x -\frac{1}{2} \delta, \bar x + \frac{1}{2} \delta \right) \nonumber \\
			&=& \frac{1}{2 \pi} \int_{\delta= -\infty}^{\infty}  d \delta  \cdot e^{-ip \delta}\rho^{\dagger} \left( \bar x +\frac{1}{2} \delta, \bar x - \frac{1}{2} \delta \right) \nonumber \\
			&=& \frac{1}{2 \pi} \int_{\delta= -\infty}^{\infty}  d \delta  \cdot e^{-ip \delta}\rho \left( \bar x +\frac{1}{2} \delta, \bar x - \frac{1}{2} \delta \right) \nonumber \\
			&=& W(\bar x, p),
\end{eqnarray}
where we used eqn. \ref{eqn:adjointapp} for the self-adjoint operator $\hat{\rho}$. Since we found $W^*(\bar x, p) = W(\bar x, p)$, we have $W(\bar x, p) \in \mathbb R$, as we claimed in eqn. \ref{eqn:reality}.
\subsection{Marginal Distributions}\label{sec:marginals}\index{Wigner Distribution!Marginal Distributions}
Based on our definition of the Wigner distribution, we note two important marginal distributions. They are \cite{hillery}
\begin{boxedeqn}{}
\int dp \cdot W(\bar x,p) = \left< \bar x \right| \hat \rho \left| \bar x \right> \label{eqn:marginal1}
\end{boxedeqn}
and
\begin{boxedeqn}{}
\int d\bar x \cdot W(\bar x,p) = \left< p \right|  \hat{\rho} \left| p \right>.
\end{boxedeqn}
To show these results, we recall the definition of the Wigner distribution. We have

\begin{eqnarray}
\int dp \cdot W(\bar x,p) &=& \int dp \cdot  \frac{1}{2 \pi} \int d \delta \cdot e^{-ip \delta}\rho \left( \bar x +\frac{1}{2} \delta, \bar x - \frac{1}{2} \delta \right) \nonumber \\
&=&  \frac{1}{2 \pi}   \int d \delta \cdot \rho \left( \bar x +\frac{1}{2} \delta, \bar x - \frac{1}{2} \delta \right) \int dp \cdot e^{-ip \delta}\nonumber \\
&=&  \frac{1}{2 \pi}   \int d \delta \cdot \rho \left( \bar x +\frac{1}{2} \delta, \bar x - \frac{1}{2} \delta \right) 2 \pi \delta_D(\delta) \nonumber \\
&=&    \int d \delta \cdot \delta_D(\delta) \rho \left( \bar x +\frac{1}{2} \delta, \bar x - \frac{1}{2} \delta \right) \nonumber \\
&=&   \rho \left( \bar x , \bar x  \right) \nonumber \\
&=& \left< \bar x \right| \hat{\rho} \left| \bar x \right>,
\end{eqnarray}
where $\delta_D$ is called the $\textbf{Dirac delta}$\index{Dirac Delta},\footnote{The Dirac delta is roughly a sharp spike at a point, and zero elsewhere. Technically, it is not quite a function, but it is a very useful construct in theoretical physics. For more information, see ref. \cite{riley}.} and has the important properties \cite{riley}
\begin{equation}
\int dx \cdot \delta_D(y) f(x+y) \equiv f(x)
\end{equation}
and
\begin{equation}
\int dx \cdot e^{-i x y} \equiv 2 \pi \delta_D(y).
\end{equation}
In preparation for calculating the postion marginal distribution, it is useful to discuss the momentum representation of the state operator\index{State Operator!in Momentum Basis}. Analogous to the position representation, we define
\begin{boxedeqn}{}
\tilde{\rho}(p,p') \equiv \left< p \right| \hat  p \left| p \right>,
\end{boxedeqn}
where the $(\, \tilde{} \, )$ is used to distinguish position matrix elements from momentum matrix elements. In terms of the momentum representation, the Wigner distribution is\index{Wigner Distribution!in Momentum Basis}
\begin{boxedeqn}{eqn:wigmomrep}
W( \bar x,  p ) \leftrightarrow W_P( x, \bar p ) \equiv \frac{1}{2 \pi} \int d \lambda \cdot e^{-i x \lambda}\tilde{\rho} \left( \bar p +\frac{1}{2} \lambda, \bar p - \frac{1}{2} \lambda \right),
\end{boxedeqn} 
where $\bar p = \frac{p + p'}{2}$ and $\lambda = p - p'$ are the average and difference momentum coordinates, in direct analogy to $\bar x$ and $\delta$. We are now ready to calculate the position marginal distribution. We have
\begin{eqnarray}
\int d\bar{x} W(\bar x, p) &\leftrightarrow&\int d\bar{x} W_P( x, \bar p) \nonumber \\ 
&=& \int dx  \frac{1}{2 \pi} \int d \lambda \cdot e^{-i x \lambda}\tilde{\rho} \left( \bar p +\frac{1}{2} \lambda, \bar p - \frac{1}{2} \lambda \right)\nonumber \\
&=&  \frac{1}{2 \pi} \int d \lambda \cdot \tilde{\rho} \left( \bar p +\frac{1}{2} \lambda, \bar p - \frac{1}{2} \lambda \right)  \int dx e^{-i x \lambda}\nonumber \\
&=&  \frac{1}{2 \pi} \int d \lambda \cdot \tilde{\rho} \left( \bar p +\frac{1}{2} \lambda, \bar p - \frac{1}{2} \lambda \right)  2 \pi \delta_D(\delta) \nonumber \\
&=&  \int d \lambda \cdot \delta_D(\delta)  \tilde{\rho} \left( \bar p +\frac{1}{2} \lambda, \bar p - \frac{1}{2} \lambda \right) \nonumber \\
&=&  \tilde \rho (\bar p , \bar p) \nonumber \\
&\leftrightarrow&  \tilde \rho (p , p) \nonumber \\
&=&  \left< p \right| \hat \rho \left| p \right>,
\end{eqnarray}
which is what we claimed.
\section{Wigner distributions of combined systems} \label{sec:combinedwig} 
Recall that in section \ref{sec:composite}, we defined the state operator of a composite system as
\begin{equation}
\hat{\rho}_{1+2} = \hat{\rho}_1 \otimes \hat{\rho}_2,
\end{equation}
where $( \otimes )$ is the tensor product. In analogy to this, we define the Wigner distribution of a composite system to be \cite{halliwell}
\begin{equation}\label{eqn:combinedwig}
W_{1+2}(x_1,x_2,p_1,p_2) \equiv W_1(x_1,p_1)W_2(x_2,p_2). \index{Wigner Distribution!Composite systems}
\end{equation}
In section \ref{sec:composite}, we also developed the partial trace, which was a method for extracting information about a single sub-state operator in a composite state operator. Not surprisingly, we define an analogous operation, which effectively annihilates one of the sub-Wigner distributions in a composite distribution by integrating out the degrees of freedom of the sub-distribution. Formally, we call this the projection function  $\mathcal A:W_{1+2}\rightarrow W_1$ and define it as
\begin{equation}
\mathcal A \left( W_{1+2} \right) \equiv \int dx_2 dp_2 W_{1+2}.  \label{eqn:annihilator}\index{Wigner Distribution!Projection Function}
\end{equation}
To understand how it works, we evaluate it on the initial total Wigner distribution. This is
\begin{eqnarray}
\mathcal A \left(W_{1+2} \right)
&=&	\mathcal A \left(W_1(x_1,p_1 ) W_2 (x_2,p_2 )\right) \nonumber \\
&=&	 \int dx_2 dp_2 W_1(x_1,p_1 ) W_2 (x_2,p_2 ) \nonumber \\
&=&	 W_1(x_1,p_1 ) \int dx_2 dp_2  W_2 (x_2,p_2 ) \nonumber \\
&=&	 W_1(x_1,p_1 ) \int dx_2 \rho_2(x_2,x_2) \nonumber \\
&=&	 W_1(x_1,p_1 ) \mathrm{Tr}\left( \rho_2 \right) \nonumber \\
&=&	 W_1(x_1,p_1 ),
\end{eqnarray}
where we have used eqns. \ref{eqn:marginal1} and definition \ref{defn:trace} to integrate $W_2$ and perform the full trace of $\rho_2$. Thus, $\mathcal A$ behaves as desired, in direct analogy to the partial trace on composite state operators.

\section{Equation of Motion for a Free Particle} \label{sec:freesys}
Now that we have laid out the basic properties of the Wigner distribution, we need to understand how to use it to describe a physical system. In this section, we investigate how a Wigner distribution evolves in time in the absence of a potential. Recall that in section \ref{sec:freeparticle}, we established the Hamiltonian of a free system as
\begin{equation}
\hat H=\frac{\hat P^2}{2m}.
\end{equation}
Given the Hamiltonian, we can calculate the time evolution of the state operator of the system via the commutator relation
\begin{equation}
\partial_t  \hat{\rho} = - i \left[ \hat H, \hat{\rho} \right],
\end{equation}
developed in eqn. \ref{eqn:heispic}, to obtain 
\begin{equation}
\partial_t  \hat{\rho} = i\left(\hat{\rho} \hat H - \hat H \hat{\rho} \right)= \frac{i}{2m} \left( \hat{\rho} \hat P^2 - \hat P^2 \hat{\rho} \right),	
\end{equation}
noting that $m$ is a scalar. So far, we have a general operator equation for the evolution of the system. If we want to know more specific information about its motion, we need to choose a basis onto which we may project our equation. Choosing momentum, we multiply both sides of the equation by $\left< p \right|$  from the left and $\left| p' \right>$ from the right, where $p$ and $p'$ are two arbitrary momentum states of our system. This gives us
\begin{equation}
\left< p \right| \partial_t  \hat {\rho} \left| p' \right>= \left< p\right| \frac{i}{2m} \left( \hat{\rho} \hat P^2 - \hat P^2 \hat{\rho} \right) \left| p' \right>.
\end{equation}
Since $p$ and $p'$ are states of definite momentum, they are eigenvalues of $\hat P$. Hence, $\hat{P} \left| p \right>= p\left| p \right>$, $\left<p \right| \hat P = \left<p \right| p$, and likewise for $p'$. So, our equation of motion becomes
\begin{eqnarray}
\left< p \right| \partial_t  \hat{\rho} \left| p ' \right>	&=&
 \left< p \right| \frac{i}{2m} \left( \hat{\rho} \hat p^2 - \hat p^2 \hat{\rho} \right) \left| p' \right> \nonumber \\
 &=&  \frac{i}{2m} \left< p \right|  \hat{\rho} \hat p^2 \left| p' \right>  -  \frac{i}{2m} \left< p \right|  \hat p^2 \hat{\rho} \left| p' \right> \nonumber \\
  &=&  \frac{i}{2m} \left< p \right|  \hat{\rho} \hat p  p' \left| p' \right>  -  \frac{i}{2m} \left< p \right|  p \hat p \hat{\rho} \left| p' \right> \nonumber \\
    &=&  \frac{i}{2m} \left< p \right|  \hat{\rho} p' \left| p' \right>  p'  -  \frac{i}{2m}   \left< p \right|  p \hat{\rho} \left| p' \right> p\nonumber \\
    &=&  \frac{i}{2m} \left< p \right|  \hat{\rho}  \left| p' \right>  \left( {p'}^2-p^2 \right) \nonumber \\ 
        &=&  \frac{i}{2m} \left< p\right|  \hat{\rho}  \left| p' \right> \left( p'-p \right) \left( p' +p \right) .      
\end{eqnarray}
Next, we substitute difference and mean variables, $\lambda$ and $\bar{p}$, for $p$ and $p'$ by defining  $\lambda\equiv p-p' $ and $2 \bar{p} \equiv p+p'$. This substitution is algebraically equivalent to $p=\bar{p}+\lambda/2$ and $p'=\bar{p}-\lambda/2$, so
\begin{equation}
\left< \bar{p}+\lambda/2 \right| \partial_t  \hat {\rho} \left| \bar{p}-\lambda/2 \right> =  \frac{i}{2m} \left< \bar{p}+\lambda/2\right|  \hat{\rho}  \left| \bar{p}-\lambda/2 \right> 2\bar{p}\lambda.
\end{equation}
We multiply both sides of the equation by $d\lambda\cdot e^{-i\lambda x}/(2 \pi)$ (the kernel of the Fourier transform) and integrate from $\lambda=- \infty$ to $\lambda= +\infty$. The left hand side is
\begin{eqnarray}
LHS
&=&\int d\lambda \cdot \frac{1}{2 \pi} e^{-i\lambda x}\left< \bar{p}+\lambda/2 \right| \partial_t  \hat {\rho} \left| \bar{p}-\lambda/2 \right>  \nonumber \\
&=& \int d\lambda \cdot \partial_t  \frac{1}{2 \pi} e^{-i\lambda x}\left< \bar{p}+\lambda/2 \right|  \hat {\rho} \left| \bar{p}-\lambda/2 \right> \nonumber \\
&=& \partial_t \frac{1}{2 \pi}\int  d\lambda \cdot e^{-i\lambda x}\left< \bar{p}+\lambda/2 \right|  \hat {\rho} \left| \bar{p}-\lambda/2 \right> \nonumber \\
&=& \partial_t W_P(x,\bar p,t) \nonumber \\
&\leftrightarrow& \partial_t W(\bar x, p,t),
\end{eqnarray}
where we use the fact that the only explicitly time dependent piece of the integrand is $\hat{\rho}$. We also assume that the integral converges and, in the last step, we use eqn. \ref{eqn:wigmomrep} for the Wigner distribution in the momentum basis. Proceeding in a similar fashion on the right hand side, we get
\begin{eqnarray}
RHS
&=& \int  d\lambda\cdot \frac{1}{2 \pi} e^{-i\lambda x} \frac{i}{2m} \left< \bar{p}+\lambda/2\right|  \hat{\rho}  \left| \bar{p}-\lambda/2 \right> 2\bar{p}\lambda \nonumber \\
&=& \frac{i \cdot i\bar{p}}{m 2 \pi} \int d\lambda\cdot  \left( \frac{\lambda}{i}e^{-i\lambda x} \right) \left< \bar{p}+\lambda /2\right|  \hat{\rho}  \left| \bar{p}-\lambda /2 \right> \nonumber \\
&=& - \frac{\bar{p}}{m} \frac{1}{2 \pi} \int d\lambda \cdot \left( -ie^{-i\lambda x} \right) \left< \bar{p}+\lambda /2\right|  \hat{\rho}  \left| \bar{p}-\lambda /2 \right>  \nonumber \\
&=& - \frac{\bar{p}}{m} \frac{1}{2 \pi} \int  d\lambda \cdot  \partial_{x} \left(e^{-i\lambda x} \right) \left< \bar{p}+\lambda /2\right|  \hat{\rho}  \left| \bar{p}-\lambda /2 \right>  \nonumber \\
&=& - \frac{\bar{p}}{m} \partial_{x} \frac{1}{2 \pi} \int  d\lambda  \cdot e^{-i\lambda x}  \left< \bar{p}+\lambda /2\right|  \hat{\rho}  \left| \bar{p}-\lambda /2 \right>  \nonumber \\
&=& - \frac{\bar{p}}{m} \partial_{x} W_P(x,\bar{p},t) \nonumber \\
&\leftrightarrow& - \frac{\bar{p}}{m} \partial_{\bar x} W(\bar x,p,t),
\end{eqnarray}
where we use the fact that $e^{-i\lambda x}$ was the only factor in the integrand that explicitly depended on $x$. We again assume that the integral converges and use eqn. \ref{eqn:wigmomrep} for the Wigner distribution in the momentum basis. Thus, equating the right hand and left hand sides in the position representation leaves \cite{hillery}
\begin{boxedeqn}{eqn:wigfree}
\partial_t W(\bar x,p,t) =  - \frac{p}{m} \partial_{\bar x} W(\bar x,p,t),\index{Wigner Distribution!Free Evolution}
\end{boxedeqn}
which is the equation of motion for a free system in terms of its Wigner distribution.

Although it might seem convoluted to introduce the Wigner form of this equation rather than using the evolution of a free particle in terms of its state operator, the power of the Wigner distribution is that it allows us to treat position and momentum simultaneously.

\section{Associated Transform and Inversion Properties}
Now that we have determined some of the properties of the Wigner distribution, it is useful to define the Wigner transform of an arbitrary distribution of two variables. 

\begin{boxeddefn}{The Wigner transform\index{Wigner Transform!Defintion}}{def:WignerTrans}
Let $D(x,y)$ be an arbitrary distribution of two variables, $x$ and $y$, and possibly have an implicit temporal dependence.  Then, the \textbf{Wigner transform} $\mathcal W$ of $D$, a special case of the Fourier transform, is defined as
\begin{equation}
\mathcal W \left( D (x,y ) \right) \equiv \frac{1}{2 \pi} \int d \delta  \cdot \cdot e^{-ip \delta } D(x,y),
\end{equation}
where $\delta=x-y$, as identified in definition \ref{def:WignerDist}.
\end{boxeddefn}

By definition, we know the Wigner transform of $\rho(x,y)$ immediately. It is
\begin{equation}
\mathcal W \left( \rho (x,y) \right) = \frac{1}{2 \pi} \int d \delta  \cdot e^{-ip \delta} \rho(x,y) = W(\bar x, p).
\end{equation}
We arrive at a more interesting result by considering
\begin{equation}
\mathcal W \left( \partial_t \rho(x,y) \right)=\int d \delta  \cdot e^{-ip \delta } \partial_t \rho(x,y).
\end{equation}
Clearly, neither $\delta$ nor $e^{-ip \delta }$ depend explicitly on time. Assuming that the integral converges, we have
\begin{equation}
\int d \delta  \cdot e^{-ip \delta } \partial_t \rho(x,y) = \partial_t \int d \delta  \cdot e^{-ip \delta }  \rho(x,y) = \partial_t W(\bar x,p),
\end{equation}
where we have applied definition \ref{def:WignerTrans}. So,
\begin{equation}
\mathcal W \left( \partial_t \rho(x,y) \right) = \partial_t W(\bar x,p),
\end{equation}
as desired.

In the following sections, we work out some of the Wigner transforms of functions that we will need later. The results of these derivations are summarized in table \ref{tab:inversions} below. 

\begin{table}[h] \caption{Wigner transforms of important quantities, where $\bar x = \frac{x+y}{2}$ and $\delta = x - y$. \index{Wigner Transform!of Important Quantities}} \label{tab:inversions}\centering  
\begin{tabular}{|cc|}
\hline Expression & Transform \\ 
\hline 
$ \rho (x,y)$ 			& $W(\bar x, p)$ \\
$\partial_t \rho(x,y)$		& $\partial_t W(\bar x,p)$ \\
$ \frac{i}{2} \left( \partial_x^2 - \partial_y^2 \right) \rho(x,y)$	& $-p \partial_{\bar x} W \left( \bar x , p\right)$\\
$(x-y) \left( \partial_x - \partial_y \right) \rho (x,y)$ 			& $- 2 \partial_p \left( p \cdot W( \bar x, p) \right)$ \\
$ \left(x - y  \right)^2 \rho( x, y)$							& $- \partial_p^2W( \bar x, p)$ \\
\hline
\end{tabular} 
\end{table}

\subsection{The Wigner transform of $\left(i/2 \right)\left( \partial_x^2 - \partial_y^2 \right) \rho(x,y)$}

By definition \ref{def:WignerTrans}, 
\begin{equation}
V \equiv \mathcal W \left( \frac{i}{2} \left( \partial_x^2 - \partial_y^2 \right) \rho(x,y)\right)=\int d \delta  \cdot e^{-ip \delta }  \frac{i}{2} \left( \partial_x^2 - \partial_y^2 \right) \rho(x,y),
\end{equation}
where we have defined $V$ for our convenience. Then, we know from Clairaut's theorem that since all partial derivatives of the state operator are continuous, $\left[ \partial _x \rho (x,y) , \partial_y \rho (x,y) \right]=0$ \cite{stewart}. $V$ then expands to
\begin{equation}
V =  \frac{i}{2} \int d \delta  \cdot e^{-ip \delta } \left( \partial_x^2 - \partial_y^2 + \partial_x \partial_y - \partial_y\partial_x \right) \rho(x,y).
\end{equation}
We next note that by definition \ref{def:WignerDist}, $x=\bar x + 1/2 \cdot \delta$ and $y = \bar x - 1/2 \cdot \delta$, which implies $\partial_{\delta}x = 1/2$ and $\partial_{\delta} y = -1/2$. Hence, $2 \partial_{\delta} x = 1$ and $2 \partial_{\delta} y = -1$. We rework $V$ to be
\begin{eqnarray}
V
 &=& \frac{i}{2}\cdot 2 \int d \delta  \cdot e^{-ip \delta } \left(  \left( \partial_{\delta}x\right)\partial_x^2 + \left( \partial_{\delta}y\right)\partial_y^2 +  \left( \partial_{\delta}x\right)\partial_x \partial_y + \left( \partial_{\delta}y\right) \partial_y\partial_x \right) \rho(x,y) \nonumber \\
  &=& i \int d \delta  \cdot e^{-ip \delta }\left(  \left( \frac{ \partial \rho(x,y) / \partial x}{\partial x} \right) \frac{\partial x}{\partial \delta} + \left( \frac{ \partial \rho(x,y) / \partial y}{\partial y} \right) \frac{\partial y}{\partial \delta} +  \left( \frac{ \partial \rho(x,y) / \partial y}{\partial x} \right) \frac{\partial x}{\partial \delta}  +\left( \frac{ \partial \rho(x,y) / \partial x}{\partial y} \right) \frac{\partial y}{\partial \delta}   \right)\nonumber \\
  &=& i \int d \delta  \cdot e^{-ip \delta } \partial _{\delta} \left( \partial_x \rho(x,y) + \partial_y \rho (x,y) \right).
\end{eqnarray} 
Now, by definition \ref{def:WignerDist}, $\partial_{\bar x } x = \partial_{\bar x } y=1$, so
\begin{equation}
\frac{\partial \rho(x,y)}{\partial \bar x } = \frac{ \partial \rho (x,y) }{\partial x} \frac{\partial x}{\partial \bar x} + \frac{ \partial \rho (x,y) }{\partial y} \frac{\partial y}{\partial \bar x} = \frac{ \partial \rho (x,y) }{\partial x}+ \frac{ \partial \rho (x,y) }{\partial y} = \partial_x \rho(x,y) + \partial_y \rho (x,y).
\end{equation} 
Thus, we have
\begin{equation}
V =  i \int d \delta  \cdot e^{-ip \delta } \partial _{\delta} \partial_{\bar x} \rho(x,y).
\end{equation}
Next, we integrate by parts to get
\begin{equation}
V = \left( e^{-ip \delta } \partial_{\bar x} \rho(x,y) \right)\Big |_{\delta = - \infty}^{\infty} - i \int d \delta  \cdot \left( \partial _{\delta} e^{-ip \delta } \right)  \partial_{\bar x} \rho(x,y).
\end{equation}
Noting that the state operator is continuous in the position basis, we find
\begin{eqnarray}
\lim_{\delta \rightarrow \pm \infty}   \partial_{\bar x} \rho(x,y) 
&=& \lim_{\delta \rightarrow \pm \infty}  \partial_{\bar x} \rho \left( \bar x + \frac{1}{2} \delta, \bar x - \frac{1}{2} \delta \right) \nonumber \\
 &=& \partial_{\bar x}\lim_{\delta \rightarrow \pm \infty}   \rho \left( \bar x + \frac{1}{2} \delta, \bar x - \frac{1}{2} \delta \right)  \nonumber \\
  &=& \partial_{\bar x}\lim_{\delta \rightarrow \pm \infty}   \rho \left( \frac{1}{2} \delta, - \frac{1}{2} \delta \right)  \nonumber \\
    &=& 0.
\end{eqnarray}
Further, 
\begin{equation}
0 \leq \left| e^{-ip \delta} \right| \leq 1 \, \forall \delta,
\end{equation}
so
\begin{equation}
 \left( e^{-ip \delta } \partial_{\bar x} \rho(x,y) \right)\Big |_{\delta = - \infty}^{\infty}  = 0.
\end{equation}
Hence,
\begin{equation}
V =  - i \int d \delta  \cdot \left( \partial _{\delta} e^{-ip \delta } \right)  \partial_{\bar x} \rho(x,y)= - i \int d \delta  \cdot \left(-ip e^{-ip \delta } \right)  \partial_{\bar x} \rho(x,y)= -p \partial_{\bar x} \int d \delta  \cdot e^{-ip \delta }  \rho(x,y).
\end{equation}
That is, by definition \ref{def:WignerDist},
\begin{boxedeqn}{}
 \mathcal W \left( \frac{i}{2} \left( \partial_x^2 - \partial_y^2 \right) \rho(x,y)\right) = -p \partial_{\bar x} \int d \delta  \cdot e^{-ip \delta }  \rho(x,y) = -p \partial_{\bar x} W \left( \bar x , p\right),
\end{boxedeqn}
which is what we wanted to show.

\subsection{The Wigner transform of $ (x-y) \left( \partial_x - \partial_y \right) \rho (x,y) $}
By definition \ref{def:WignerTrans}, 
\begin{equation}
V \equiv \mathcal W \left(  (x-y) \left( \partial_x - \partial_y \right) \rho (x,y)  \right)=\int d \delta  \cdot e^{-ip \delta }   (x-y) \left( \partial_x - \partial_y \right) \rho (x,y) ,
\end{equation}
where we have again defined $V$ for our convenience. Since $\delta = x - y $, we have
\begin{equation}
V =  \int d \delta \cdot \delta e^{-ip \delta }  \left( \partial_x - \partial_y \right) \rho (x,y).
\end{equation}
As we did in the previous section, we note that $\partial_{\delta} x = 1/2$ and $\partial_{\delta} y = - 1/2$, so $2 \partial_{\delta} x = 1$ and $2 \partial_{\delta} y = - 1$. Thus,
\begin{equation}
V = 2  \int d \delta \cdot \delta e^{-ip \delta }  \left( \frac{\partial \rho(x,y)}{\partial x} \frac{\partial x}{\partial \delta}+ \frac{\partial \rho(x,y)}{\partial y} \frac{\partial y}{\partial \delta} \right).
\end{equation}
Next, we use the chain rule to find
\begin{equation}
V = 2  \int d \delta \cdot \delta e^{-ip \delta }  \partial_{\delta} \rho(x,y).
\end{equation}
After that, we use integration by parts to get 
\begin{equation}
V = 2 \left(   \delta e^{-ip \delta } \rho(x,y) \right)\Big |_{\delta = - \infty}^{\infty} - 2  \int d \delta \cdot \rho(x,y) \partial_{\delta} \left( \delta e^{-ip \delta }  \right) .
\end{equation}
As before, we investigate the boundary term. The non-oscillatory component follows
\begin{equation}
\lim_{\delta \rightarrow \pm \infty}  \delta \rho(x,y) = 0,
\end{equation}
since $\rho(x,y)$ goes to zero rapidly off the diagonal, as we noted in section \ref{sec:projonbasis}. Since
\begin{equation}
0 \leq \left| e^{-ip \delta} \right| \leq 1 \, \, \forall \delta,
\end{equation}
we have 
\begin{equation}
\lim_{  \delta \rightarrow \pm \infty} e^{-ip \delta} \delta \rho(x,y) = 0,
\end{equation}
so
\begin{equation}
V =  - 2  \int d \delta \cdot \rho(x,y) \partial_{\delta} \left( \delta e^{-ip \delta }  \right).
\end{equation}
Finally, note that 
\begin{equation}
 \partial_{\delta} \left( \delta e^{-ip \delta }  \right) =  \partial_{p} \left( p e^{-ip \delta }  \right),
\end{equation}
hence
\begin{equation}
V =  - 2  \int d \delta \cdot \rho(x,y) \partial_{p} \left( p e^{-ip \delta }  \right) =  - 2  \partial_{p} \left( p \int d \delta \cdot e^{-ip \delta }  \rho(x,y) \right) = - 2 \partial_p \left( p \cdot W( \bar x, p) \right).
\end{equation}
That is,
\begin{boxedeqn}{}
\mathcal W \left(  (x-y) \left( \partial_x - \partial_y \right) \rho (x,y)  \right) = - 2 \partial_p \left( p \cdot W( \bar x, p) \right),
\end{boxedeqn}
as desired.

\subsection{The Wigner transform of $\left(x - y  \right)^2 \rho( x, y) $}
By definition \ref{def:WignerTrans}, 
\begin{equation}
V \equiv \mathcal W \left(  \left(x - y  \right)^2 \rho( x, y)   \right) =\int d \delta  \cdot e^{-ip \delta } \left(x - y  \right)^2 \rho( x, y)  ,
\end{equation}
where, as in the past two sections, we have defined $V$ for our convenience. By definition \ref{def:WignerDist}, $\delta = x-y$, so
\begin{equation}
V = \int d \delta  \cdot e^{-ip \delta } \delta^2 \rho( x, y)=\int d \delta \cdot  \delta^2 e^{-ip \delta } \rho( x, y).
\end{equation}
Now, since
\begin{equation}
\delta^2 e^{-ip \delta }= -\left(i^2 \delta^2 \right) e^{-ip \delta } =- \partial_p^2 e^{-ip \delta }  ,
\end{equation}
we have
\begin{equation}
V= \int d \delta  \cdot \left( - \partial_p^2  e^{-ip \delta } \right) \rho( x, y) = -  \partial_p^2 \int d \delta   \cdot e^{-ip \delta }  \rho( x, y)  =  - \partial_p^2W( \bar x, p),
\end{equation}
or
\begin{boxedeqn}{}
 \mathcal W \left(  \left(x - y  \right)^2 \rho( x, y)   \right) = - \partial_p^2W( \bar x, p),
 \end{boxedeqn}
which is what we wanted to show.

\section{Example: The Wigner Distribution of a Harmonic Oscillator}\label{sec:wig_harmonic}
We will next develop the Wigner distribution for the quantum harmonic oscillator. The Hamiltonian is \cite{griffiths}
\begin{equation}
\hat H = \frac{\hat P^2}{2m} + \frac{1}{2} k x^2,
\end{equation}
where $k$ is the spring constant, and the angular frequency is
\begin{equation}
\omega = \sqrt{\frac{k}{m}}.
\end{equation}
From the time-independent Schr\"odinger equation, eqn. \ref{eqn:schrotimeind}, we have
\begin{equation}
 - \frac{\hbar^2}{2M} \check d _x^2 \psi (x) + \frac{1}{2} k x^2 \psi(x) = E \psi (x).
\end{equation}
This equation is readily solved using power series, and has the well-known family of solutions \cite{ballentine,griffiths,cohtan}
\begin{equation}
\psi_n(x) = \frac{1}{\sqrt{n!}}\left( \frac{1}{\sqrt{2 m \omega}}\left( m \omega x - \partial_x \right) \right)^n \left( \frac{m \omega}{\pi} \right)^{1/4}e^{-\frac{m \omega}{2}x^2},
\end{equation}
\begin{figure}[tb] 
\begin{center} 
\includegraphics[width=0.9 \linewidth]{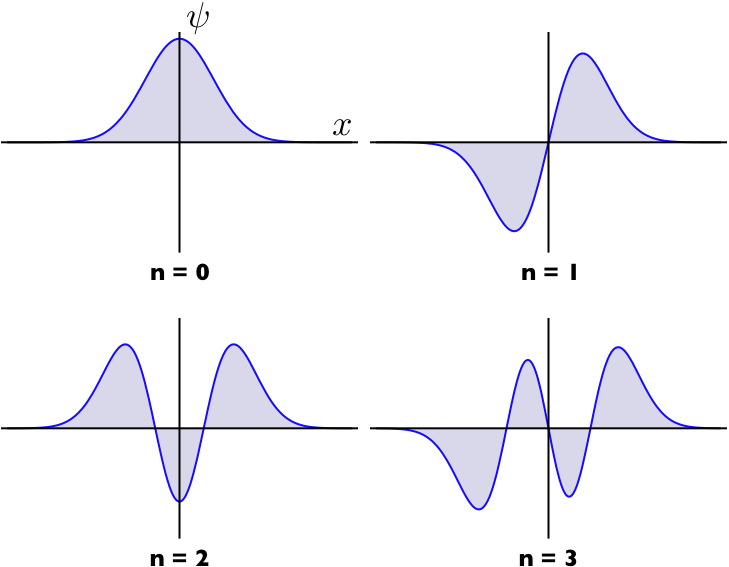}
\end{center} 
\caption[Wavefunctions of the quantum harmonic oscillator]{The first four energy states, $\psi_n(x)$, of the harmonic oscillator.}\label{fig:harmonic_wavefunctions} 
\end{figure}
which correspond to states of constant energy \cite{griffiths}
\begin{equation}
E_n=\left( n + \frac{1}{2} \right) \omega.
\end{equation}
For the purposes of this example, we will concentrate on the ground state ($n=0$) and the first three excited states, shown in figure \ref{fig:harmonic_wavefunctions}. In order to calculate the Wigner distribution of these states, we must use eqn. $\ref{eqn:wigalt}$, so we need explicit forms for the matrix elements of the state operators in the position basis. Fortunately, since the harmonic oscillator is pure, we easily obtain these by
\begin{equation}
\rho_n (x,y) = \left< x \right| \hat{\rho}_n \left| y \right> = \left< x \big| \psi_n \right> \left< \psi_n \big| y \right> = \psi_n^*(x) \psi_n(y),
\end{equation}
where we used eqn. \ref{eqn:wavefunction} to identify the wavefunction $\psi_n(y)$ and its complex conjugate $\psi_n^*(x)$. Since $\psi(x)$ is real we have,
\begin{boxedeqn}{eqn:stateopharmonic}
\rho_n(x,y) =  \frac{1}{n!} \left( \frac{m \omega}{\pi} \right)^{1/2} \left( \frac{1}{2 m \omega}\right)^n\left( m \omega x - \partial_x  \right)^n \left( m \omega y - \partial_y  \right)^n  e^{-\frac{m \omega}{2}x^2} e^{-\frac{m \omega}{2}y^2}.
\end{boxedeqn}
\begin{figure}[tb] 
\begin{center} 
\includegraphics[width=0.9 \linewidth]{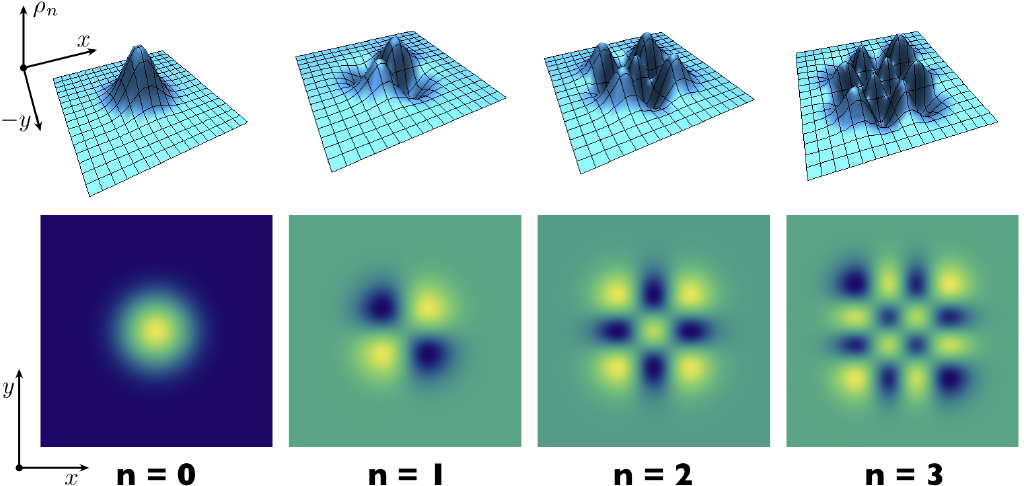}
\end{center} 
\caption[State operators of the quantum harmonic oscillator]{The position representation of the state operator, $\rho_n(x,y)$, for the first four energy states of the harmonic oscillator. In the density plots, yellow indicates maximum values, while blue indicates minimum.}\label{fig:harmonic_stateops} 
\end{figure}
Particularly, for $n=0$ through $n=3$, in units where $m=\omega = \hbar = 1$, we have
\begin{eqnarray}
\rho_0(x,y)&=& \frac{1}{\sqrt{ \pi}}e^{ - \frac{ x^2+y^2}{2}} \nonumber, \\
\rho_1(x,y)&=& 2 x y \frac{1}{\sqrt{ \pi}}e^{ - \frac{ x^2+y^2}{2}} \nonumber, \\
\rho_2(x,y)&=& \frac{1}{8}\left( -2 e^{-x^2}+4e^{-x^2}x^2\right) \left( -2 e^{-y^2}+4e^{-y^2}y^2\right) \frac{1}{\sqrt{ \pi}}e^{ \frac{ x^2+y^2}{2}} \nonumber, \\
\rho_3(x,y)&=& \frac{1}{48} \left( 12 e^{-x^2} x - 8e^{-x^2}x^3\right) \left( 12 e^{-y^2} y - 8e^{-y^2}y^3 \right) \frac{1}{\sqrt{ \pi}}e^{ + \frac{ x^2+y^2}{2}} \nonumber, \\
\end{eqnarray}
which we plot in figure \ref{fig:harmonic_stateops}. Now that we have the general form of the state operator matrix elements, it is just a matter of evaluating eqn. \ref{eqn:wigalt} to get the corresponding Wigner distributions. Starting with $n=0$, we have
\begin{eqnarray}
W_0(\bar x,p) 
&=& \frac{1}{2 \pi} \int d \delta \cdot e^{-ip \delta} \psi_0 \left( \bar x +\frac{1}{2} \delta \right) \psi_0 \left( \bar x - \frac{1}{2} \delta \right) \nonumber \\
&=& \frac{1}{2 \pi} \int d \delta \cdot e^{-ip \delta} \left( \left( \frac{\omega m }{\pi}\right) ^{1/4} e^{-\frac{1}{2} \omega m  \left(  \bar x +\frac{1}{2} \delta  \right)^2 }\right)  \left( \left( \frac{\omega m }{\pi}^{1/4}\right)  e^{-\frac{1}{2} \omega m  \left(  \bar x -\frac{1}{2} \delta  \right)^2 }\right)   \nonumber \\
&=& \frac{1}{2 \pi} \left( \frac{\omega m }{\pi}\right) ^{1/2} \int d \delta \cdot e^{-ip \delta}  e^{-\frac{1}{4} \omega m\left( \bar x - \frac{1}{2} \delta \right)^2 }    \nonumber \\ 
&=& \frac{1}{\pi} e^{- p^2 - \bar x^2},
\end{eqnarray}
\begin{figure}[tb] 
\begin{center} 
\includegraphics[width=0.9 \linewidth]{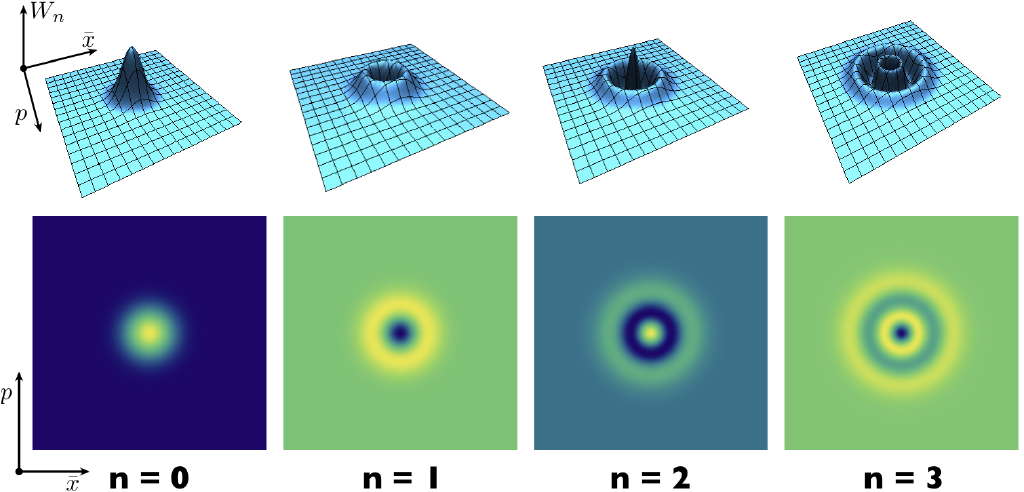}
\end{center} 
\caption[Wigner distributions of the quantum harmonic oscillator]{The Wigner distribution, $W_n(\bar x, p )$, of the first four energy states of the harmonic oscillator with their well-known shape \cite{zachos2}. In the density plots, yellow indicates maximum values, while blue indicates minimum.}\label{fig:harmonic_wigs} 
\end{figure}
which is just a three-dimensional Gaussian distribution. The calculations involved for the excited states are similar, but the algebra is significantly less trivial. They are easily performed using a computer algebra system, so we state the result. The Wigner distributions are\index{Harmonic Oscillator!Wigner Distribution Solutions}
\begin{eqnarray}
W_0(\bar x,p)  &=&  \frac{1}{\pi} e^{- p^2 - \bar x^2} \nonumber \\
W_1(\bar x,p)  &=&  \frac{2p^2+2\bar x^2 - 1}{ \pi} e^{- p^2 - \bar x^2} \nonumber \\
W_2(\bar x,p)  &=&  \frac{2 p^4 + 2 \bar x^4 +4 p^2 \bar x^2 -4 p^2 -4 \bar x^2+1}{ \pi} e^{- p^2 - \bar x^2} \\
W_3(\bar x,p)  &=&  \frac{4 \bar x^6 + 12 p^2 \bar x^ 4 - 18 \bar x^4 +12 p^4 \bar x^2 - 36 p^2 \bar x^2 +18 \bar x^2 + 4 p^6 - 18 p^4 + 18 p^2 - 3}{3 \pi} e^{- p^2 - \bar x^2}, \nonumber
\end{eqnarray}
\begin{figure}[tb] 
\begin{center} 
\includegraphics[width=0.9 \linewidth]{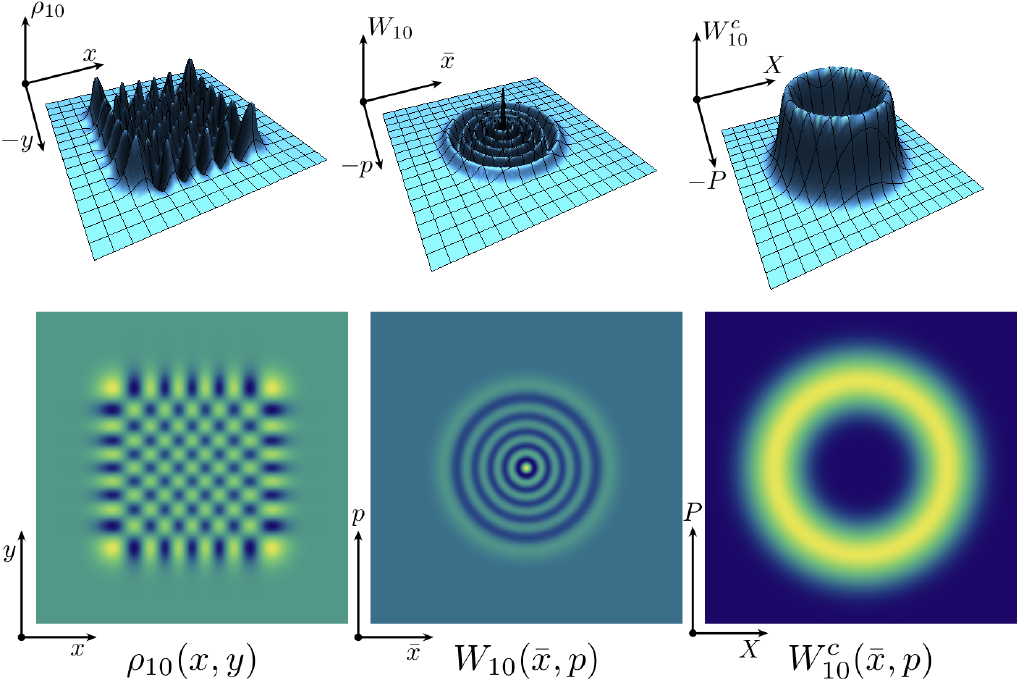}
\end{center} 
\caption[Classical correspondence of the quantum harmonic oscillator]{The position representation of the state operator, $\rho_10(x,y)$, at $n=10$, its associated Wigner distribution, $W_{10}(\bar x, p)$, and the smoothed Wigner distribution generated by convolution with a Gaussian, $W_{10}^c(X,P)$.}\label{fig:convolution} 
\end{figure}
which are plotted in figure \ref{fig:harmonic_wigs}. Note how $W_0 > 0$ for all values of $x$ and $p$, but the higher energy states are sometimes negative. As we mentioned briefly before, the Wigner distribution is motivated by classical phase-space probability distributions, but is permitted to have negative values. These ``negative probabilities'' are a weird signature of a quantum mechanical system. To make this analogy more concrete, we consider $W_{10}(\bar x, p)$, shown in figure \ref{fig:convolution}. At the high energy of $n=10$, the oscillations inside the Wigner distribution become increasingly rapid. In the classical limit, as $n \rightarrow \infty$, we expect the negative portions to overlap and cancel with the positive components, giving us a positive-definite, classical probability distribution. In order to force this for $n=10$, we perform a careful function smoothing, known as a convolution\index{Convolution}, of $W_{10}$ with a simple Gaussian. Mathematically, this is \cite{hecht}
\begin{equation}
W^c_{10}(X,P) \equiv \int d \bar x d p \cdot W_{10}(\bar x, p) e^{-(X - \bar x)^2 -(P - p)^2}.
\end{equation}
As shown in figure \ref{fig:convolution}, this averages the inner oscillations to zero, but retains a large, outer, positive ring. This is what we expect, since a classical simple harmonic oscillator has eliptical orbits in phase-space.


\chapter{The Master Equation for Quantum Brownian Motion}\label{text}
\lettrine[lines=2, lhang=0.33, loversize=0.1]{I}n this chapter, we develop the fundamental equation of quantum decoherence, the master equation for quantum brownian motion. The master equation dictates the time-evolution of a \textbf{system} and an \textbf{environment} with which the system interacts (these terms will be precisely defined later). To facilitate this, we use the formalism of the Wigner distribution developed in chapter \ref{chap:wigner}, since it incorporates both position and momentum simultaneously, and consider how the system's Wigner distribution changes with time. Then, we invert the Wigner transformation to get the master equation, written in terms of the the system's state operator. After the equation is developed in this chapter, we examine its physical meaning and work through an example in chapter \ref{chap:applications}.

\section{The System and Environment}

\begin{figure}[t] 
\begin{center} 
\includegraphics[width=0.9 \linewidth]{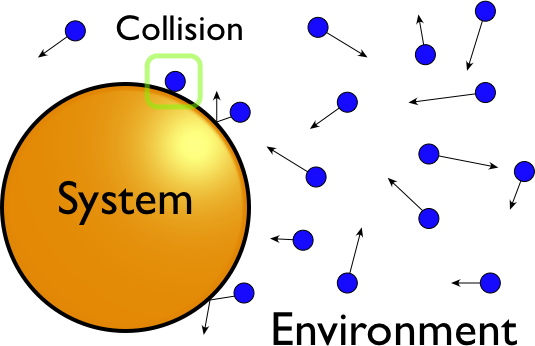}
\end{center} 
\caption[System particle interacting with environment particles]{A graphic representation of the system and environment. Note that one of the environment particles is undergoing a collision with the system. For simplicity we consider the corresponding one-dimensional problem.}\label{fig:collision} 
\end{figure}

The idea of collisions between a system and environment can be represented intuitively in a physical picture, as shown in figure \ref{fig:collision}. However, before we begin, we need to define precisely the notion of \textbf{system} and \textbf{environment}. Further, we need to specify what we mean by an interaction or \textbf{collision} between the system and environment.

\begin{boxeddefn}{System\index{System}}{def:system}
A \textbf{system}, or system particle, denoted as $\mathcal S$, is a single, one-dimensional point particle. It has momentum $p_S$, mass $m_S$, and position $x_S$. 
\end{boxeddefn}

\begin{boxeddefn}{Environment\index{Environment}}{def:environment}
An \textbf{environment} of a system $\mathcal S$ is denoted $\mathcal E_S$, and consists of an ideal one-dimensional gas of light particles. Each of these particles has momentum $p_E$, mass $m_E$, and position $x_E$. We will often abbreviate $\mathcal E_S$ to $\mathcal E$ if it is clear to what system $\mathcal S$ the environment belongs.
\end{boxeddefn}

\begin{boxeddefn}{Collision\index{Collision}}{def:collision}
A \textbf{collision} between a particle of an environment $\mathcal E$ and a system $\mathcal S$ is defined as an instantaneous transfer of momentum that conserves both kinetic energy and momentum. 
\end{boxeddefn}
It is important to note that the system we are considering is very large (massive) when compared to the individual environmental particles. Precisely, we take \cite{halliwell}
\begin{equation}\label{eqn:massapprox}
\frac{m_E}{m_S} \ll 1,
\end{equation}
and we will typically neglect terms of second or higher order in this factor.

Now that we have defined the key objects treated by the master equation, we begin to investigate its structure. As stated above, we first want to consider how the Wigner Distribution of the system, $W_S$, changes with time. Quantum mechanically, we separate this change into two pieces. First, $W_S$ undergoes standard unitary time evolution, with the system treated as a free particle. Second, $\mathcal S$ collides with environment particles, and the collisions alter the system's energy and momentum. In section \ref{sec:freesys}, we considered the change in the Wigner distribution of a particle due to its free evolution, which we will make use of later. Now, we begin to consider the influence of an environment on a system.
\section{Collisions Between Systems and Environment Particles}
Before we begin to examine how a system behaves in the presence of an environment, we first consider the collision between a system particle and one particle from an environment. For each collision, we derive equations for momentum and position change. First, we address momentum change.

Let $p_S$ and $p_E$ denote the initial momenta of a system and an environment particle. By definition \ref{def:collision}, the interaction between the two particles is totally elastic. That is, both kinetic energy and momentum are conserved. We write kinetic energy conservation as \cite{hrw}
\begin{equation}
\frac{p_S^2}{2m_S}+\frac{p_E^2}{2m_E}=\frac{\bar p_S^2}{2m_S}+\frac{\bar p_E^2}{2m_E}, 
\end{equation}
which is equivalent to
\begin{equation}\label{eqn:kecons}
m_S\left( p_E- \bar p_E \right) \left( p_E + \bar p_E \right)=-m_E\left( p_S- \bar p_S \right) \left( p_S + \bar p_S \right),
\end{equation}
and momentum conservation as \cite{hrw}
\begin{equation}
p_s+p_E=\bar p_S + \bar p_E,
\end{equation}
which is also written as
\begin{equation}\label{eqn:momcons}
\left( p_E- \bar p_E \right) =-\left(p_S - \bar p_S \right).
\end{equation}
We then assume that, since a collision occurred, the momenta of both the system and environment particle have changed, i.e. $ p_E- \bar p_E \neq 0$ and $p_S - \bar p_S \neq 0$. So, we divide eqn. \ref{eqn:kecons} by eqn. \ref{eqn:momcons} to get 
\begin{equation}
\frac{m_S\left( p_E- \bar p_E \right) \left( p_E + \bar p_E \right)}{ \left( p_E- \bar p_E \right)}=\frac{-m_E\left( p_S- \bar p_S \right) \left( p_S + \bar p_S \right)}{-\left(p_S - \bar p_S \right)},
\end{equation}
which implies
\begin{equation}\label{eqn:redkecons}
m_S\left( p_E + \bar p_E \right)=m_E \left( p_S + \bar p_S \right).
\end{equation}
Then, we solve eqns. \ref{eqn:momcons} and \ref{eqn:redkecons} simultaneously for both $\bar p_S$ and $\bar p_E$. We have \cite{hrw}
\begin{eqnarray}\label{eqn:pebar}
m_S\left( p_E + \bar p_E \right)=m_E \left( p_S + p_S+p_E-\bar p_E \right) 
&\Rightarrow& -(m_S-m_E)p_E+2m_Ep_S=(m_E+m_S)\bar p_E \nonumber \\
&\Rightarrow& \bar p_E = -\frac{m_S-m_E}{m_S+m_E}p_E+\frac{2m_E}{m_S+m_E}p_S
\end{eqnarray}
and
\begin{eqnarray}\label{eqn:psbar}
m_S\left( p_E + p_E+p_S -\bar p_S \right)=m_E \left( p_S + \bar p_S\right) 
&\Rightarrow& (m_S-m_E)p_S+2m_Sp_E=(m_S+m_E)\bar p_S\nonumber \\
&\Rightarrow& \bar p_S=\frac{m_S-m_E}{m_S+m_E}p_S+\frac{2m_S}{m_S+m_E}p_E ,
\end{eqnarray}
which are the changes in the momenta of the environment particle and the system.

Now that we have investigated the momentum change that results from a collision, we will develop the corresponding position change. To do this, we use the plane wave treatment for the total system we developed in section \ref{sec:freeparticle} and note how changes in momentum imply changes in position.

The wavefunction of the composite system containing both the system and environment particle, a product state, is given by\footnote{Since the product state vector is $\left| \phi \right> = \left| \phi_S \right> \otimes \left| \phi_E \right>$, the wavefunction form of the composite state takes ordinary multiplication.}
\begin{equation}
\phi = \phi_S \phi_E.
\end{equation}
Using equation \ref{eqn:planewave}, we can form the composite plane wave, $\phi_i$, from the individual incident plane wave of each particle.\footnote{Remember that plane waves are states of definite momentum. We are using them in this case because we are conserving the momentum in the collision.} This is\begin{equation}
\phi_i=e^ {ip_Sx_S} e^ {ip_Ex_E}.
\end{equation}
After collision, using the momentum representation, the plane wave, $\phi_f$, becomes
\begin{equation}
\phi_f=e^ {i \bar p_S x_S}e^{i \bar p_E x_E}.
\end{equation}
By eqns. \ref{eqn:pebar} and \ref{eqn:psbar}, this can be written as
\begin{eqnarray}
\left( \text{Exponent of $\phi_f$} \right)
&=& \left(i \left(  \frac{m_S-m_E}{m_S+m_E}p_S+\frac{2m_S}{m_S+m_E}p_E\right) x_S \right)  + \left( i\left(-\frac{m_S-m_E}{m_S+m_E}p_E+\frac{2m_E}{m_S+m_E}p_S\right) x_E \right)\nonumber \\
&=&  i \left(  \frac{m_S-m_E}{m_S+m_E}p_S x_S+\frac{2m_S}{m_S+m_E}p_E x_S-\frac{m_S-m_E}{m_S+m_E}p_E x_E +\frac{2m_E}{m_S+m_E}p_S x_E \right)  \nonumber \\
&=& \left( ip_S\left(\frac{m_S-m_E}{m_S+m_E}x_S+\frac{2m_E}{m_S+m_E}x_E\right)\right) + \left( ip_E\left( \frac{2m_S}{m_S+m_E}x_S-\frac{m_S-m_E}{m_S+m_E}x_E \right) \right). \nonumber 
\end{eqnarray}
We define
\begin{equation}
\phi_f=e^ {i \bar p_S x_S} e^{i \bar p_E x_E} \equiv e^{ip_S\bar x_S }e^{ip_E\bar x_E},
\end{equation}
where \cite{halliwell}
\begin{equation}
 \bar x_S = \frac{m_S-m_E}{m_S+m_E}x_S+\frac{2m_E}{m_S+m_E}x_E
\end{equation} 
and 
\begin{equation}
\bar x_E =  \frac{2m_S}{m_S+m_E}x_S-\frac{m_S-m_E}{m_S+m_E}x_E.
\end{equation}

This way, we now have position and momentum representations of the collision. As is common in physics, we need to require that these collision interactions are local\index{Collision!Locality}.\footnote{It is important to emphasize that locality is \textit{not} an approximation, but is necessary to include in our treatment. Ideally, we would work this into our equations formally. However, for simplicity, we can achieve local interactions by requiring this condition.} Thus, throughout the collision, we take 
\begin{equation} \label{eqn:locality1}
 \left| x_S-x_E  \right| \ll \left| x_S \right|
\end{equation}
and 
\begin{equation}
\left| x_S-x_E  \right| \ll \left| x_E \right|,
\end{equation}
since the potential energy, $V \left(x_S-x_E \right) \rightarrow 0$ as $\left| x_S-x_E  \right| \rightarrow \infty$. Recalling that from eqn. \ref{eqn:massapprox}
\begin{equation}
\frac{m_E}{m_S} \ll 1,
\end{equation}
it is reasonable to ignore contributions to distances of order
\begin{equation}
\frac{m_E}{m_S} (x_S-x_E) \ll x_S,x_E.
\end{equation}
Enforcing the locality of collision, we find
\begin{eqnarray}
\bar x_S 
&=& \frac{m_S-m_E}{m_S+m_E}x_S+\frac{2m_E}{m_S+m_E}x_E \nonumber \\
&=& \left( \frac{m_S+m_E}{m_S+m_E}-\frac{2m_E}{m_S+m_E} \right)x_S+\frac{2m_E}{m_S+m_E}x_E \nonumber \\
&=& \left( 1-\frac{2m_E}{m_S+m_E} \right)x_S+\frac{2m_E}{m_S+m_E}x_E \nonumber \\
&=& x_S-\frac{2m_E}{m_S+m_E} x_S+\frac{2m_E}{m_S+m_E}x_E \nonumber \\
&=& x_S+ \frac{2m_E}{m_S+m_E}\left( x_E- x_S \right) \nonumber \\
&=& x_S +2 \frac{m_E}{m_S}\left( x_E- x_S \right) - 2 \left( \frac{m_E}{m_S} \right)^2\left( x_E- x_S \right) + ... \nonumber \\
&\sim& x_S 
\end{eqnarray}
and
\begin{eqnarray}
\bar x_E 
&=&\frac{2m_S}{m_S+m_E}x_S-\frac{m_S-m_E}{m_S+m_E}x_E\nonumber \\
&=&\left( \frac{2m_S+2m_E}{m_S+m_E}-\frac{2m_E}{m_S+m_E} \right)x_S+\left( \frac{2m_E}{m_S+m_E}-\frac{m_S+m_E}{m_S+m_E} \right) 
x_E\nonumber \\
&=&\left( 2-\frac{2m_E}{m_S+m_E}\right)x_S+\left(\frac{2m_E}{m_S+m_E}-1 \right) x_E\nonumber \\
&=&2x_S-x_E+\frac{2m_E}{m_S+m_E}\left(x_E-x_S\right) \nonumber \\
&=& 2x_S-x_E+2 \frac{m_E}{m_S}\left( x_E- x_S \right) - 2 \left( \frac{m_E}{m_S} \right)^2\left( x_E- x_S \right) + ... \nonumber \\
&\sim&2x_S-x_E,
\end{eqnarray}
which amounts to a phase shift in our plane wave state. We have now worked out all the position and momentum components we will need to treat the full case of a system coupled to an environment. In summary, we have\index{Two-body Collisions}
\begin{eqnarray} \label{eqn:summary}
\bar p_S	&=&	\frac{m_S-m_E}{m_S+m_E}p_S+\frac{2m_S}{m_S+m_E}p_E, \nonumber \\
\bar p_E 	&=& -\frac{m_S-m_E}{m_S+m_E}p_E+\frac{2m_E}{m_S+m_E}p_S, \nonumber \\
\bar x_S 	&\sim&	x_S, \nonumber \\
\bar x_E 	&\sim&	2x_S-x_E.
\end{eqnarray}

\section{Effect of Collision on a Wigner Distribution}
In this section, we consider the change in the Wigner distribution of the system, $W_S$, from one collision with an environment particle. Since we have a composite state of environment and system particle, we use equation \ref{eqn:combinedwig} to write the Wigner distribution for the system and environment as
\begin{equation}
W_{S+E}=W_SW_E.
\end{equation}
It follows that the change in the total Wigner distribution for the system and environment, $\Delta W_{S+E}$, due to one collision is \cite{halliwell}
\begin{eqnarray}
\Delta W _{S+E}
&=& \overline{W}_{S+E}-W_{S+E} \nonumber \\
&=&\overline{W}_{S}\overline{W}_{E}-W_{S}W_{E} \nonumber \\
&=& W_S\left(\bar x_S,\bar p_S \right) W_E \left(\bar x_E, \bar p_E \right) - W_S(x_S,p_S ) W_E (x_E,p_E ).
\end{eqnarray}
Now that we have the change in the total (system and environment) Wigner distribution, we use eqn. \ref{eqn:annihilator} developed in section \ref{sec:combinedwig} to deduce $\Delta W$, the change in the system's Wigner distribution, by summing (integrating) over all environmental configurations. We have \cite{ballentine}
\begin{eqnarray}\label{eqn:deltawintegrals}
\Delta W	&=& \mathcal A \left(\Delta W _{S+E} \right) \nonumber \\
 &=& \int dp_E dx_E \cdot\left( W_S\left(\bar x_S,\bar p_S \right) W_E \left(\bar x_E, \bar p_E \right) - W_S(x_S,p_S ) W_E (x_E,p_E ) \right) \nonumber \\
		&=& \int dp_E dx_E \cdot W_S\left(\bar x_S,\bar p_S \right) W_E \left(\bar x_E, \bar p_E \right) - \int dp_E dx_E \cdot W_S(x_S,p_S ) W_E (x_E,p_E ).
\end{eqnarray}
To evaluate these integrals, we need to perform some algebraic manipulations on the first term in eqn. \ref{eqn:deltawintegrals}. From the eqn. \ref{eqn:summary}, we know that the first term of eqn. \ref{eqn:deltawintegrals} is (approximately) given by
\begin{equation}\label{eqn:deltawfirstterm}
\int dp_E dx_E \cdot W_S\left(x_S,\frac{m_S-m_E}{m_S+m_E}p_S+\frac{2m_S}{m_S+m_E}p_E  \right) W_E \left(2x_S-x_E,-\frac{m_S-m_E}{m_S+m_E}p_E+\frac{2m_E}{m_S+m_E}p_S \right).
\end{equation}
We make the substitution
\begin{eqnarray}\label{eqn:subs}
u & \equiv& 2x_S-x_E \nonumber \\
v & \equiv & -\frac{m_S-m_E}{m_S+m_E}p_E+\frac{2m_E}{m_S+m_E}p_S ,
\end{eqnarray}
from which it follows that
\begin{eqnarray}\label{eqn:subsdiff}
dx_E \cdot&=&-du \nonumber \\
dp_E&=&-\frac{m_S+m_E}{m_S-m_E} dv.
\end{eqnarray}
Further, since 
\begin{equation}
p_E=\left( \frac{m_S+m_E}{m_S-m_E} \right) \left( \frac{2m_E}{m_S+m_E}p_S - v \right), 
\end{equation}
we have
\begin{eqnarray} \label{eqn:substaylor}
\frac{m_S-m_E}{m_S+m_E}p_S+\frac{2m_S}{m_S+m_E}p_E 
&=& \frac{m_S-m_E}{m_S+m_E}p_S+\frac{2m_S}{m_S+m_E}\left( \frac{m_S+m_E}{m_S-m_E} \right) \left( \frac{2m_E}{m_S+m_E}p_S - v \right)   \nonumber \\
&=&  \frac{m_S-m_E}{m_S+m_E}p_S+ \frac{4 m_E m_S}{\left(m_S - m_E\right) \left( m_E +m_S \right)} p_S - \frac{2 m_S}{m_S - m_E} v \nonumber \\
&=& \frac{m_E+m_S}{m_S-m_E} p_S - \frac{2 m_S}{m_S - m_E} v \nonumber \\
&=& p_S + \frac{2 \left( m_E p_S - m_S u \right)}{m_S-m_E}. 
\end{eqnarray}
Thus, substituting eqns. \ref{eqn:subs}, \ref{eqn:subsdiff}, and \ref{eqn:substaylor} into eqn. \ref{eqn:deltawfirstterm} gives
\begin{equation}
\frac{m_S+m_E}{m_S-m_E} \int dv du W_S\left(x_S, p_S + \frac{2 \left( m_E p_S - m_S u \right)}{m_S-m_E} \right) W_E \left(u,v \right).
\end{equation}
Next, we make the substitution $u \equiv x_E$ and $v \equiv p_E$, so eqn. \ref{eqn:deltawfirstterm} becomes
\begin{equation}
\frac{m_S+m_E}{m_S-m_E} \int dp_E dx_E \cdot W_S\left(x_S, p_S + \frac{2 \left( m_E p_S - m_S p_E \right)}{m_S-m_E} \right) W_E \left(x_E,p_E \right).
\end{equation}
Now, eqn. \ref{eqn:deltawintegrals} is
\begin{eqnarray}\label{eqn:deltawintegrals2}
\Delta W &=& \frac{m_S+m_E}{m_S-m_E} \int dp_E dx_E \cdot W_S\left(x_S, p_S + \frac{2 \left( m_E p_S - m_S p_E \right)}{m_S-m_E} \right) W_E \left(x_E,p_E \right) \\
&-& \int dp_E dx_E \cdot W_S(x_S,p_S ) W_E (x_E,p_E ) \nonumber \\
&=&  \int dp_E dx_E \cdot \left( \frac{m_S+m_E}{m_S-m_E} W_S\left(x_S, p_S + \frac{2 \left( m_E p_S - m_S p_E \right)}{m_S-m_E} \right) -W_S(x_S,p_S ) \right)W_E \left(x_E,p_E \right). \nonumber
\end{eqnarray}
Next, we expand 
\begin{equation}
W_S\left(x_S, p_S + \frac{2 \left( m_E p_S - m_S p_E \right)}{m_S-m_E} \right)
\end{equation}
using a Taylor series expansion in momentum about $p=p_S$. This is
\begin{eqnarray}\label{eqn:taylor1}
W_S\left(x_S, p_S + \frac{2 \left( m_E p_S - m_S p_E \right)}{m_S-m_E} \right) &=& W_S(x_S,p_S) +  \frac{2 \left( m_E p_S - m_S p_E \right)}{m_S-m_E} \frac{\partial W_S}{ \partial p_S}(x_S,p_S) \nonumber \\
&+& \frac{1}{2} \left( \frac{2 \left( m_E p_S - m_S p_E \right)}{m_S-m_E} \right)^2 \frac{\partial ^2 W_S}{ \partial p_S^2}(x_S,p_S)+...
\end{eqnarray}
In order to justify dropping the high-order terms of the expansion, we need to show that
\begin{equation}
\frac{2 \left( m_E p_S - m_S p_E \right)}{m_S-m_E}  \ll \left| p_S \right| ,
\end{equation} 
which is not readily apparent. If we expand the term in $m_E/m_S$, we have
\begin{equation}
\frac{2 \left( m_E p_S - m_S p_E \right)}{m_S-m_E}  = -2 p_E + 2 \left( p_S - p_E \right) \frac{m_E}{m_S} + 2 \left( p_S - p_E \right) \left( \frac{m_E}{m_S} \right)^2 + ...
\end{equation}
Recalling eqn. \ref{eqn:massapprox}, it is obvious that while the terms of first order and higher in $m_E/m_S$ are small compared to $p_S$, the first term, $-2 p_E$, is not necessarily small with respect $p_S$. Fortunately, since we are expanding in an integrand and the average value of $p_E$ is zero, we can neglect this term and so we are justified in dropping high order terms in our Taylor expansion.\footnote{The fact that the average value of $p_E$ is zero is dealt with explicitly in eqn. \ref{eqn:peavgiszero}.} 
Simplifying coefficients and dropping terms of third order or higher, eqn. \ref{eqn:taylor1} is approximately
\begin{equation}
W_S(x_S,p_S) +  \frac{2 \left( m_E p_S - m_S p_E \right)}{m_S-m_E} \frac{\partial W_S}{\partial p_S}(x_S,p_S)  +\left( \frac{2m_S^2 p_E^2-4m_Em_Sp_Ep_S+2m_E^2p_S^2}{\left( m_E - m_S \right)^2} \right) \frac{\partial ^2 W_S}{\partial p_S^2}(x_S,p_S).
\end{equation}
Hence, we write $\Delta W$ as \cite{halliwell}
\begin{equation} \label{eqn:deltawap1}
\Delta W \sim  \int dp_E dx_E \cdot\left(  A W_S(x_S,p_S) W_E \left(x_E,p_E \right) + B \partial_{p_S}W_S(x_S,p_S) W_E \left(x_E,p_E \right)+ C\partial_{p_S}^2 W_S(x_S,p_S) W_E \left(x_E,p_E \right)\right), 
\end{equation}
for some $A$, $B$, and $C$. We now work out the values of these coefficients, starting with $A$, which is
\begin{eqnarray}
A &=& \frac{m_S+m_E}{m_S-m_E}  - 1 \nonumber \\
	&=& \frac{2 m_E}{m_S-m_E} \nonumber \\
	&=&  \frac{2 m_E}{m_S-m_E} \cdot \frac{1/m_S}{1/m_S} \nonumber \\
	&=&	\frac{2m_E}{m_S} \cdot \frac{1}{1-m_E/m_S} \nonumber \\
	&=& \frac{2m_E}{m_S} \left( 1 + \frac{m_E}{m_S} + \left( \frac{m_E}{m_S} \right)^2+... \right) \nonumber \\
	&=& 2 \frac{m_E}{m_S} +  2 \left( \frac{m_E}{m_S} \right)^2 + ... \nonumber \\
	&\sim& 2 \frac{m_E}{m_S},
\end{eqnarray}
where we used the approximation in eqn. \ref{eqn:massapprox} to neglect the terms of order two or higher in $m_E/m_S$. We now turn to $B$, given by
\begin{equation}
B 	=		\left(  \frac{m_S+m_E}{m_S-m_E} \right) \frac{2 \left( m_E p_S - m_S p_E \right)}{m_S-m_E}.
\end{equation}
Anticipating a series expansion, we change variables to $r = m_E/m_S$ so that $m_E=r m_S$. $B$ is then
\begin{equation}
B =\left(  \frac{m_S+r m_S}{m_S-r m_S} \right) \frac{2 \left( r m_S p_S - m_S p_E \right)}{m_S-r m_S} = \left(  \frac{1+r}{1-r} \right) \frac{2 \left( r p_S - p_E \right)}{1-r} .
\end{equation}
We also calculate the first and second derivatives of $B$ with respect to $r$. They are
\begin{equation}
\frac{d B}{dr} = \frac{2 p_E (3 - r) - 2 (p_S +3 p_S r)}{(r-1)^3}
\end{equation}
and
\begin{equation}
\frac{d^2 B}{dr^2}= \frac{4 p_E (5+r) - 12 p_S (1+r)}{(r-1)^4}.
\end{equation}
Taking the Taylor series expansion of $B$ in $r$ about $r=0$, we find
\begin{eqnarray}
B &=& B\big|_{r=0}+ \frac{d B}{dr}\Big|_{r=0}  \cdot r + \frac{d^2 B}{dr^2}\Big|_{r=0} \cdot \frac{r^2}{2}+... \nonumber \\
	&=& -2p_E +\left(2p_S-6p_E \right) \cdot r + \left( 12p_S - 20 p_E \right) \cdot \frac{r^2}{2} + ... \nonumber \\
	&=& -2p_E +\left(2p_S-6p_E \right) \cdot \frac{m_E}{m_S} + \left( 6p_S - 10 p_E \right) \cdot \left( \frac{m_E}{m_S}\right)^2+ ... \nonumber \\
	&\sim& -2p_E +\left(2p_S-6p_E \right) \cdot \frac{m_E}{m_S} \nonumber \\
	&=& 2 p_S \frac{m_E}{m_S} - \left( 2 + 6 \frac{m_E}{m_S} \right) p_E,
\end{eqnarray}
where we used eqn. \ref{eqn:massapprox} to neglect the terms of order two or higher in $m_E/m_S$. Finally, we consider the coefficient $C$, given by
\begin{equation}
C = \left(  \frac{m_S+m_E}{m_S-m_E} \right)  \frac{2m_S^2 p_E^2-4m_Em_Sp_Ep_S+2m_E^2p_S^2}{\left( m_E - m_S \right)^2}.
\end{equation}
In the same way we worked out coefficient $B$, we make the substitution $m_E = rm_S$, which gives us
\begin{equation}
C= \left(  \frac{1+r}{1-r} \right)  \frac{2 p_E^2-4rp_Ep_S+2r^2p_S^2}{\left( r - 1 \right)^2},
\end{equation}
\begin{equation}
\frac{d C}{dr} = \frac{4 \left( p_E^2 (2+r) + p_S^2 r (1+2 r) - p_E p_S \left(1+4r +r^2\right) \right)}{(r-1)^4},
\end{equation}
and
\begin{equation}
\frac{d^2 C}{dr^2} = \frac{4 \left( 2 p_E p_S \left( 4 + 7r +r^2 \right) - 3 p_E^2 (3+r ) - p_S^2 \left( 1 + 7r +4r^2 \right) \right)}{(r-1)^5}.
\end{equation}
When we take the Taylor series expansion of $C$ in $r$ about $r=0$, we have
\begin{eqnarray}
C &=& C\big|_{r=0}+ \frac{d C}{dr}\Big|_{r=0}  \cdot r + \frac{d^2 C}{dr^2}\Big|_{r=0} \cdot \frac{r^2}{2}+... \nonumber \\
	&=& 2p_E^2 +\left(8 p_E^2- 4 p_E p_S \right) \cdot r + \left( 36 p_E^2 - 32 p_E p_S + 4 p_S^2 \right) \cdot \frac{r^2}{2} + ... \nonumber \\
	&=& 2p_E^2 +\left(8 p_E^2- 4 p_E p_S \right) \cdot \frac{m_E}{m_S} +\left( 18 p_E^2 - 16 p_E p_S + 2 p_S^2 \right) \cdot \left( \frac{m_E}{m_S}\right)^2+ ... \nonumber \\
	&\sim&  2p_E^2 +\left(8 p_E^2- 4 p_E p_S \right)\cdot \frac{m_E}{m_S} \nonumber \\
	& = & \left( 2 + 8 \frac{m_E}{m_S} \right) p_E^2 - 4 p_E p_S \frac{m_E}{m_S}.
\end{eqnarray}
Thus, using eqn. \ref{eqn:deltawap1}, we can write $\Delta W$ as 
\begin{equation} \label{eqn:wignersimple}
\Delta W \sim X + Y+ Z,
\end{equation}
where 
\begin{equation}
X =\left(2 \frac{m_E}{m_S} \right) \int dp_E dx_E \cdot W_E(x_E,p_E) W_S(x_S,p_S),
\end{equation}
\begin{equation}
Y = \left(2 p_S \frac{m_E}{m_S} \right)  \int dp_E dx_E \cdot W_E(x_E,p_E) \partial_{p_S}W_S(x_S,p_S)  - \left(2+6 \frac{m_E}{m_S} \right) \int dp_E dx_E \cdot p_E W_E(x_E,p_E)  \partial_{p_S}W_S(x_S,p_S), 
\end{equation}
and
\begin{equation}
Z =\left( 2 + 8 \frac{m_E}{m_S} \right)  \int dp_E dx_E \cdot p_E^2 W_E(x_E,p_E) \partial_{p_S}^2W_S(x_S,p_S)  - 4 p_S \frac{m_E}{m_S} \int dp_E dx_E \cdot p_E W_E(x_E,p_E)  \partial_{p_S}^2W_S(x_S,p_S). 
\end{equation}
Now, we recall from our preliminary discussion on the marginal distributions of the Wigner distribution in section \ref{sec:marginals} that
\begin{eqnarray}
\int dp_E dx_E \cdot O(p_E) W_E(x_E,p_E) W_S(x_S,p_S) 
&=& W_S(x_S,p_S) \int dp_E \cdot O(p_E) \int  dx_E \cdot W_E(x_E,p_E) \nonumber \\
&=& W_S(x_S,p_S) \int dp_E \cdot O(p_E) \tilde \rho\left( p_E, p_E \right)  \nonumber \\
&=& W_S(x_S,p_S) \mathrm{Tr} \left(\hat O \hat{\rho} \right) \nonumber \\
&=& W_S(x_S,p_S) \left< \hat O \right>,
\end{eqnarray}
where $O$ is an observable. Hence, our previous calculations yield
\begin{equation}
X = \left(2 \frac{m_E}{m_S} \right) \left< 1 \right> W_S(x_S,p_S)= 2 \frac{m_E}{m_S}W_S(x_S,p_S),
\end{equation}
\begin{equation}
Y = \left(2 p_S \frac{m_E}{m_S} \right) \partial_{p_S} W_S(x_S,p_S) - \left(2+6 \frac{m_E}{m_S} \right) \left< p_E \right> \partial_{p_S} W_S(x_S,p_S),
\end{equation}
and 
\begin{equation}
Z = \left( 2 + 8 \frac{m_E}{m_S} \right) \left< p_E^2 \right>  \partial_{p_S}^2 W_S(x_S,p_S)- 4 p_S \frac{m_E}{m_S}\left< p_E\right> \partial_{p_S}^2 W_S(x_S,p_S).
\end{equation}
However, originally we considered the environment as an ideal (one-dimensional) gas of environment particles, so it is reasonable to assume that any measurement\index{Measurement} of an environment particle momentum is equally likely to be in the opposite direction, i.e. $\left< p_E \right> = 0$. Eqn. \ref{eqn:wignersimple} then becomes
\begin{equation}\label{eqn:peavgiszero}
\Delta W \sim 2 \frac{m_E}{m_S} W_S(x_S,p_S) + \left(2 p_S \frac{m_E}{m_S} \right) \partial_{p_S} W_S(x_S,p_S)+\left( 2 + 8 \frac{m_E}{m_S} \right) \left< p_E^2 \right>  \partial_{p_S}^2 W_S(x_S,p_S).
\end{equation}
We notice that 
\begin{eqnarray}
2 \frac{m_E}{m_S} W_S(x_S,p_S) + \left(2 p_S \frac{m_E}{m_S} \right) \partial_{p_S} W_S(x_S,p_S) &=& 2 \frac{m_E}{m_S} \left( W_S(x_S,p_S) + p_S \partial_{p_S} W_S(x_S,p_S) \right) \nonumber \\
&=& 2 \frac{m_E}{m_S} \partial_{p_S} \left( p_S W_S(x_S,p_S) \right),
\end{eqnarray}
so we write the change in the Wigner distribution of the system due to one environmental collision as
\begin{boxedeqn}{eqn:environcol}
\Delta W \sim 2 \frac{m_E}{m_S} \partial_{p_S} \left( p_S W_S(x_S,p_S) \right)+\left( 2 + 8 \frac{m_E}{m_S} \right) \left< p_E^2 \right>  \partial_{p_S}^2 W_S(x_S,p_S).
\end{boxedeqn}

\section{The Master Equation for Quantum Brownian Motion}
In our simple model, the system is only under the influence of environmental particles, and is free otherwise. Thus, the total change in the system's Wigner distribution with time is given by its free particle term added to some contribution due to the environment. Since the environment acts on the system through collisions, if we define $\Gamma$ to be the statistical number of collisions per unit time between the system and environmental particles, we combine eqns. \ref{eqn:wigfree} and \ref{eqn:environcol} to get \cite{halliwell}\index{Master Equation!Wigner Form}
\begin{equation}
\frac{\partial W_S}{\partial t} = - \frac{p_s}{m_S} \partial_{x_S} W(x_S,p_S,t) + \Gamma \left( 2 \frac{m_E}{m_S} \partial_{p_S} \left( p_S W_S(x_S,p_S) \right)+\left( 2 + 8 \frac{m_E}{m_S} \right) \left< p_E^2 \right>  \partial_{p_S}^2. W_S(x_S,p_S) \right),
\end{equation}
an expression for the total change in the system's Wigner distribution with time. We use table \ref{tab:inversions} to convert our equation for the Wigner distribution to an equation for the state operator of the system. This is
\begin{eqnarray}
\mathcal W \left( \partial_t \rho_S(x,y) \right) &=& \frac{1}{m_S} \mathcal W \left( \frac{i}{2} \left( \partial_x^2 - \partial_y^2 \right) \rho_S(x,y) \right) - \Gamma \frac{m_E}{m_S} \mathcal W \left(  (x-y) \left( \partial_x - \partial_y \right) \rho_S (x,y)  \right) \nonumber \\ &-& \Gamma \left( 2 + 8 \frac{m_E}{m_S} \right)\left< p_E^2 \right> \mathcal W \left(  \left(x - y  \right)^2 \rho_S( x, y)   \right). 
\end{eqnarray}
Noting that this is true for all $\rho_S (x,y)$, we have
\begin{equation}
\partial_t \rho_S(x,y) =  \frac{i}{2m_S} \left( \partial_x^2 - \partial_y^2 \right) \rho_S(x,y) - \Gamma \frac{m_E}{m_S}  (x-y) \left( \partial_x - \partial_y \right) \rho_S (x,y)  - \Gamma \left( 2 + 8 \frac{m_E}{m_S} \right) \left< p_E^2 \right> \left(x - y  \right)^2 \rho_S( x, y).
\end{equation}
We take the standard definition for the dissipation\index{Dissipation} rate $\gamma$ to be \cite{halliwell}
\begin{equation} \label{eqn:dissipation}
\gamma \equiv \frac{m_E}{m_S} \Gamma,
\end{equation}
so 
\begin{equation}
\partial_t \rho_S(x,y) =  \frac{i}{2m_S} \left( \partial_x^2 - \partial_y^2 \right) \rho_S(x,y) - \gamma  (x-y) \left( \partial_x - \partial_y \right) \rho_S (x,y)  -\gamma \frac{m_S}{m_E} \left( 2 + 8 \frac{m_E}{m_S} \right) \left< p_E^2 \right>  \left(x - y  \right)^2\rho_S( x, y).
\end{equation}
To express this result in standard form, we use the definition for temperature in one dimension from statistical mechanics, which is \cite{kittelkroemer}\index{Temperature}
\begin{equation} \label{eqn:temp}
\frac{1}{2} k T\equiv \frac{\left< p_E^2 \right> }{2 m_E},
\end{equation} 
 where $T$ is temperature and $k$ is the Boltzmann constant. Using this definition, we examine the last term more closely and find
\begin{eqnarray}
\gamma \frac{m_S}{m_E} \left( 2 + 8 \frac{m_E}{m_S} \right) \left< p_E^2 \right>  \left(x - y  \right)^2\rho_S( x, y)
&=& \gamma \frac{m_S}{m_E} \left( 2 + 8 \frac{m_E}{m_S} \right) m_E k T  \left(x - y  \right)^2\rho_S( x, y) \nonumber \\
&=& \gamma \left( 2m_S + 8 m_E \right) k T  \left(x - y  \right)^2\rho_S( x, y) \nonumber \\
&\sim& \gamma   2m_S k T  \left(x - y  \right)^2\rho_S( x, y),
\end{eqnarray}
where we have used the fact that $m_E \ll m_S$. Thus, our final result is \cite{halliwell}\index{Master Equation!State Operator Form}
\begin{boxedeqn}{eqn:masterequation}
\partial_t \rho_S(x,y) =  \frac{i}{2m_S} \left( \partial_x^2 - \partial_y^2 \right) \rho_S(x,y) - \gamma  (x-y) \left( \partial_x - \partial_y \right) \rho_S (x,y)  -  2m_S \gamma k T  \left(x - y  \right)^2\rho_S( x, y),
\end{boxedeqn}
which is the accepted master equation for quantum Brownian motion \cite{omnes, zurek, halliwell}. Using dimensional analysis, we can reinsert $\hbar$ to bring the master equation into SI units. This is
\begin{boxedeqn}{eqn:masterequationSI}
\partial_t \rho_S(x,y) =  \frac{i}{2m_S \hbar } \left( \partial_x^2 - \partial_y^2 \right) \rho_S(x,y) - \gamma  (x-y) \left( \partial_x - \partial_y \right) \rho_S (x,y)  -  \frac{2m_S}{\hbar^2} \gamma k T  \left(x - y  \right)^2\rho_S( x, y).
\end{boxedeqn}
The assumptions used to derive this equation are listed in table \ref{tab:asum}.

\begin{table}[h] \caption{Assumptions used for the derivation of eqn. \ref{eqn:masterequation} \label{tab:asum}}\centering  
\begin{tabular}{|ccc|}
\hline Assumption & Equation & Label \\ 
\hline Small mass ratio & $m_E/m_S \ll 1$ &  \ref{eqn:massapprox} \\
Locality & $\left| x_S-x_E  \right| \ll \left| x_S \right| $ & \ref{eqn:locality1} \\ 
Statistical environment & $\left<p_E\right> = 0$ & \ref{eqn:peavgiszero} \\
Dissipation & $ \gamma = m_E/m_S \cdot  \Gamma$ & \ref{eqn:dissipation}\\
Temperature  & $1/2 \cdot k T = \left< p_E^2 \right>/(2m_E)$ & \ref{eqn:temp} \\
\hline 
\end{tabular} 
\end{table}

\chapter{Consequences of the Master Equation}\label{chap:applications}
\lettrine[lines=2, lhang=0.33, loversize=0.1]{W}e now explore the physical ramifications of the master equation for quantum Brownian motion, developed in the previous chapter. First, we investigate its physical meaning term by term. Next, we consider the simple example of a quantum harmonic oscillator undergoing decoherence. Finally, we offer some closing remarks on decoherence theory in general and suggestions for further reading.
\section{Physical Significance of the first two terms}
In the realm of master equations, eqn. \ref{eqn:masterequation} for quantum Brownian motion actually is \textit{simple} \cite{zurek}. Even so, the purpose of each term is not immediately obvious. In this section, we examine the physical meaning of the first and second terms. The first term is the free system evolution, as it is the transform of eqn. \ref{eqn:wigfree}. It does not hurt to verify this explicitly, without employing the Wigner distribution. If we switch to SI units via eqn. \ref{eqn:masterequationSI}, the first term is
\begin{eqnarray}
 \frac{i\hbar}{2m_S} \left( \partial_x^2 - \partial_y^2 \right) \rho_S(x,y). 
 &=& \frac{i\hbar}{2m_S} \left( \partial_x^2 - \partial_y^2 \right) \left< x \right| \hat{\rho_S} \left| y \right> \nonumber \\
 &=& \frac{i\hbar}{2m_S} \partial_x^2 \left< x \right| \hat{\rho_S} \left| y \right> - \frac{i\hbar}{2m_S} \partial_y^2 \left< x \right| \hat{\rho_S} \left| y \right> \nonumber \\
&=& \frac{i\hbar}{2m_S}\frac{-1}{\hbar^2} \left( \frac{\hbar}{i}  \partial_x \right) ^2 \left< x \right| \hat{\rho_S} \left| y \right> - \frac{i\hbar}{2m_S} \frac{-1}{\hbar^2} \left( \frac{\hbar}{i}  \partial_y \right) ^2 \left< x \right| \hat{\rho_S} \left| y \right> \nonumber \\
&=& \frac{i\hbar}{2m_S}\frac{-1}{\hbar^2} \check P_x^2 \left< x \right| \hat{\rho_S} \left| y \right> - \frac{i\hbar}{2m_S} \frac{-1}{\hbar^2} \check P_y ^2 \left< x \right| \hat{\rho_S} \left| y \right> \nonumber \\
&=&- \frac{i}{2m_S\hbar}\left(\check P_x^2 \left< x \right|\right) \left( \hat{\rho_S} \left| y \right>\right) +\frac{i}{2m_S\hbar}   \left( \left< x \right| \hat{\rho_S} \right) \left(\check P_y ^2 \left| y \right> \right) \nonumber \\
&=&- \frac{i}{2m_S\hbar}\left( \left< x \right| \hat P^2 \right) \left( \hat{\rho_S} \left| y \right>\right) +\frac{i}{2m_S\hbar}   \left( \left< x \right| \hat{\rho_S} \right) \left(\hat P ^2 \left| y \right> \right) \nonumber \\
&=&- \frac{i}{\hbar}\left(  \left< x \right|  \frac{\hat P^2}{2m_S} \hat{\rho_S} \left| y \right> - \left< x \right| \hat{\rho_S}  \frac{\hat P ^2}{2m_S} \left| y \right> \right)\nonumber \\
&=& -\frac{i}{ \hbar  }\left< x \right| \left(    \frac{\hat P^2}{2m_S} \hat{\rho_S} - \hat{\rho}_S  \frac{\hat P ^2}{2m_S} \right)  \left| y \right>.
  \end{eqnarray}
  By eqn. \ref{eqn:eop}, the free system (for which $V=0$) has a Hamiltonian of
  \begin{equation}
  \hat H_f = \frac{\hat P^2}{2m_s},
 \end{equation}
so our equation becomes
 \begin{equation}
 -\frac{i}{ \hbar  }\left< x \right| \left(   \hat H_f \hat{\rho_S} - \hat{\rho_S}  \hat H_f \right)  \left| y \right> = \left< x \right|  \frac{i}{ \hbar} \left[ \hat{\rho}_S ,\hat H_f \right] \left| y \right>,
 \end{equation}
 which is 
 \begin{equation}
 \left< x \right| \hat{\partial}_t \hat{\rho}_S  \left| y \right> = \check{\partial}_t \rho_S(x,y)
 \end{equation}
 by eqn. \ref{eqn:heispic}. Thus, we confirm that the first term in the master equation is the free evolution of the state operator.

The second term is not so obvious, and turns out to be responsible for damping our system's motion. To explain this, we use the master equation to calculate the rate of change of the expectation value of momentum due to the second term. In the position basis, the second term reduces to \cite{omnes}
\begin{eqnarray}
\partial_t \left< \hat P \right>_2 
&=&  \partial_t  \mathrm{Tr}\left(\hat P \hat \rho \right)_2 \nonumber \\
&=&   \mathrm{Tr}\left(\hat P  \partial_t \hat \rho \right)_2 \nonumber \\
&=& - \gamma \mathrm{Tr} \left( \check P _x \gamma (x-y) \left( \partial_x - \partial_y \right) \rho(x,y) \right) \nonumber \\
&=& - \gamma \mathrm{Tr} \left( \frac{1}{i} \partial_x\left[  \gamma(x-y) \left( \partial_x - \partial_y \right) \rho(x,y)\right] \right) \nonumber \\
&=& - \gamma \mathrm{Tr} \left( \frac{1}{i} \left( \partial_x - \partial_y \right) \rho(x,y) \right)- \gamma \mathrm{Tr} \left( \frac{1}{i} (x-y) \left( \partial_x^2 - \partial_x \partial_y \right) \rho(x,y) \right)  \nonumber \\
&=&- \gamma \int dx \cdot \frac{1}{i} \left( \partial_x - \partial_y \right) \rho(x,x) + \gamma \int dx \cdot \frac{1}{i} (x-x) \left( \partial_x^2 - \partial_x \partial_y \right) \rho(x,x)  \nonumber \\
&=&- \gamma  \int dx \cdot \frac{1}{i} \left( \partial_x - 0 \right) \rho(x,x) + 0  \nonumber \\
&=& - \gamma \mathrm{Tr} \left( \frac{1}{i} \partial_x \rho(x,y) \right) \nonumber \\
&=& - \gamma \mathrm{Tr} \left(\hat P  \hat \rho \right) \nonumber \\
&=& - \gamma \left< \hat P \right> ,
\end{eqnarray}
which is
\begin{boxedeqn}{}
\partial_t \left< \hat P \right>_2  = - \gamma \left< \hat P \right>.
\end{boxedeqn}
Hence, the contribution to the rate of change of momentum of the second term is the dissipation\index{Dissipation} (a scalar) times the momentum, pointed in the opposite direction as the momentum. This is precisely a damping effect, which is what we wanted to show \cite{thornton}.

\section{The Decoherence Term}
\begin{figure}[p] 
\begin{center} 
\includegraphics[width=0.9 \linewidth]{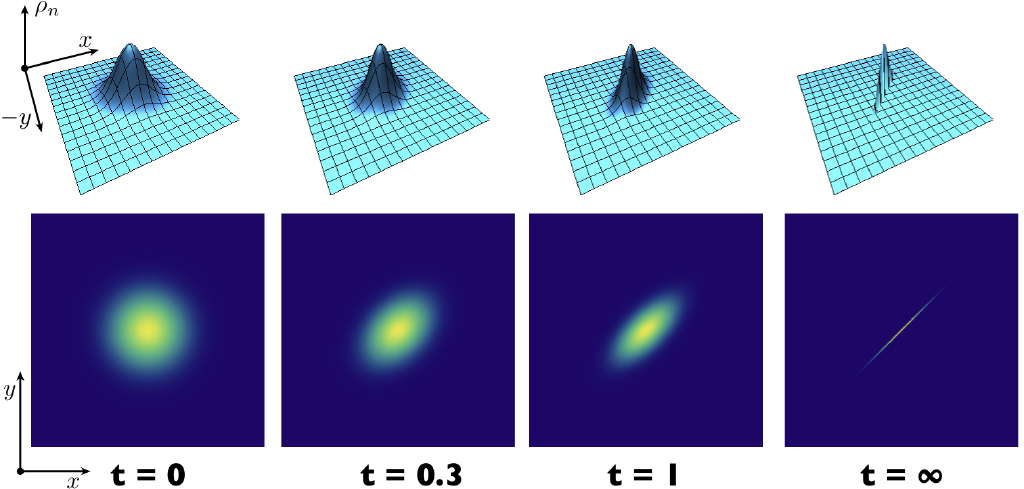}
\end{center} 
\caption[Decoherence in the ground state of the quantum harmonic oscillator]{The decoherence of the ground state of the quantum harmonic oscillator under the simplified master equation \ref{eqn:shodecoh}. In the density plots, yellow indicates maximum values, while blue indicates minimum.}\label{fig:decoherence0}
\end{figure}
The last term of the master equation turns out to cause decoherence of the system, so it is central to our discussion. To interpret it, we will make some reasonable approximations. To get a better idea of the relative size of the terms, we use the SI version of the master equation, eqn. \ref{eqn:masterequationSI}. Notice that the last term contains a numerical factor of $1/\hbar^2 \approx 10^{68}$, while the other terms are either first or zeroth order in $1/\hbar$. Thus, we surmise that for sufficiently large $\left| x - y \right|$, the last term will dominate equation.\footnote{In the matrix representation of a state operator, this corresponds to the \textit{off-diagonal} elements. Recall that the totally mixed state in eqn. \ref{eqn:thisisamixture} had a diagonal state operator. This confirms that decoherence works on the off-diagonal elements of the state operator.} Hence, our drastically simplified master equation is \cite{omnes}
\begin{boxedeqn}{eqn:simplifiedmaster}
\partial_t \rho_S(x,y) \sim -  \frac{2m_S \gamma k T}{\hbar^2}  \left(x - y  \right)^2\rho_S( x, y),
\end{boxedeqn}
which has the standard solution
\begin{equation} \label{eqn:approxdecoherence}
\rho_S(x,y,t) = \rho_S(x,y,0) e^{ -\frac{2m_S \gamma k T}{\hbar^2}\left(x - y  \right)^2t}.
\end{equation}
Since the argument of the exponential must be dimensionless, $\frac{\hbar^2}{2m_S \gamma k T\left(x - y  \right)^2}$ has units of time. Customarily, we identify \cite{omnes}\index{Decoherence Time}
\begin{boxedeqn}{}
t_d \equiv \frac{\hbar^2}{2m_S \gamma k T\left(x - y  \right)^2}
\end{boxedeqn}
as the (characteristic) decoherence time of the system, which is its $e$-folding time.\footnote{When $t=t_d$, $\rho(x,y,t_d) = \frac{1}{e} \rho(x,y,t)$.} Notice also that the decoherence time varies with location in state-space, as it depends on both $x$ and $y$. Thus, we are not surprised to find that some regions decay faster than others. Further, since $\hbar^2 \approx 10^{-68}$ in SI units, the decoherence time for any reasonably large system is incredibly small.\footnote{For example, if we suppose our environment is an ideal, one dimensional gas at room temperature with a mass of $10^{-26}$ kg per particle and a collision rate with the system of $\Gamma \approx 10^{10}$ collisions per second (atmospheric conditions), we find the decoherence time of the system for length scales of nanometers to be of order $t_d =\frac{\hbar^2}{2 m_E \Gamma k T (x-y)^2} \approx \frac{10^{-68}}{2 \cdot 10^{-26} \cdot 10^{10} \cdot 10^{-23} \cdot 300 \cdot 10^{-9}} \approx 10^{-19	}$ s.} Next, we consider an example to show how decoherence operates on a simple situation.
\section{Example: The Harmonic Oscillator in a Thermal Bath}
\begin{figure}[p] 
\begin{center} 
\includegraphics[width=0.9 \linewidth]{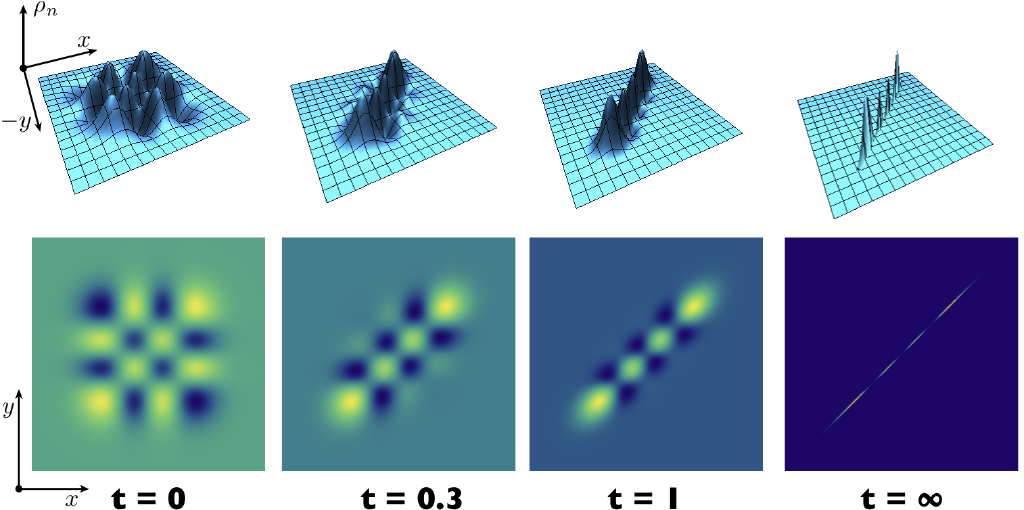}
\end{center} 
\caption[Decoherence in the third excited state of the quantum harmonic oscillator]{The decoherence of the third excited state of the simple harmonic oscillator under the simplified master equation \ref{eqn:shodecoh}. In the density plots, yellow indicates maximum values, while blue indicates minimum.}\label{fig:decoherence3} 
\end{figure}
So far, we have supposed that $\rho_S$ is a free particle. However, note that our simplified master equation, eqn. \ref{eqn:simplifiedmaster}, does not explicitly depend on the system's Hamiltonian (this was contained in the first term), so we are free to replace our initial state operator with some other state operator of a different system. We choose, due to its utility and familiarity, the harmonic oscillator. From our work in section \ref{sec:wig_harmonic}, we know that state operator for the harmonic oscillator, eqn. \ref{eqn:stateopharmonic}, is
\begin{equation}
\rho_n(x,y,t=0) =  \frac{1}{n!} \left( \frac{m \omega}{\pi} \right)^{1/2} \left( \frac{1}{2 m \omega}\right)^n\left( m \omega x - \partial_x  \right)^n \left( m \omega y - \partial_y  \right)^n  e^{-\frac{m \omega}{2}x^2} e^{-\frac{m \omega}{2}y^2}.
\end{equation}
If we place this state operator in a thermal bath, we expect the system to evolve approximately according to eqn. \ref{eqn:approxdecoherence}, so the time dependent state operator of the harmonic oscillator is\index{Harmonic Oscillator!Decoherence of}
\begin{boxedeqn}{eqn:shodecoh}
\rho_n(x,y,t) =  \frac{1}{n!} \left( \frac{m \omega}{\pi} \right)^{1/2} \left( \frac{1}{2 m \omega}\right)^n\left( m \omega x - \partial_x  \right)^n \left( m \omega y - \partial_y  \right)^n  e^{-\frac{m \omega}{2}x^2} e^{-\frac{m \omega}{2}y^2}e^{ -2m_S \gamma k T\left(x - y  \right)^2t}.
\end{boxedeqn}
In figures \ref{fig:decoherence0} and \ref{fig:decoherence3}, we plot the state operators for $n=0$ and $n=3$. As is evident from the form of eqn. \ref{eqn:shodecoh}, the off-diagonal matrix elements (when $x \neq y$) quickly vanish with time. Physically, the off-diagonal elements of the state operator represent the quantum interference terms, terms that can interact only with other quantum systems. These interference terms are what give the entangled states we explored in sections \ref{sec:quantumsup} and \ref{sec:bellstate} their interesting qualities. By zeroing the off-diagonal elements, we take a quantum mechanical system and force it into a classical distribution. 

As it turns out, this interpretation becomes obvious as $t \rightarrow \infty$. By eqn. \ref{eqn:shodecoh}, this is
\begin{figure}[b] 
\begin{center} 
\includegraphics[width=0.9 \linewidth]{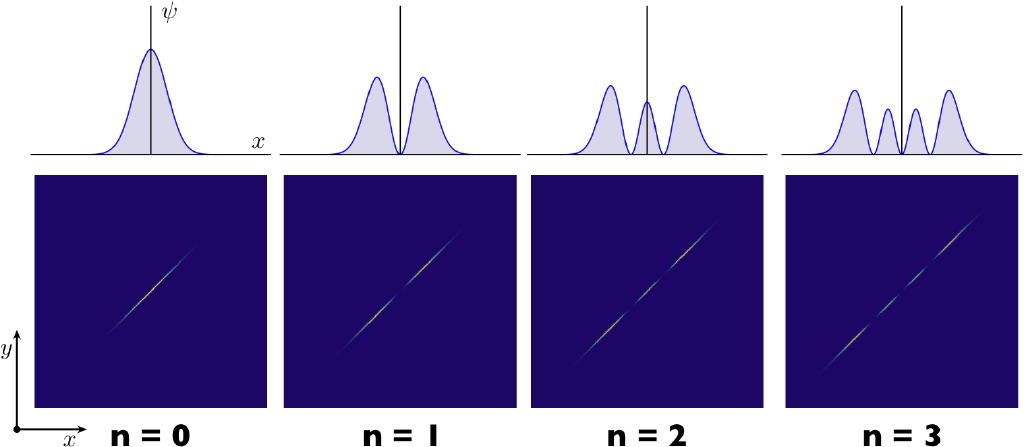}
\end{center} 
\caption[Decoherence of the quantum harmonic oscillator as $t \rightarrow \infty$]{The final state reached by the first four energy states of the simple harmonic oscillator under the simplified master equation \ref{eqn:shodecoh}. The two-dimensional plots coincide with the diagonal of the density plots.}\label{fig:decoherenceall} 
\end{figure}
\begin{equation}
\lim_{t \rightarrow \infty} \rho(x,y,t) = 
\begin{cases} 
0 & \text{if $x\neq y$}, \\ 
\psi^*(x)\psi(x) = \left| \psi(x) \right|^2 & \text{if $x =y$}, 
\end{cases} 
\end{equation}
as shown in figure \ref{fig:decoherenceall}. This quantity is a statistical probability distribution, and as we saw with the roulette wheel at the beginning of this thesis, decoherence has effectively blocked us from accessing any of the quantum mechanical information present in our initial system.

\section{Concluding Remarks}
We have now developed and applied the master equation for quantum Brownian motion, and used it to clarify how a macroscopic, classical object might emerge from quantum mechanics. We started by setting the stage with the mathematics and formalism we would need to develop quantum mechanics. Then, we used the tools we made to derive the Schr\"odinger equation and the equation of motion for the state operator. 

We then shifted and considered quantum mechanics in phase-space, where the central object is the Wigner distribution. Next, we explored some of its key properties and described and example of its application using the harmonic oscillator. After that, we used it to derive the simple master equation for one-dimensional quantum Brownian motion. We explained each of the terms physically, and finally considered an example of decoherence, where the master equation transformed a quantum harmonic oscillator into a classical probability distribution.

The debate still rages in the physics community; does decoherence theory \textit{solve} the philosophical problems brought about by paradoxes like Schr\"odinger's cat\index{Schr\"odinger's cat}, or does it merely postpone the problem, pushing the fundamental issue into an environmental black box \cite{meuffels,hobson,schlosshauer}?  Regardless, it provides a practical framework for performing objective measurements\index{Measurement} without an observer, which is of key importance to the emerging fields of quantum computation and quantum information. Current efforts are underway to probe decoherence directly, both experimentally and theoretically. Through the use of mesoscopic systems, scientists have been able to manufacture tiny oscillators that are getting very close to the quantum regime \cite{lahaye, blencowe}. Theoretical predictions of what should be observed at the quantum-classical barrier have also been made, with the promise of experimental feasibility within a few years \cite{katz}. 

Just last year, scientists performed experiments involving ultra-cold chlorophyll, confirming that even photosynthesis is a quantum-emergence phenomenon, and thus governed by decoherence theory \cite{engel}. The group went so far as to suggest that chloroplasts were actually performing quantum computation algorithms on themselves to speed-up reaction times. This idea of selective self-measurement is intriguing, but largely undeveloped theoretically. It, along with the many other application areas of quantum decoherence theory, are sure to occupy physicists for years to come.

\index{Mixture|see{State, Impure}}
\index{Mixed State|see{State, Impure}}
\index{Dual Vector|see{Linear Functional}}
\index{Energy Operator|see{Hamiltonian}}
\index{Operator|see{Linear Operator}}
\index{Vector Space|see{Linear Vector Space}}
\index{Basis!of Eigenvectors|see{Spectral Theorem}}
\index{Free Particle!Wigner Distribution|see{Wigner}}

\backmatter




\renewcommand\bibname{References} 

\nocite{*} 
\bibliographystyle{apsrev.bst} 
\bibliography{gamble} 

%
%
%

\if@xetex
	\cleardoublepage
	\phantomsection
	\addcontentsline{toc}{chapter}{Index}
\else
	\ifpdf
		\cleardoublepage
		\phantomsection
		\addcontentsline{toc}{chapter}{Index}
	\else
		\cleardoublepage
		\addcontentsline{toc}{chapter}{Index}
	\fi
\fi
\printindex
\end{document}